\documentclass[%
11pt,tightenlines,
onecolumn,aip,jmp,amsmath,amssymb,eqsecnum,amsfonts,
nofootinbib
]{revtex4-2}

\usepackage{graphicx}
\usepackage{dcolumn}
\usepackage{bm}

\begin{document}

\title{Power law duality in classical and quantum mechanics}

\author{Akira Inomata}
\email{ainomata@albany.edu}
\affiliation{%
Department of Physics, State University of New York at Albany,\\
Albany, NY 12222, USA
}%

\author{Georg Junker}\email{gjunker@eso.org}\altaffiliation[Also at: ]{Institut f\"ur Theoretische Physik I, Universit\"at Erlangen- \\
\indent N\"urnberg, Staudtstra{\ss}e 7, D-91058 Erlangen, Germany; georg.junker@fau.de}
\affiliation{European Organization for Astronomical Research in the Southern Hemisphere,
Karl-Schwarzschild-Stra{\ss}e 2, D-85748 Garching, Germany
}%
\date{\today}

\begin{abstract}
The Newton--Hooke duality and its generalization to arbitrary power laws in classical, semiclassical and quantum mechanics are discussed. We pursue a view that the power-law duality is a symmetry of the action under a set of duality operations. The power dual symmetry is defined by invariance and reciprocity of the action in the form of Hamilton's characteristic function. We find that the power-law duality is basically a classical notion and breaks down at the level of angular quantization. We propose an ad hoc procedure to preserve the dual symmetry in quantum mechanics. The energy-coupling exchange maps required as part of the duality operations that take one system to another lead to an energy formula that relates the new energy to the old energy. The transformation property of {the} Green function satisfying the radial Schr\"odinger equation yields a formula that relates the new Green function to the old one.
The energy spectrum of the linear motion in a fractional power potential is semiclassically evaluated. We find a way to show the Coulomb--Hooke duality in the supersymmetric semiclassical action. We also study the confinement potential problem with the help of the dual structure of a two-term power potential.
\end{abstract}
\keywords{power-law duality; classical and quantum mechanics; semiclassical quantization; supersymmetric quantum mechanics; quark confinement}
\maketitle

\newcommand{\rmi}{\textrm{i}}
\newcommand{\rmd}{\textrm{d}}
\newcommand{\balpha}{\bm{\alpha}}
\newcommand{\bsigma}{\bm{\sigma}}
\newcommand{\rme}{\textrm{e}}
\newcommand{\JPA}{J.\ Phys.\ A}
\newcommand{\RMP}{Rev.\ Mod.\ Phys.\ }
\newcommand{\PRL}{Phys.\ Rev.\ Lett.\ }
\newcommand{\JMP}{J.\ Math.\ Phys.\ }
\renewcommand*{\thefootnote}{\fnsymbol{footnote}}

\section{Introduction}\label{Section1}
In recent years, numerous exoplanets have been discovered. One of the best Doppler spectrographs to discover low-mass exoplanets using the radial velocity method are HARPS (High Accuracy Radial Velocity Planet Searcher) installed on ESO's 3.6 m telescope at La Silla and ESPRESSO (Echelle Spectrograph for Rocky Exoplanet- and Stable Spectroscopic Observations) installed on  ESO's VLT at Paranal Observatory in Chile. See, e.g.,~\cite{ESO1,ESO2}. NASA's Kepler space telescope has discovered more than half of the currently known exoplanets using the so-called transit method. {See, e.g.,~\cite{NASA1,NASA2}.
For some theoretical work on planetary systems see, e.g., \cite{Theo}.
In~exoplanetary research it is a generally accepted view that Newton's law of gravitation holds in extrasolar systems~\cite{Wald}.  Orbit mechanics of exoplanets, as~is the case of solar planets and satellites, is classical mechanics of the Kepler problem under small perturbations. The~common procedure for the study of perturbations to  the Kepler motion is the so-called {regularization}, introduced by Levi--Civita (1906) for the planar motion~\cite{LevCiv06,LevCiv20} and generalized by Kustaanheimo and Stiefel (1965) to the spatial motion~\cite{KustStief}. The~regularization in celestial mechanics is a transformation of the singular equation of motion for the Kepler problem to the non-singular equation of motion for the harmonic oscillator problem with or without perturbations. It identifies the Kepler motion with the harmonic oscillation, assuring the dual relation between Newton's law and Hooke's law here, following the tradition, we mean by Newton's law the inverse-square force law of gravitation and by Hooke's law the linear force law for the harmonic oscillation. Although~Hooke found the inverse square force law for gravitation prior to Newton, he was short of skills in proving that the orbit of a planet is an ellipse in accordance with Kepler's first law, while Newton was able not only to confirm that the inverse square force law yields an elliptic orbit but also to show conversely that the inverse square force law follows Kepler's first law. History gave Newton the full credit of the inverse square force law for gravitation. For~a detailed account, see, e.g.,~Arnold's book~\cite{Arno}}). The Newton--Hooke duality has been discussed by many authors from various aspects~\cite{Szeb,StiefelSchleifele}. The~basic elements of regularization are: (i) a transformation of space variables, \, (ii) interpretation of the conserved energy as the coupling constant, and~\, (iii) a transformation of time parameter. The~choice of space variables and time parameter is by no means unique.  The~transformation of space variables has been represented in terms of parabolic coordinates~\cite{LevCiv06,LevCiv20}, complex numbers~\cite{Sund, Bohl}, spinors~\cite{KustStief}, quaternions~\cite{Wald,Viva,Vrbik}, etc. The~time transformation used by Sundman~\cite{Sund,Saari} and by Bohlin~\cite{Bohl} (for Bohlin's theorem see also reference~\cite{Arno}) is essentially based on Newton's finding~\cite{Chand} that the areal speed $\rmd A/\rmd t$ is constant for any central force motion. It takes the form $\rmd s = Cr\rmd t$ where $s$ is a {fictitious time} related to the eccentric anomaly. To~improve numerical integrations for the orbital motion, a~family of time transformations $\rmd s = C_{\eta}r^{\eta} \rmd t$, called generalized Sundman transformations, has also been discussed~\cite{JanBon}, in~which $s$ corresponds to the mean anomaly if $\eta = 0$, the~eccentric anomaly if $\eta = 1$, the~true anomaly if $\eta = 2$, and~intermediate anomalies~\cite{Naco} for other values of $\eta$. Even more generalizing, a~transformation of the form $\rmd s = Q(r) \rmd t$ has been introduced in the context of regularization~\cite{FFS}.

As has been pointed out in the literature~\cite{Arno, Need, Need2, Chand, GranRosn},  the~dual relation between the Kepler problem and the harmonic oscillator was already known in the time of Newton and Hooke. What Newton posed in their {Principia} was more general. According to Chandrasekhar's reading~\cite{Chand} out of the propositions and corollaries (particularly Proposition VII, Corollary III) in the {Principia}, Newton established the duality between the centripetal forces of the form, $r^{\alpha}$ and $r^{\beta}$, for~the pairs $(\alpha, \beta) = (1, -2), (-1, -1)$ and $(-5, -5)$. Revisiting the question on the duality between a pair of arbitrary power forces, Kasner~\cite{Kasn} and independently Arnol'd~\cite{Arno} obtained the condition, $(\alpha + 3)(\beta + 3)=4$, for~a dual pair. There are a number of articles on the duality of arbitrary power force laws~\cite{HallJosi, Grandati}. Now on, for~the sake of brevity, we shall refer to the duality of general power force laws as the {power duality}. The~power duality includes the Newton--Hooke duality as a {special} case.

The quantum mechanical counterpart of the Kepler problem is the hydrogen atom problem. In~1926, Schr\"{o}dinger~\cite{Schr, Schr2} solved their equation for the hydrogen atom and successively for the harmonic oscillator. Although~it must have been known that both radial equations for the hydrogen atom and for the harmonic oscillation are reducible to confluent hypergeometric equations~\cite{Mann}, there was probably no particular urge to relate the Coulomb problem to the Hooke problem, before~the interest in the accidental degeneracies arose~\cite{Fock, Fock2, LapoRain}. Fock~\cite{Fock, Fock2} pointed out that for the bound states the hydrogen atom has a hidden symmetry $SO(4)$ and an appropriate representation of the group can account for the degeneracy.  In~connection with Fock's work, Jauch and Hill~\cite{JaucHill} showed that the $2-D$ harmonic oscillator has an algebraic structure of $su(2)$ which is doubly-isomorphic to the $so(3)$ algebra possessed by the $2-D$ hydrogen atom. The~transformation of the radial equation from the hydrogen atom to that of the harmonic oscillator or vice verse was studied by Schr\"{o}dinger~\cite{Schro41} and others, see Johnson’s article \cite{John} and references therein. The~same problem in arbitrary dimensions has also been discussed from the supersymmetric interest~\cite{KNT}.
In the post-{Kustaanheimo--Stiefel (KS)} era, the~relation between the three dimensional Coulomb problem and the four dimensional harmonic oscillator was also investigated by implementing the KS transformation or its variations in the Schr\"{o}dinger equation. See ref.\ \cite{Kibler} and references therein. The~duality of radial equations with multi-terms of power potentials was studied in connection with the quark confinement~\cite{QR,Gaze,John}.

The time transformation of the form $\rmd s= C_{\eta}r^{\eta} \rmd t$ used in classical mechanics is in principle integrable only along a classical trajectory. In~other words, the~fictitious time $s$ is globally meaningful only when the form of $r(t)$ as a function of $t$ is known. In~quantum mechanics, such a transformation is no longer applicable due to the lack of classical paths. Hence it is futile to use any kind of time transformation formally to the time-dependent Schr\"odinger equation. The~Schr\"odinger equation subject to the duality transformation is a time-independent radial equation possessing a fixed energy and a fixed angular momentum. The~classical time transformation is replaced in quantum mechanics by a renormalization of the time-independent state function~\cite{Junk}. In~summary, the~duality transformation applicable to the Schr\"odinger equation consists of (i) a change of radial variable, (ii) an exchange of energy and coupling constant, and~(iii) a transformation of state function. Having said so, when it comes to Feynman's path integral approach, we should recognize that the classical procedure of regularization~prevails.

Feynman's path integral is based on the $c$-number Lagrangian and, as~Feynman asserted~\cite{Feyn}, the~path of a quantum particle for a short time $\rmd t$ can be regarded as a classical path. Therefore, the~local time transformation associated with the duality transformation in classical mechanics can be revived in path integration. In~fact, the~Newton--Hooke duality plays an important role in path integration. Feynman's path integral in the standard form~\cite{Feyn,Schu} provides a way to evaluate the transition probability from a point to another in space (the propagator or the Feynman kernel). The~path integral in the original formulation gives exact solutions only for quadratic systems including the harmonic oscillator, but~fails in solving the hydrogen atom problem. However, use of the KS transformation enables to convert the path integral for the hydrogen atom problem to that of the harmonic oscillator if the action of Feynman's path integral is slightly modified with a fixed energy term. In~1979, Duru and Kleinert~\cite{DuruKlein}, formally applying the KS transformation to the Hamiltonian path integral, succeeded to obtain the energy-dependent Green function for the hydrogen atom in the momentum representation. Again, with~the help of the KS transformation, Ho and Inomata (1982) \cite{HoIno} carried out detailed calculations of Feynman's path integral with a modified action to derive the energy Green function in the coordinate representation. In~1984, on~the basis of the polar coordinate formulation of path integral (1969) \cite{Peak}, without~using the KS variables, the~radial path integral for the hydrogen atom was transformed to that for the radial harmonic oscillator by Inomata for three dimensions~\cite{Ino1984} and by Steiner for arbitrary dimensions~\cite{Stein1984, Stein1984a}.  Since then a large number of examples have been solved by path integration~\cite{IKG1992,GrosSein}.  Applications of the Newton--Hooke duality in path integration include those to the Coulomb problem on uniformly curved spaces~\cite{BIJ1987, BIJ1990}, Kaluza--Klein monopole~\cite{IJ1991}, and~many others~\cite{GrosSein}. The~idea of classical regularization also helped to open a way to look at the path integral from group theory and harmonic analysis~\cite{BJ87,IJ1994,IKG1992}. The~only work that discusses a confinement potential in the context of path integrals is Steiner's~\cite{Stein1986}.

As has been briefly reviewed above, the~Newton--Hooke duality and its generalizations have been extensively and exhaustively explored.
In the present paper we pursue the dual relation (power-duality) between two systems with arbitrary power-law potentials from the symmetry point of view. While most of the previous works deal with equations of motion, we focus our attention on the symmetry of action integrals under a set of duality operations. Our duality discussion covers the classical, semiclassical and quantum-mechanical cases.
In Section~\ref{Section2}, we define the dual symmetry by invariance and reciprocity of the classical action in the form of Hamilton's characteristic function and specify a set of duality operations. Then we survey comprehensively the properties of the power-duality. The~energy-coupling exchange relations contained as a part of the duality operations lead to various energy formulas.
In Section~\ref{Section3}, we bring the power-duality defined for the classical action to the semiclassical action for quantum mechanical systems. We argue that the power-duality is basically a classical notion and breaks down at the level of angular quantization. To~preserve the basic idea of the dual symmetry in quantum mechanics, we propose as an ad hoc procedure to treat angular momentum $L$ as a continuous parameter and to quantize it only after the transformation is completed. A~linear motion in a fractional power-law potential is solved as an example to find the energy spectrum by extended use of the classical energy formulas. We also discussed the dual symmetry of the supersymmetric (SUSY) semiclassical action. Although~we are unable to verify general power duality, we find a way to show the Coulomb--Hooke symmetry in the SUSY semiclassical action. Section~\ref{Section4} analyzes the dual symmetry in quantum mechanics on the basis of an action having wave functions as variables. The~energy formulas, eigenfunctions and Green functions for dual systems are discussed in detail, including the Coulomb--Hooke problem. We also explore a quark confinement problem as an application of multi-power potentials, showing that the zero-energy bound state in the confinement potential is in the power-dual relation with a radial harmonic oscillator. Section~\ref{Section5} gives a summary of the present paper and an outlook for the future work.  Appendix \ref{SectionAPP} presents the Newton--Hooke--Morse triality that relates the Newton--Hooke duality to the Morse~oscillator.
\clearpage

\section{Power-Law Duality as a~Symmetry}\label{Section2}
Duality is an interesting and important notion in mathematics and physics, but~it has many faces~\cite{Atiyah2007}. In~physics it may mean equivalence, complementarity, conjugation, correspondence, reciprocity, symmetry and so on. Newton's law and Hooke's law may be said dual to each other in the sense that a given orbit of one system can be mapped into an orbit of the other (one-to-one correspondence), whereas they may be a dual pair because the equation of motion of one system can be transformed into the equation of motion for the other (equivalence).

In this section, we pursue a view that the power duality is a symmetry of the classical action in the form of Hamiltonian's characteristic function, and~discuss the power duality in classical, semiclassical and quantum mechanical~cases.

\subsection{Stipulations}\label{Section2.1}
Let us begin by proposing an operational definition of the power duality. We consider two distinct systems, $A$ and $B$. System $A$ (or $A$ in short), characterized by an index or a set of indices $a$, consists of a power potential $V_{a}(r) \sim r^{a}$ and a particle of mass $m_{a}$ moving in the potential with fixed angular momentum $L_{a}$ and energy $E_{a}$. Similarly, system $B$ ($B$ in short), characterized by an index or a set of indices $b$, consists of a power potential $V_{b}(r) \sim r^{b}$ and a particle of mass $m_{b}$ moving in the potential with fixed angular momentum $L_{b}$ and energy $E_{b}$.

If there is a set of invertible transformations $\Delta(B, A)$ that takes $A$ to $B$, then we say that $A$ and $B$
are {equivalent}. Naturally, the~inverse of $\Delta(B, A)$ denoted by $\Delta(A, B)=\Delta^{-1}(B, A)$ takes $B$ to $A$.

Let $X(a, b)$ and $X(b, a)=X^{-1}(a, b)$ be symbols for replacing the indices $b$ by $a$ and $a$ by $b$, respectively. If~$B$ becomes $A$ under $X(a, b)$ and $A$ becomes $B$ under $X(b, a)$, then we say that $A$ and $B$ are {reciprocal} to each other with respect to $\Delta(B, A)$. If~$A$ and $B$ are equivalent and reciprocal, we say they are {dual} to each other. Since each of the two systems has a power potential, we regard the duality so stipulated as the {power duality}.

The successive applications of $\Delta(A, B)$ and $X(a, b)$ transform $A$ to $B$ and change $B$ back to $A$. Consequently the combined actions leave $A$ unchanged. In~this sense we can view that the set of operations, $\{\Delta(A, B), X(a, b)\}$, or~its inverse, $\{\Delta(B, A), X(b, a)\}$, is a symmetry operation for the power~duality.

If a quantity $Q_{a}$ belonging to system $A$ transforms to $Q_{b}$ while $\Delta(B, A)$ takes system $A$ to system $B$, then we write $Q_{b}=\Delta(B, A)Q_{a}$. If~$Q_{b}$ can be converted to $Q_{a}$ by $X(a, b)$, then we write $Q_{a}=X(a, b)Q_{b}$ and say that $Q_{a}$ is {form-invariant} under $\Delta(B, A)$. If~$Q_{a}=Q_{b}$, then $Q_{a}$ is an {invariant} under $\Delta(B, A)$. If~every $Q_{a}$ belonging to system $A$ is an invariant under $\Delta(B, A)$, then $\Delta(B, A)$ is an identity~operation.

\subsection{Duality in the Classical~Action}\label{Section2.2}
The power duality in classical mechanics may be most easily demonstrated by considering the action integral of the form of Hamilton's characteristic function, $W(E)=S(t) + Et$, where $S$ is the Hamilton's principal function and $E$ is the energy of the system in question.
The action is usually given by Hamilton's principal function,
\begin{equation}\label{Action1}
S(\tau) =\int^{\tau} \rmd t\,{{\cal L}}=\int^{\tau} \rmd t\left[\frac{m}{2}\dot{\vec{r}}^{\,2}-V(\vec{r})\right]
\end{equation}
which leads to the Euler--Lagrange equations via Hamilton's variational principle. If~the system is spherically symmetric, that is, if~the potential $V(\vec{r})$ is independent of angular variables, then the action remains invariant under rotations. If~the system is conservative, that is, if~the Lagrangian is not an explicit function of time, then the action is invariant under time translations. In~general, if~the action is invariant under a transformation, then the transformation is often called a symmetry~transformation.

For a conserved system, we can choose as the action Hamilton's characteristic function,
\begin{equation}\label{Action2}
W(E) =\int^{\tau} \rmd t\, \{{{\cal L}} + E\}=S(\tau)+E\tau\,,\qquad E=-\frac{\partial S(\tau)}{\partial\tau}.
\end{equation}

Insofar as the system is conservative, both the principal action $S(\tau)$ and the characteristic action $W(E)$ yield the same equations of motion.
For the radial motion of a particle of mass $m$ with a chosen value of energy $E$ and a chosen value of angular momentum $L$ in a spherically symmetric potential $V(r)$, the~radial action has the form,
\begin{equation}\label{act1}
W_{(r, t)}(E) = \int_{\mathrm{I}_{t}}\,\rmd t \, \left\{\frac{m}{2}\left(\frac{\rmd r}{\rmd t}\right)^{2} - \frac{L^{2}}{2m r^{2}} - V(r) + E\right\},
\end{equation}
where $\mathrm{I}_{t}=\tau(E) \ni t$ is the range of $t$.  We let a system with a specific potential $V_{a}$ be system A and append the subscript $a$ to every parameter involved. In~a similar manner, we let a system with $V_{b}$ be system B whose parameters are all marked with a subscript $b$. For~system A with a radial potential $V_{a}(r)$, we rewrite the action (\ref{act1}) in the form,
\begin{equation}\label{act2}
W_{(r, t)}(E_{a}) = \int_{\mathrm{I}_{\varphi}}\, \rmd\varphi \, {\left(\frac{\rmd t}{\rmd\varphi} \right)} \left\{\frac{m_{a}}{2}{\left(\frac{\rmd t}{\rmd\varphi} \right)^{-2}}\left(\frac{\rmd r}{\rmd \varphi}\right)^{2} {-} \frac{L_{a}^{2}}{2m_{a} r^{2}} - U_{a}(r) \right\},
\end{equation}
with
\begin{equation}\label{Ua}
U_{a}(r)=V_{a}(r) - E_{a},
\end{equation}
where $\varphi $ is some fiducial time {and} $\mathrm{I}_{\varphi} \ni \varphi $ is the range of integration.

In (\ref{act2}), as~is often seen in the literature~\cite{Gaze,John,Junk}, we change the radial variable from $r$ to $\rho$ by a bijective differentiable map,
\begin{equation}\label{change1}
\mathfrak{R}_{f}:\, \quad \, r=f(\rho) \, \quad \, \Leftrightarrow \, \quad \, \rho = f^{-1}(r),
\end{equation}
where $f$ is a positive differentiable function of $\rho$, $0 < r < \infty$ and $0 < \rho < \infty$. With~this change of variable we associate a change of time derivative from ${(\rmd t/\rmd\varphi)}$ to ${(\rmd s/\rmd\varphi)}$ by a bijective differentiable  map,
\begin{equation}\label{change2}
{{\mathfrak{T}}_{g}}: \, \quad \, {(\rmd t/\rmd\varphi)} = g(\rho){(\rmd s/\rmd\varphi)} \, \quad \, \Leftrightarrow \, \quad \,  {(\rmd s/\rmd\varphi)} = \frac{{(\rmd t/\rmd\varphi)}}{g(f^{-1}(r))}.
\end{equation}

In the above, we assume that both $r$ and $\rho$ are of the same dimension and that $s$ has the dimension of time as $t$ does. As~a result of operations $\mathfrak{R}_{f}$ and ${{\mathfrak{T}}_{g}}$ on the action (\ref{act2}), we~obtain
\begin{equation}\label{act3}
W_{(r, t)}( E_{a}) = \int_{\mathrm{I}_{\varphi}}\, \rmd\varphi \,{\left(\frac{\rmd s}{\rmd\varphi} \right)} \left\{\frac{m_{a}}{2}\frac{f'^{2}}{g}{\left(\frac{\rmd s}{\rmd\varphi} \right)^{-2}}\left(\frac{\rmd\rho}{\rmd\varphi}\right)^{2} {-} \frac{gL_{a}^{2}}{2m_{a} f^{2}} - gU_{a}(f(\rho)) \right\},
\end{equation}
whose implication is obscure till the transformation functions $f$ and $g$ are appropriately~specified.

Suppose there is a set of operations $\Delta$, including $\mathfrak{R}_{f}$ and ${{\mathfrak{T}}_{g}}$ as a subset, that can convert $W_{(r, t)}(E_{a})$ of (\ref{act3}) to the form,
\begin{equation}\label{act4}
W_{(\rho, s)}( E_{b}) = \int_{\mathrm{I}_{\varphi}}\, \rmd\varphi \,{\left(\frac{\rmd s}{\rmd\varphi} \right)} \left\{\frac{m_{b}}{2}{\left(\frac{\rmd s}{\rmd\varphi} \right)^{-2}}\left(\frac{\rmd\rho}{\rmd\varphi}\right)^{2} {-}  \frac{L_{b}^{2}}{2m_{b} \rho^{2}} - U_{b}(\rho) \right\},
\end{equation}
with
\begin{equation}\label{Ub}
U_{b}=V_{b}(\rho) - E_{b},
\end{equation}
where $V_{b}(\rho)$ is a real function of $\rho$, and~$E_{b}$ is a constant having the dimension of energy. Then we identify the new action (\ref{act4}) with the action of system $B$ representing a particle of mass $m_{b}$ which moves in a potential $V_{b}(\rho)$ with fixed values of angular momentum $L_{b}$ and energy $E_{b}$.
If $W_{\xi_{a}}(E_{a})=X(a, b)W_{\xi_{b}}(E_{b})$ where $\xi_{a}=(r,t)$ and $\xi_{b}=(\rho,s)$, then $W_{\xi_{a}}(E_{a})$ is form-invariant under $\Delta$. Since $W_{(\rho, s)}(E_{a})$ is physically identical with $W_{(r, t)}(E_{a})$, if~$W_{(\rho, s)}(E_{a}) = X(a, b) W_{(\rho, s)}(E_{b})$, then we say that system $A$ represented by $W_{(r, t)}(E_{a})$ is {dual} to system $B$ represented by $W_{(\rho, s)}(E_{b})$ with respect to $\Delta$.

\subsection{Duality~Transformations}\label{Section2.3}
In an effort to find such a set of operations $\Delta$, we wish, as~the first step, to~determine the transformation functions $f(\rho)$ of (\ref{change1}) and $g(\rho)$ of (\ref{change2}) by demanding that the set of space and time transformations $\{\mathfrak{R}_{f}, {{\mathfrak{T}}_{g}}\}$ preserves the form-invariance of each term of the action. In~other words, we determine $f(\rho)$ and $g(\rho)$ so as to retain (i) form-invariance of the kinetic term, (ii) form-invariance of the angular momentum term and (iii) form-invariance of the shifted potential~term.

In the action $W_{(r,t)}(E_{a})$ of (\ref{act3}), the~functions $f(\rho)$ and $g(\rho)$ are arbitrary and independent of each other. To~meet the condition (i), it is necessary that $g=\mu f'^{2}$ where $\mu$ is a positive constant. Then the kinetic term expressed in terms of the new variable can be interpreted as the kinetic energy of a particle with mass
\begin{equation}\label{changem}
\mathfrak{M}: \, \, \quad \, m_{b}=m_{a}/\mu.
\end{equation}

In order for the angular momentum term to keep its inverse square form as required by (ii), the~transformation functions are to be chosen as
\begin{equation}\label{fg}
f(\rho)=C_{\eta} \rho^{\eta}, \,\, \quad \, \, g(\rho)=\mu C_{\eta}^{2}\eta^{2} \rho^{2\eta - 2},
\end{equation}
where $\eta$ is a non-zero real constant and $C_{\eta}$ is an $\eta$ dependent positive constant which has the dimension of $r^{1 - \eta}$ as $r$ and $\rho$ have been assumed to possess the same dimension. \mbox{With~(\ref{fg})}, the~angular momentum term of (\ref{act3}) takes the form, $L_{b}^{2}/(2m_{b}\rho^{2})$, when the mass changes by $\mathfrak{M}$ of (\ref{changem}), and~the angular momentum $L_{a}$ transforms to
\begin{equation}\label{changeL}
\mathfrak{L}: \, \, \quad \, L_{b}=\eta L_{a}.
\end{equation}

To date, the forms of $f(\rho)$ and $g(\rho)$ in (\ref{fg}) have been determined by the asserted conditions (i) and (ii), even before the potential is specified. This means that (iii) is a condition to select a potential $V(r)$ pertinent to the given form of $g(\rho)$. More explicitly, (iii)~demands that $gU_{a}(r)$ must be of the form,
\begin{equation}\label{gU1}
gU_{a} = V_{b}(\rho) - E_{b},
\end{equation}
where $V_{b}(\rho)$ is such that $V_{a}(\rho) = X(a, b)V_{b}(\rho)$.
Therefore, the~space-time transformation $\{\mathfrak{R}_{f}, {{\mathfrak{T}}_{g}}\}$ subject to the form-invariance conditions (i)--(iii) is only applicable to a system with a limited class of~potentials.

The simplest potential that belongs to this class is the single-term power potential $V_{a}(r)=\lambda_{a}r^{a}$ where $\lambda_{a} \in \mathbb{R}$ and $a \in \mathbb{R}$. The~corresponding shifted potential is given by
\begin{equation}\label{Ua2}
U_{a}(r)=\lambda_{a} r^{a} - E_{a}
\end{equation}
which transforms with (\ref{fg}) into
\begin{equation}\label{gU2}
gU_{a}(r)=\mu \lambda_{a}C_{\eta}^{a + 2}\eta^{2} \rho^{a\eta + 2\eta -2} - \mu C_{\eta}^{2}\eta^{2}\rho^{2\eta -2}E_{a}.
\end{equation}

Under the condition (iii) the expected form of the shifted potential is
\begin{equation}\label{Ub2}
U_{b}(\rho) = gU_{a}(r)=\lambda_{b} \rho^{b} - E_{b},
\end{equation}
where $\lambda_{b} \in \mathbb{R}$ and $b \in \mathbb{R}$.
Comparison of (\ref{gU2}) and (\ref{Ub2}) gives us only two possible combinations for the new exponents and the new coupling and energy,
\begin{eqnarray}
&& b=a \eta + 2\eta - 2 \, \quad  \mbox{and} \,\quad  \,2\eta - 2 = 0, \label{comb1a} \\
&&\lambda_{b} = \mu C_{\eta}^{a + 2}\eta^{2} \lambda_{a} \, \quad  \mbox{and}\, \quad \, E_{b}=\mu C_{\eta}^{2}\eta^{2} E_{a} \label{comb1b}
\end{eqnarray}
and
\begin{eqnarray}
&& b = 2 \eta -2 \,\quad \,  \mbox{and}\,\quad \,  a~\eta + 2 \eta - 2 = 0  , \qquad (a \neq -2), \label{comb2a} \\
&& \lambda_{b} =- \mu C_{\eta}^{2}\eta^{2}\rho^{2\eta -2}E_{a}  \quad \mbox{and}\, \quad  E_{b} = - \mu C_{\eta}^{a + 2}\eta^{2} \lambda_{a}\, , \label{comb2b}
\end{eqnarray}

Note that $a=-2$ is included in the first combination but excluded from the second~combination.

In the following, we shall examine the two possible combinations in more detail by expressing the admissible transformations in terms of the exponents,
\begin{equation}\label{eta1a}
\eta_{1}=1, \, \, \quad \, \eta_{a} = 2/(a+2)\, \qquad (a \neq 0, -2),
\end{equation}
and separating the set of $\eta_{a}$ into two as
\begin{equation}\label{etapm}
\eta_{+} = \{\eta_{a}| a > -2\} , \, \, \quad \, \eta_{-}=\{ \eta_{a}| a < -2 \}.
\end{equation}

Chandrasekhar in their book~\cite{Chand} represents a pair of dual {forces} by $(a - 1, b - 1)$. In~a way analogous to their notation, we also use the notation $(a, b)$ via $\eta$ for a pair of the exponents of power {potentials} when system $A$ and system $B$ are related by a transformation with $\eta$. We shall put the subscript {F} to differentiate the pairs of dual {forces} from those of dual {potentials} as $(a -1, b-1)_{F} = (a, b)$ whenever needed. Caution must be exercised in interpreting $(0, 0)$ which may mean $\lim_{\varepsilon \rightarrow 0}(\pm \varepsilon, \pm \varepsilon)$, $\lim_{\varepsilon \rightarrow 0}(\pm \varepsilon, \mp \varepsilon)$ and purely $(0, 0)$ (see the comments in below Subsections). We shall refer to the sets of pairs $(a, b)$ related to the first combination (\ref{comb1a})--(\ref{comb1b}) and the second combination (\ref{comb2a})--(\ref{comb2b}) as {Class I} and \mbox{{Class II},~respectively}.

\subsubsection{Class~I}\label{Section2.3.1}
{Class I} is the supplementary set of self-dual pairs. Equation~(\ref{comb1a}) of the first combination~implies
\begin{equation}\label{eta0}
\mathfrak{C}_{1}: \,\, \quad \, \eta_1 = 1, \, \, \quad \, a~= b \in \mathbb{R},
\end{equation}
which is denoted by $(a, a)$ via $\eta_{1}$. In~this case, (\ref{fg}) yields $f(\rho) = C_{1}\rho$ and $g(\rho) = \mu C_{1}^{2}$ where $C_{1}$ and $\mu$ are arbitrary dimensionless constants. With~these transformation functions,  \mbox{(\ref{change1}) and (\ref{change2})} lead to a set of space and time transformations whose scale factors depend on neither space nor time,
\begin{equation}\label{spacescale}
\mathfrak{R}_{1}:\, \, \quad \, r=C_{1}\rho ,
\end{equation}
and
\begin{equation}\label{timescale1}
{{\mathfrak{T}}_{1}}: \, \, \quad \, {(\rmd t/\rmd\varphi)} = \mu C_{1}^{2}{(\rmd s/\rmd\varphi)}.
\end{equation}

Associated with the space and time transformations (\ref{spacescale}) and (\ref{timescale1}) are the scale changes in coupling and energy, as~shown by (\ref{comb1b}),
\begin{equation}\label{scaleE}
 \mathfrak{E}_{1}: \, \, \quad \, \lambda_{a} \rightarrow \lambda_{b}=(\mu C_{1}^{a + 2}) \lambda_{a}, \, \, \quad \, E_{a} \rightarrow E_{b}=(\mu C_{1}^{2})E_{a}.
 \end{equation}

According to (\ref{changem}), the~mass also changes its scale,
\begin{equation}\label{mass0}
\mathfrak{M}_{1}: \,\, \quad \, m_{b}=m_{a}/\mu.
\end{equation}

From (\ref{changeL}) and (\ref{eta0}) follows the scale-invariant angular momentum ({{}{we}
 use the subscript 0 for trivial transformations representing an identity}),
\begin{equation}\label{scaleL}
{\mathfrak{L}_{0}}:\, \, \quad \, L_{b}=L_{a}.
\end{equation}

In this manner we obtain a set of operations $\Delta_{1}=\{\mathfrak{C}_{1}, \mathfrak{R}_{1}, {{\mathfrak{T}}_{1}}, \mathfrak{E}_{1}, \mathfrak{M}_{1}, \mathfrak{L}_{0}\}$ that leaves form-invariant the action for the power potential system. System $B$ reached from system $A$ by $\Delta_{1}$ can go back to system $A$ by $X(a,b)$. Hence, system $A$ is dual to system $B$. Notice, however, that $\Delta_{1}$ leads to a self-dual pair $(a, a)$ via $\eta_{1}$ for any given $a \in \mathbb{R}$. In~particular, $(0, 0) = \lim_{\varepsilon \rightarrow 0} (\pm \varepsilon, \mp \varepsilon)$.\\

\noindent {\bf Remark 1:}
Class I consists of self-dual pairs $(a, a)$ via $\eta_{1}$ for all $a \in \mathbb{R}$. All pairs in this class are {supplemental} in the sense that they are not traditionally counted as dual pairs. Since $\Delta_{1}$ is a qualified set of operations for preserving the form-invariance of the action,
we include self-dual pairs of {Class I} in order to extend slightly the scope of the duality~discussion.\\

\noindent {\bf Remark 2:}
The space transformation $\mathfrak{R}_{1}$ of (\ref{spacescale}) is a simple scaling of the radial variable as $C_{1} > 0$. The~scaling is valid for any chosen positive value of $C_{1}$. Hence it can be reduced, as~desired, to~the identity transformation $r=\rho$ by letting $C_{1}=1$. Those dual pairs linked by scaling may be regarded as trivial.\\

\noindent {\bf Remark 3:}
The scale transformation with $C_{1} > 0$ induces the time scaling ${{\mathfrak{T}}_{1}}$ whereas the time has its own scaling behavior. The~change in time (\ref{timescale1}) integrates to $t = C_{1}\mu s + \nu$ where $\nu$ is a constant of integration. The~resulting time equation may be understood as consisting of a time translation $t = t' + \nu$, a~scale change due to the space scaling $t'= C_{1}s'$, and~an intrinsic time scaling $s'=\mu s$. The~time translation, under~which the energy has been counted as conserved, is implicit in ${{\mathfrak{T}}_{1}}$. The~scale factor $\mu$ of time scaling, independent of space scaling, can take any positive value. If~$C_{1}=1$ and $\mu=1$, then ${{\mathfrak{T}}_{1}}$  becomes the identity transformation of time, ${(\rmd t/\rmd\varphi)}= {(\rmd s/\rmd\varphi)}$.\\

\noindent {\bf Remark 4:}
The scale change in mass $m_{b}=\mu m_{a}$ is only caused by the intrinsic time scaling $t = \mu s$. If~$\mu=1$, then the mass of the system is conserved. Conversely, if~$m_{a}=m_{b}$ is preferred, the~time scaling with $\mu =1$ must be chosen. The~time scaling in classical mechanics has no particular significance. In~fact, it adds nothing significant to the duality study. Therefore, in~addition to the form-invariant requirements (i)--(iii), we demand (iv) the mass invariance $m_{a}=m_{b}=m$ by choosing $\mu = 1$. In~this setting the time scaling occurs only in association with the space-scaling. In~accordance with the condition (iv), we shall deal with systems of an invariant mass $m$ for the rest of the present~paper.\\

\noindent {\bf Remark 5:}
If $C_{1}=1$ and $\mu = 1$, then operations, $\mathfrak{E}_{1}$, $\mathfrak{M}_{1}$, and~$\mathfrak{L}_{0}$, become identities of respective quantities. Thus, $\Delta_{1}$ for $C_{1}=1$ and $\mu=1$ is the set of identity operations, which we denote $\Delta_{0}$. The~set of operations $\Delta_{1}$ for $C_{1} > 0$ is {trivial} in the sense that it is reducible to the set of identity operations $\Delta_{0}$.\\

\noindent {\bf Remark 6:}
If {Class I} is based only on the scale transformation, it may not be worth pursuing. As~will be discussed in the proceeding sections, there are some examples that do not belong to the list of traditional dual pairs ({Class II}). In~an effort to accommodate those exceptional pairs within the present scheme for the duality discussion, we look into the details hidden behind the space identity transformation $r=\rho$. The~radial variable as a solution of the orbit equations, such as the Binet equation, depends on an angular variable and is characterized by a coupling parameter. In~application to orbits, the~identity transformation $r=\rho$ means $r(\theta; \lambda_{a}) = \rho(\tilde{\theta}; \lambda_{b})$, which occurs when $\theta \rightarrow \tilde{\theta}$. The~angular transformation $\tilde{\theta}=\theta + \theta_{0}$ where $-2\pi < \theta_{0} < 2\pi$ causes a rotation of a given orbit $\rho(\tilde{\theta}; \lambda_{b})=r(\theta;\lambda_{a})=r(\tilde{\theta} - \theta_{0};\lambda_{a})$ about the center of force by $\theta_{0}$. For~instance, the~cardioid orbit $r=r_{0}\cos^{2}(\theta/2)$ in a potential with power $a=-3$ maps into $\rho=r_{0}\sin^{2}(\tilde{\theta}/2)$ by a rotation $\tilde{\theta} = \theta + \pi$. This example belongs to the self-dual pair $(-3, -3)$ via $\eta = 1$. In~this regard, we argue that the identity transformation includes rotations about the center of forces. Of~course, the~rotation with $\theta_{0}=0$ is the {bona fide} identity~transformation.\\

\noindent {\bf Remark 7:}
Suppose two circular orbits pass through the center of attraction. It is known that the attraction is an inverse fifth-power force. If~the radii of the two circles are the same, then the inverse fifth-power force is self-dual under a rotation. If~the radii of the two circles are different, the~two orbiting objects must possess different masses. A~map between two circles with different radius, passing through the center of the same attraction, is precluded from possible links for the self-dual pair $(-4, -4)$ by the mass invariance requirement (iv).\\

\noindent {\bf Remark 8:}
If $C_{1} < 0$ in (\ref{spacescale}), either $r$ or $\rho$ must be negative contrary to our initial assumption. However, when we consider the mapping of orbits, as~we do in {{}{Remark 6}
}, we recognize that  there is a situation where the angular change $\theta \rightarrow \tilde{\theta}$ induces $\rho(\tilde{\theta}; \lambda_{b}) = - r(\theta; \lambda_{a}) = r(\theta; -\lambda_{a})$. For~instance, consider an orbit given by a conic section $r=p/(1 + e \cos \theta)$ where $p > 0$ and $- 1/e < \cos\theta \leq 1$. If~$e > 1$, then it is possible to find $\tilde{\theta}$ such that $  -1 \leq \cos\tilde{\theta} < - 1/e$ by $\theta \rightarrow \tilde{\theta}$. Consequently the image of the given orbit is $\rho(\tilde{\theta}; p) =r(\tilde{\theta}; p) = - r(\theta; p)< 0$. Certainly the result is unacceptable. The~latus rectum $p$ is inversely proportional to $\lambda_{a}$. Hence in association with the sign change in coupling $\lambda_{a} \rightarrow \lambda_{b} = - \lambda_{a}$, we are able to obtain a passable orbit $\rho(\tilde{\theta}, -p) = r(\tilde{\theta}; -p) =- r(\theta; -p) > 0$. The~orbit mapping of this type cannot be achieved by a rotation. To~include the situation like this in the space transformation, we formally introduce the {inversion},
\begin{equation}\label{reverseR}
\mathfrak{R}_{i}:\, \, \quad \, r \rightarrow  - \rho,
\end{equation}
and treat it as if the case of $C_{1}=-1$. Then we interpret the negative sign of the radial variable as a result of a certain change in the angular variable $\theta$ involved in the orbital equation by associating it with a sign change in coupling so that both $r$ and $\rho$ remain positive. If~$\mu = 1$, the~inversion causes no change in time, mass, energy, and~angular momentum, but~entails, as~is apparent from (\ref{scaleE}), a~change in coupling,
\begin{equation}\label{lambdaab}
\lambda_{a} \rightarrow \lambda_{b} =(-1)^{a}\lambda_{a}.
\end{equation}

The inversion set $\Delta_{1}$ with $C_{1}=-1$ and $\mu = 1$, denoted by $\Delta_{i}$, is partially qualified as
a duality transformation. The~reason why $\Delta_{i}$ is ''partially'' qualified is that it is admissible only when $a$ is an integer. Notice that $(-1)^{a}$ appearing in (\ref{lambdaab}) is a complex number unless $a$ is an integer. As~$\lambda_{a}$ and $\lambda_{b}$ are both assumed to be real numbers, $a$ must be integral. Having said so, in~the context of the inversion, we need a further restriction on $a$. The~sign change in coupling is induced by the inversion only when $a$ is an odd number. Since $\Delta_{i}$ is not generally reducible to the identity set $\Delta_{0}$, it is non-trivial.

\subsubsection{Class~II}\label{Section2.3.2}
{Class II} is the set of proper (traditional) dual pairs. Equation~(\ref{comb2a}) of the second combination can be expressed as
\begin{equation}\label{eta21a}
\mathfrak{C}_{2}: \, \, \quad \, \eta = 2/(a + 2)\,\, \quad \mbox{with}\,\, \quad  b= -2a/(a + 2) \,, \qquad (a \neq -2).
\end{equation}
which implies that a pair $(a, b)=(a, -2a/(a+2))$ is linked by $\eta_{a}$ when $a \neq -2$. The~above operation $\mathfrak{C}_{2}$ may as well be given by
\begin{equation}\label{eta21b}
\mathfrak{C}_{2}': \, \, \quad \, \eta =(b + 2)/2\,\, \quad \mbox{with}\,\, \quad  a= -2b/(b + 2) \,, \qquad (b \neq -2),
\end{equation}
which means a pair $(a, b) = (- 2b/(b+2), b)$ linked via $\eta = (b+2)/2$.  Another expression for $\mathfrak{C}_{2}$ is
\begin{equation}\label{eta21c}
\mathfrak{C}_{2}'':\, \, \quad \, \eta = (b+2)/2, \,\, \quad \mbox{with} \,\,\quad (a + 2) (b + 2) = 4 \,, \qquad \, (a \neq -2, \, b \neq -2),
\end{equation}
which is a version of what Needham~\cite{Need,Need2} calls the {Kasner--Arnol'd theorem} for dual forces. If~$a \neq 0$ and $b \neq 0$,
\begin{equation}\label{eta22}
 \eta = 2/(a +2) = (b+2)/2 = - b/a\, , \qquad \, (a \neq -2, \, b \neq -2),
\end{equation}
from which follows that to every $(a, b)$ via $\eta_{a}$ there corresponds $(b, a)$ via $\eta_{a}^{-1}$ if $a \neq 0, -2$. If~$ |a| \ll 1$, then $b \approx - a$ and $(a, b) \approx (a, -a)$. Hence $(0,0) = \lim_{a \rightarrow 0}(a, -a)$ via $\eta_{+}$, which overlaps with $(0, 0) = \lim_{a \rightarrow 0}(a, a)$ of {Class I} in the limit but differs in approach. In~the above $\eta_{a}$ stand for $\eta$ with a fixed $a$.

In this case, the~transformation functions of (\ref{fg}) can be written as $f(\rho)=C_{a} \rho^{\eta_{a}}$ and  $g(\rho)=\mu C_{a}^{2} \eta_{a}^{2}\rho^{2\eta_{a} - 2}$ where $C_{a}=C_{\eta_{a}}$. Here we choose $\mu=1$ by the reason stated in {{}{Remark 4}
}. The~change of radial variable (\ref{change1}) and the change of time derivative (\ref{change2}) become, respectively,
\begin{equation}\label{change12}
\mathfrak{R}_{a}: \, \, \quad \, r =C_{a} \rho^{\eta_{a}},
\end{equation}
and
\begin{equation}\label{change22}
{{\mathfrak{T}}_{a}}: \, \, \quad \, {(\rmd t/\rmd\varphi)} = C_{a}^{2} \eta_{a}^{2}\rho^{2\eta_{a}-2} {(\rmd s/\rmd\varphi)}.
\end{equation}

Equation~(\ref{comb2b}) of the second combination, associated with $\{\mathfrak{R}_{a}, {{\mathfrak{T}}_{a}}\}$, yields the coupling-energy exchange operation,
\begin{equation}\label{Elam}
\mathfrak{E}_{a}:\, \, \quad \, \lambda_{b} =- C_{a}^{2}\eta_{a}^{2}E_{a} \, , \,\quad \,
E_{b} = - C_{a}^{a + 2}\eta_{a}^{2} \lambda_{a} \,, \, \quad (a \gtrless -2).
\end{equation}

The time scaling has been chosen so as to preserve the mass invariance (\ref{changem}),
\begin{equation}\label{mass2}
\mathfrak{M}_{0}: \,\, \quad \, m_{b}=m_{a}=m,
\end{equation}
and the scale change in the angular momentum follows from (\ref{changeL}) with $\eta_{a}$,
\begin{equation}\label{changeL2}
\mathfrak{L}_{a}:\, \, \quad \, L_{b}=\eta_{a}L_{a}.
\end{equation}

Now we see that each of the sets $\Delta_{a} = \{\mathfrak{C}_{a}, \mathfrak{R}_{a}, {{\mathfrak{T}}_{a}}, \mathfrak{E}_{a}, \mathfrak{M}_{0}, \mathfrak{L}_{a}\}$ preserves the form-invariance of  the action (\ref{act2}) with a power potential. The~form-invariance warrants that $X(a, b)\Delta_{a} = \Delta_{b}$. Hence system $B$ is dual to system $A$ with respect to $\Delta_{a}$. Let $\Delta_{\pm}=\{\Delta_{a}; \, a~\gtrless -2 \}$.
The set $\Delta_{+}$ links $a>-2$ and $b>-2$ of $(a, b)$, whereas $\Delta_{-}$ relates $a <-2$ to $b<-2$.
No $\Delta_{a}$ links $a \gtrless -2$ to $b \lessgtr -2$. Hence there is no pair $(a, b)$ consisting of $a \gtrless -2$ and $b \lessgtr -2$.\\

\noindent {\bf Remark 9:}
{Class II} consists of {proper} dual pairs $(a, b)$ linked by $\Delta_{\pm}$, which have been widely discussed in the literature~\cite{Arno, Chand, Need, Need2, GranRosn, Gaze, John}. Here $a$ and $b$ are distinct except for two self-dual pairs, $(0, 0)$ via $\eta_{+}$ and $(-4, -4)$ via $\eta_{-}$.\\

\noindent {\bf Remark 10:}
Note that the time transformation (\ref{change22}) is not integrable unless the time-dependence of the space variable (i.e., the~related orbit) is~specified.\\

\noindent {\bf Remark 11:}
The scale factor $C_{1}$ appeared in {Case I} was dimensionless.  A~space transformation of (\ref{fg}) for a given value of $\eta_{a}$ contains a constant $C_{\eta_{a}}$ which has a dimension of $r^{a/(a+2)}$. Let $C_{\eta_{a}}=C_{a}d_{a}$ where $C_{a}$ and $d_{a}$ are a dimensionless magnitude and the dimensional unit of $C_{\eta_{a}}$, respectively. Use of an appropriate scale transformation which is admissible as seen in {Case I} enables $C_{a}$ to reduce to unity. More over, the~dimensional unit may be suppressed to $d_{a} = 1$. Therefore, if~desirable, the~space transformation (\ref{change12}) may simply be written as $r = \rho^{\eta_{a}}$ without altering physical~contents.\\

\noindent {\bf Remark 12:}
Let $(a, b)$ be a dual pair satisfying the relation $(a + 2)(b + 2)=4$.
Then the left element $(a, \,)$ of $(a, a)$ maps via $(a, b)$ into $(b, \,)$, and~the right element $(\, , a)$ into
$(\, ,b)$. Hence the self-dual pair $(a, a)$ can be taken by $(a, b)$ to the self-dual pair $(b, b)$. Schematically,
\[
(a, a) \, \stackrel{(a, b)}{\longrightarrow} \, (b, a) \, \stackrel{(a, b)}{\longrightarrow}\, (b, b).
\]
We call $((a, a), (b, b))$ a {grand dual pair}.
\subsection{Graphic Presentation of Dual~Pairs}\label{Section2.4}

\begin{figure}
\includegraphics{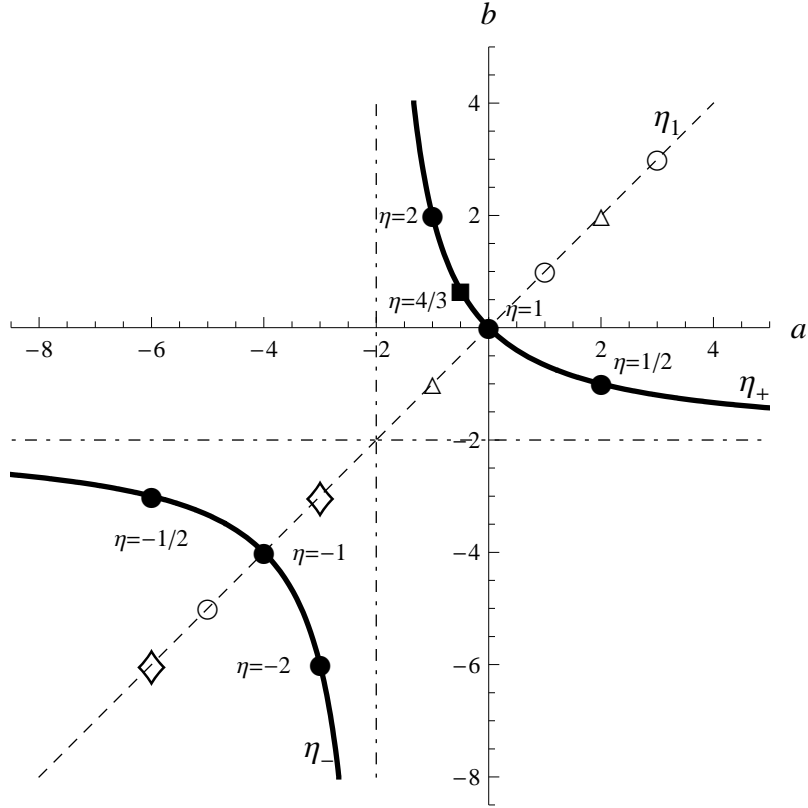}
 \caption{\label{Fig1} The solid line shows the allowed combinations of dual pairs $(a,b)$ of power laws. The~dashed line indicates the symmetry axis $(a,b)\leftrightarrow(b,a)$. The~bullets show the only dual pairs where both $a$ and $b$ are integers representing the Newton--Hook duality. The~square represents the duality pair discussed in Section~\ref{Section4.4}.}
\end{figure}

A dual pair $(a, b)$ is presented as a point in a two-dimensional $a-b$ plane as shown in Figure~\ref{Fig1}.
All self-dual pairs $(a, a)$ of  {Class I} are on a dashed straight line $a=b$ denoted by $\eta_{1}$. Every dual pair $(a, b)$ of {Class II} is shown as a point on two branches $\eta_{\pm}$ of a hyperbola described by the equation $(a+2)(b+2)=4$ of (\ref{eta21c}).  The~graph for {Class II} is similar to the one given by Arnol'd for dual {forces} \cite{Arno}.

Among the dual pairs of {Class I}, there are pairs $(a, a)$ linked by scale transformations (inclusive of rotations), which cover all real $a$, and~those $(a, a)$ related by the inversion, which are defined only when $a$ is an odd number. In~this regard, every pair $(a, a)$, occupying a single point on $\eta_{1}$, plays multiple roles. While the pairs linked by scale transformations admissible for all real values of $a$ form a continuous line $\eta_{1}$ indicated by a dashed line, those pairs linked by the inversion appear as discrete points on $\eta_{1}$ and are indicated by~circles.

The hyperbola representing all pairs of {Class II} has its center at $(-2, -2)$, transverse axis along $b=a$, and~asymptotes on the lines $a=-2$ and $b=-2$. The~bullets indicate all pairs $(a, b)$ via $\eta_{\pm}$ with integral $a$'s; namely,  $(-1, 2)$ via $\eta=2$, $(0, 0)$ via $\eta =1$, $(-3, -6)$ via $\eta=-2$, and~$(-4, -4)$ via $\eta=-1$. There are no integer pairs other than those listed above in {Class II}. The~square represents the dual pair $(-1/2, 2/3)$ to be discussed in Section III D. On~the branch of $\eta_{+}$, a~dual pair $(a, b)$ via $\eta_{+}$ and its inverse pair $(b, a)$ via $\eta_{+}^{-1}$ are symmetrically located about the transverse axis $\eta_{1}$. Since both $(a, b)$ and $(b, a)$ signify that system $A$ and system $B$ are dual to each other, the~curves $\eta_{\pm}$ have redundancy in describing the $A-B$ duality. An~example is the Newton--Hooke duality for which two equivalent pairs $(-1, 2)$ via $\eta =2$ and $(2, -1)$ via $\eta=1/2$ appear in symmetrical positions on $\eta_{+}$.

We notice that there are two special points on the graph. They are the intersections of $\eta_{1}$ and $\eta_{\pm}$; namely, $(0, 0)$ with $\eta =1$, and~$(-4, -4)$ with $\eta =\pm1 $.  The~former is an overlapping point of $\eta_{1}$ and $\eta_{+}$ where $\eta=1$. The~latter is like an overhead crossing of $\eta_{1}$ and $\eta_{-}$ where the pair belonging to $\eta_{1}$ is linked by a transformation with $\eta=1$ while the one belonging to $\eta_{-}$ is linked with $\eta=-1$.

In approaching the crossing of $\eta_{1}$ and $\eta_{+}$, the~pair $(0, 0)$ at $\eta_{1}=1$ has a limiting behavior as  $(0, 0) = \lim_{\varepsilon \rightarrow 0} (\pm \varepsilon, \pm \varepsilon)$, while $(0, 0)$ at $\eta_{+}=1$ behaves like $(0, 0) = \lim_{\varepsilon \rightarrow 0} (\pm \varepsilon, \mp \varepsilon)$ via $\eta =1$.  As~has been
mentioned earlier, $(a, b) = (a-1, b-1)_{F}$. However, the~counterpart of $(0, 0)$ is not exactly equal to
$(-1, -1)_{F}$. The~potential corresponding to the inverse force $F \sim 1/r$ is $V \sim \ln r$. Thus, it is more appropriate to put symbolically $(-1, -1)_{F}=(\ln, \ln)$. Yet, $(0, 0) \neq (\ln, \ln)$. Consider  $V_{a}(r) = \lambda_{a}r^{\varepsilon}$. For~$\varepsilon$ small, $V_{a}(r) \approx \lambda_{a}(1 + \varepsilon \ln r)$, which gives rise to the force $F \approx \kappa /r$ where $\kappa = \lambda_{a}\varepsilon$. As~long as $\kappa$ can be treated as finite, $(\varepsilon, -\varepsilon) \approx  (-1, -1)_{F}$. Chandrasekhar~\cite{Chand} excluded $(-1, -1)_{F}$ from the list of dual pairs on physical grounds. We exclude $(\ln, \ln)$ because the logarithmic potential, being not a power potential, lies outside our~interest.

By analyzing Corollaries and Propositions in the {Principia}, Chandrasekhar~\cite{Chand} pointed out that Newton had found not only the Newton--Hooke dual pair but also the self-dual pairs $(2, 2)$, $(-1, -1)$ and $(-4, -4)$. He also mentioned that $(-3, -6)$ was not included in the {Prinpicia}. For~$a$ integral, there are only two grand dual pairs $((-1, -1), (2, 2))$ and $((-3, -3), (-6, -6))$.  In~Figure~\ref{Fig1}, $(2, 2)$ and $(-1, -1)$ are marked with triangles on $\eta_{1}$, while $(-3, -3)$ and $(-6, -6)$ are marked with diamonds on $\eta_{1}$.

\subsection{Classical~Orbits}\label{Section2.5}
Here we discuss the orbital behaviors for the dual pairs in relation with energy and~coupling.

First, we consider self-dual pairs $(a, a)$ of {Class I}. If~an effective shifted potential is defined by $U^{eff}(r)=U(r) + L^{2}/(2mr^{2})$,  the~space transformation $r=C_{1}\rho $ induces
\begin{equation}\label{Ueff1}
U_{a}^{eff}(r)=\lambda_{a} r^{a} + \frac{L_{a}^{2}}{2mr^{2}} - E_{a}, \, \quad \Rightarrow \quad \, U_{b}^{eff}(\rho)=C_{1}^{a+2}\lambda_{a} \rho^{a} + \frac{L_{a}^{2}}{2m \rho^{2}} - C_{1}^{2}E_{a},
\end{equation}
resulting in self-dual pairs $(a, a)$ for any real $a$. The~space transformation includes scale transformations $r=C_{1}\rho$ with $C_{1} > 0$, identity transformation $r = \rho$ (inclusive of rotations), and~inversion formally defined by $r = - \rho$.

\vspace{12pt}
\noindent\textbf{Statement 1:} \emph{System $A$ and system $B$ linked by a scale transformation are physically identical but described in different scale. Typically an orbit of system $A$ maps to an orbit of system $B$ similar in shape but different in~scale.}
\vspace{12pt}

\noindent\textbf{Statement 2:} \emph{In the limit $C_{1} \rightarrow 1$, the~two orbits become congruent (identical) to each other. Any self-dual pair $(a, a)$ due to a scale transformation is reducible to a trivial pair $(a, a)$ linked by the identity transformation. However, in~dealing with the orbital behaviors, we have to look into the angular dependence of radial variables by allowing the identity transformation $r=\rho$ to contain $r(\theta)=\rho(\tilde{\theta})=r(\tilde{\theta} - \theta_{0})$ with $\theta \rightarrow \tilde{\theta}=\theta + \theta_{0}$, which represents a rotation of a given orbit about the center of force by $\theta_{0}$.}
\vspace{12pt}

The inversion $r \rightarrow - \rho $ entails $\lambda_{b}=(-1)^{a}\lambda_{a}$, as~is apparent from (\ref{Ueff1}). If~$a$ is an even number, the~sign change in coupling does not occur. Hence the inversion for even $a$ cannot properly be defined and must be precluded. Only when $a$ is odd, the~inversion is meaningful. However, we have to notice that orbits in a potential with $a > 0$ are all bounded if $\lambda_{a} > 0$ and all unbounded if $\lambda_{a} < 0$. Under~the inversion, the~sign of $\lambda_{a}$ changes, so that a bound orbit with $E_{a} > 0$ is supposed to go to an unbounded orbit with $E_{b}=E_{a} > 0$. It is uncertain whether there are such examples to which the inversion~works.

\vspace{12pt}
\noindent\textbf{Statement 3:}\emph{ If $a$ is a negative odd number, under~the inversion, an~orbit in an attractive (repulsive) potential maps to an orbit in a repulsive (attractive) potential, keeping the energy~unchanged. }
\vspace{12pt}

In the {Principia}, Newton proved that if an orbit passing through the center of attraction is a circle then the force is inversely proportional to the fifth-power of the distance from the center (Corollary I to Proposition VII). From~Corollary I of Proposition VII and other corollaries in the {Principia} Chandrasekhar~\cite{Chand} shows in essence that if an object moves on a circular orbit under centripetal attraction emanating from two different points on the circumference of the circle then the forces from the two points exerted on the orbiting object are of the same inverse fifth-power law. Then he suggests, in~this account, that the inverse fifth power law of attraction is self-dual for motion in a circle. In~contrast to Chandrasekhar's view on the self-dual pair $(-4, -4)$, we maintain that $(-4, -4)$ can be understood as a member of {Class I} and {Class II}.
The circular orbit in an attractive potential $V_{a}(r) = \lambda_{a} r^{-4}$, which occurs when $E_{a} = 0$, can be described by the equation $r=2R_{a} \cos \theta$ where $R_{a}= \sqrt{-\lambda_{a}m/(2L_{a}^{2})}$ is the radius of the circle and $-\pi /2 < \theta < \pi /2$  is the range of $\theta$. The~scale transformation $r = C_{1}\rho$ with $C_{1} > 0$ converts the orbit equation into $\rho=2 R_{b} \cos\theta $ where $R_{b}=R_{a}/C_{1}$. Apparently it is consistent with the requirements $L_{b}=L_{a}$ and $\lambda_{b} = C_{1}^{-2}\lambda_{a}$ of (\ref{Ueff1}). Thus, the radius of the circle is rescaled while the center of force is fixed at the origin and the range of $\theta$ is unaltered. The~inverse fifth-power law of attraction may be viewed as self-dual under a scale change for motion in a circle. If~the identity map $r=\rho$ may include a rotation $r(\theta) \rightarrow \rho(\tilde{\theta}) = r(\tilde{\theta} - \theta_{0})$, then $\rho(\tilde{\theta}) =2R\cos(\tilde{\theta}-\theta_{0})$ with the angular range $-\pi/2 + \theta_{0} < \tilde{\theta} < \pi/2 + \theta_{0}$. In~particular, if~$\theta_{0}=\pi$, then $\rho(\tilde{\theta}) =-2R\cos(\tilde{\theta})$ with $\pi/2 < \tilde{\theta} < 3\pi/2$. The~circular orbit maps into itself, though~rotated about the center of force.
In this sense, the~inverse fifth-power law of attraction is self-dual under a rotation for motion in a circle.
In much the same way, the~inverse fifth-power force, whether attractive or repulsive, may be considered as self-dual under a scale change and a rotation for motion in any other orbits. Hence the self-dual pair $(-4, -4)$ linked by the scale transformation (including rotations) is a member of {Class I}. The~same self-dual pair $(-4, -4)$ has another feature as a member of {Class II} which will be discussed in {{}{Remark 13}}.

Secondly, we consider dual pairs $(a, b)$ of {Class II}.

All dual pairs $(a, b)$ of {Class II} are subject to the proper dual transformation $\Delta_{II}$. The~members $a$ and $b$ of each pair obey the Kasner--Arnol'd formula $(a + 2)(b + 2) = 4$, and~are related via $\eta =2/(a + 2)$ (or $\eta = - b/a$ if $a \neq 0$). These dual pairs belong to branch $\eta_{+}$ if $ a > -2$, and~branch $\eta_{-}$ if $a < -2$.

Now the space and time transformations $r=C_{a}\rho^{2/(a+2)}$ and
${(\rmd t/\rmd\varphi)} = C_{\eta} ^{2} \eta^{2}\rho^{-2a/(a+2)} {(\rmd s/\rmd\varphi)}$ induce
the energy-coupling exchange,
\begin{equation}\label{Ecoup2}
\lambda_{b} =- C_{\eta}^{2}\eta^{2}E_{a} \, ,\qquad
E_{b} = - C_{\eta}^{a + 2}\eta^{2} \lambda_{a},
\end{equation}
where $C_{\eta} > 0$ and $a \neq -2$. Hence the effective shifted potential transforms as
\begin{equation}\label{Ueff2}
U_{a}^{eff}(r)=\lambda_{a} r^{a} + \frac{L_{a}^{2}}{2mr^{2}} - E_{a}, \, \quad \Rightarrow \quad \,
U_{b}^{eff}(\rho)=- C_{\pm}^{2} \eta_{\pm}^{2} E_{a} \rho^{b}  +  \frac{\eta_{\pm}^{2}L_{a}^{2}}{2m\rho^{2}} +
C_{\pm}^{a + 2} \eta_{\pm}^{2}  \lambda_{a}
\end{equation}
where $a \neq -2$ and $b=-2a/(a+2)$.

The two equations in (\ref{Ecoup2}) are not simply to exchange the roles of energy and coupling. They also provide a useful relation between $E_{a}$ and $E_{b}$. In~general $E_{a}$ depends on $\lambda_{a}$. So we let $E_{a}=\mathcal{E}_{a}(\lambda_{a})$, and~invert it as $\lambda_{a}=\mathcal{E}_{a}^{-1}(-\lambda_{b}/\eta^{2}C_{\eta}^{2})$ with the help of the first equation of~(\ref{Ecoup2}). Substitution of this into the second equation of (\ref{Ecoup2}) yields
\begin{equation}\label{EaEb}
  E_{b}=- \eta^{2}C_{\eta}^{a + 2} \mathcal{E}_{a}^{-1}\left(- \frac{\lambda_{b}}{\eta^{2}C_{\eta}^{2}}\right).
\end{equation}
which shows that $E_{b}$ depends on $E_{a}$ through the coupling $\lambda_{a}$.\\

\noindent\textbf{Statement 4:} \emph{For a dual pair $(a, b)$ of {Class II}, if~the coupling dependence of $E_{a}$ is explicitly known, then $E_{b}$ can be determined by (\ref{EaEb}), and~vice versa.}
\vspace{6pt}

From (\ref{Ecoup2}) there follow four possible mapping patters,
\begin{eqnarray*}
(0) \, \, & ~ & \, (E_{a} = 0, \lambda_{a} \gtreqless 0)\, \, \Longrightarrow \, (E_{b} \lesseqgtr 0, \lambda_{b}=0)\\
(1) \, \, & ~ & \, (E_{a} > 0, \lambda_{a} < 0) \, \Longrightarrow \, (E_{b} > 0, \lambda_{b} < 0) \\
(2) \, \, & ~ & \, (E_{a} < 0, \lambda_{a} < 0) \, \Longrightarrow \, (E_{b} > 0, \lambda_{b} > 0) \\
(3) \, \, & ~ & \, (E_{a} > 0, \lambda_{a} > 0) \, \Longrightarrow \, (E_{b} < 0, \lambda_{b} < 0) \\
(4) \, \, & ~ & \, (E_{a} < 0, \lambda_{a} > 0) \, \Longrightarrow \, (E_{b} < 0, \lambda_{b} > 0)
\end{eqnarray*}

In the above, pattern (0) implies that any zero energy orbit of system $A$ goes to a rectilinear orbit of system $B$ with no potential. Patterns (1)--(4) imply that any positive energy orbit of system $A$, regardless of the sign of $\lambda_{a}$, maps to an orbit of system $B$ with a coupling $\lambda_{b} < 0$, and~any negative energy orbit of system $A$, independent of $\lambda_{a}$, maps to an orbit of system $B$ with a coupling $\lambda_{b} > 0$.

The dual pairs $(a, b)$ of  {Class II} can be grouped into those on $\eta_{+}$ and those on $\eta_{-}$. Furthermore, the~pairs of the first group can be divided into two parts for $\eta_{+} > 1$ and $0 < \eta_{+} <1$.
If we let $\eta_{+}^{>}$ denote the part for $ \eta_{+} > 1$, then $\eta_{+}^{>} = \{-b/a |\, -2 < a < 0,  b > 0 \}$. Similarly, let $\eta_{+}^{<}$ denote the part for $0 < \eta_{+} < 1$. Then $\eta_{+}^{<} = \{-b/a | \,a > 0, -2 < b < 0 \} = \{ - a/b|\, -2 < a < 0,  b > 0 \}$. Thus, $\eta_{+}^{<} = [\eta_{+}^{>}]^{-1}$. It is sufficient to consider the set $\eta_{+}^{>}$. The~same can be said for the second group on $\eta_{-}$. We take up only the set
$\eta_{-}^{>}=\{- b/a|\, -4 < a < - 2, b < -4 \}$.

For the case of $\eta_{+}^{>}$, $\lambda_{a} > 0 \, ( < 0 ) $ implies a repulsion (attraction), while $\lambda_{b} > 0 \, (< 0) $ means an attraction (repulsion). There are no negative energy orbits in a repulsive potential with $\lambda_{a} > 0$ and in an attractive potential with $\lambda_{b} > 0$. For~$\eta_{-}^{>}$, both $\lambda_{a} > 0$ and $\lambda_{b} > 0$ are repulsive, and~both $\lambda_{a} < 0$ and $\lambda_{b} < 0$ are attractive. In~any repulsive potential with $\lambda_{a} >0$ or $\lambda_{b} > 0$, no negative energy orbits are present. Pattern (4)
is not physically meaningful. Taking these features of potentials into account, we can restate the implication of the relations in (\ref{Ecoup2}) as~follows.

\noindent\textbf{Statement 5:} \emph{Under the proper duality transformation $\Delta_{II}$, if~$-2 < a < 0$ (i.e., $b > 0$), then any positive energy orbit in the potential of system $A$, whether attractive or repulsive, maps to an orbit in a repulsive potential of system $B$, and~any negative energy (bound) orbit maps to a positive energy (bound) orbit in an attractive potential. If~$a > 0$ (i.e., $-2 < b < 0$), then the above situations are reversed. If~ $-4 < a < - 2$ (i.e.,
$b < -4$), then any positive orbit in an attractive potential maps to a positive orbit under attraction, any negative bound orbit in an attractive potential maps to a positive orbit under repulsion, and~any positive orbit under repulsion maps to a negative bound orbit in an attractive potential. Even for the case where $ a < - 4$ (i.e., $-4 < b < -2$), the~mapping patterns are the same as those for $ -4< a < -2$. In~all cases, zero energy orbits map to force-free rectilinear~orbits.}\\

This is a modified version of Needham's statement made in supplementing the Kasner--Arnol'd theorem~\cite{Need,Need2}.\\

\noindent {\bf Remark 13:}
The pair $(-4, -4)$ has another feature as a point on $\eta_{-}$, that is, as~a member of {Class II}. From~(\ref{Ecoup2}), it is obvious that $\lambda_{b}=0$ for the circular zero energy orbit. Hence the duality transformation $\Delta_{II}$ maps the orbit into a force-free rectilinear orbit. According to {{}{Statement 5}
}, any positive energy orbit must map to an orbit in an attractive potential, and~any negative energy orbit maps to an orbit in a repulsive potential. Therefore, the~self-dual pair $(-4, -4)$ Newton established is not a member of {Class II}. It must be $(-4, -4)$ on $\eta_{1}$, belonging to {Class I}.\\

In what follows, we make remarks on the Newton--Hooke pairs and related self-dual~pairs.\\

\noindent {\bf Remark 14:}
{Statement 5} applies to the pair $(-1, 2)$. The~mapping patterns (0)--(3) works in going from the Newton system with $a=-1$ to the Hooke system with $b=2$. Namely,
(0) the zero energy orbit of the attractive Newton system maps to a rectilinear orbit; \,
(1) a positive unbound orbit of the attractive Newton system maps to a positive unbound orbit of the repulsive Hooke system; \, (2) a negative energy bound orbit of the attractive Newton system maps to a positive energy bound orbit of the attractive Hooke system; and \,
(3) a positive unbound orbit of the repulsive Newton system maps to a negative unbound orbit of the repulsive Hooke system. Since there are no negative orbits for the repulsive Newton system and the attractive Hooke system, pattern (4) is~irrelevant.\\

\noindent {\bf Remark 15:}
In view of the orbit structure, we study in more detail the mapping process from the Newton system to the Hooke system. As~is well-known, for~the motion in the inverse-square force, the~orbit equation in polar coordinates has the form,
\begin{equation}\label{conic}
r=\frac{p}{1+e\cos\theta},
\end{equation}
where $p$ is the semi-latus rectum, $e$ the eccentricity. The~orbit is of conic sections and the origin of the coordinates is at the focus closest to the pericenter of the orbit. The~angle $\theta$ is between the position of the orbiting object and the direction to the pericenter located at $r=r_{min}$ and $\theta =0$. The~semi-latus rectum, the~semi-major axis, and~the eccentricity of the orbit are determined by $p=-L_{a}^{2}/(m\lambda_{a})$,
$\,\bar{a} = - \lambda_{a}/(m|E_{a}|)$, and~$e=\sqrt{1 + (2L_{a}^{2}E_{a}/m\lambda_{a}^{2})}$, , respectively.
If the inverse square force is attractive, i.e.,~if $\lambda_{a} < 0$, then $\bar{a} > 0$, $p > 0$, and~$ 1 > \cos\theta > -1/e$. If~repulsive, i.e.,~if $\lambda_{a} > 0$, then $\bar{a} < 0$, $p < 0$ and $ -1 < \cos\theta < -1/e$.

(i) \, For the bound motion, $E_{a} < 0$, $e < 1$ and $p=\bar{a}(1 - e^{2}) > 0$. The~Equation~(\ref{conic}) describes an elliptic orbit with semi-major axis $\bar{a}$ and eccentricity $e$. Apparently, \mbox{$r_{min}=\bar{a}(1- e)$}. For~the duality mapping, a~more suited choice is the orbit equation expressed in terms of the eccentric anomaly $\psi$,
\begin{equation}\label{ellipse}
r = \bar{a}(1 - e \cos \psi),
\end{equation}
which may be put in the form,
\begin{equation}\label{ellipse2}
r = \bar{a}\left\{(1 + e) \cos^{2}(\psi/2) +(1 - e)\sin^{2}(\psi/2)\right\}.
\end{equation}
Here $\psi$ is related to the polar angle $\theta$ by $\tan(\theta/2)=[(1+e)/(1-e)]^{1/2} \tan(\psi/2)$.
Since $r=C_{2}\rho^{2}$, use of (\ref{ellipse2}) leads to
\begin{equation}\label{rhoe}
\rho = \left[\alpha^{2} \cos^{2}(\psi/2) + \beta^{2}\sin^{2}(\psi/2)\right]^{1/2},
\end{equation}
where
\begin{equation}\label{alphbet}
\alpha = \sqrt{\bar{a}(1 + e)/C_{2}\, }, \,\, \quad \, \, \beta = \sqrt{\bar{a}(1-e)/C_{2}\, }.
\end{equation}

Let $\rho = \sqrt{ u^{2} + v^{2}\,}$  in cartesian coordinates, and~let
\begin{equation}\label{uv}
u = \alpha \cos(\psi/2), \, \, \quad \, \, v=\beta \sin(\psi/2).
\end{equation}

Then it is clear that the trajectory drawn by $\rho$ is given as an ellipse,
\begin{equation}\label{Hellip}
\frac{u^{2}}{\alpha^{2}} + \frac{v^{2}}{\beta^{2}} = 1,
\end{equation}
with semi-major axis $\alpha$ and semi-minor axis $\beta$, centered at the origin of the $u-v$ plane.
It is obvious that $\rho_{min}=\sqrt{\bar{a}(1-e)/C_{2}\, }$ is the semi-minor axis of the ellipse on the $u-v$ plane.
The above calculation shows that the elliptic Kepler orbit with semi-major axis $\bar{a}$ and eccentricity $e$ maps to an ellipse with semi-major axis $\alpha=\sqrt{\bar{a}(1 + e)/C_{2}\,}$ and eccentricity
$\epsilon = \sqrt{2e/(1+ e)\,}$. The~semi-major and semi-minor axes of the resultant ellipse depend on the scaling factor $C_{2}$. With~different values of $C_{2}$, a~Kepler ellipse of eccentricity $e$ is mapped to ellipses of different sizes having a common eccentricity $\epsilon$. In~general, the~resultant ellipse having eccentricity $\epsilon$ is not similar to the Kepler orbit with eccentricity $e$.  If~$e=0$, then $\epsilon = 0$. Namely, a~circular orbit of radius $\bar{a}$ under an inverse-square force maps to a circle with radius $\alpha = \sqrt{\bar{a}/C_{2}\,}$. With~a particular scale $C_{2}=1/\sqrt{\bar{a}}$, the~mapped circle is congruent to the original orbit. In~the limit $e \rightarrow 1$, the~Kepler orbit becomes a parabola with $E_{a}=0$, which maps to a force-free rectilinear orbit described by $(u, v)=(\rho, 0)$.

(ii) \, If $E_{a} > 0$, then $e >1$ and $\bar{a} > 0$ for $\lambda_{a} < 0$. The~semi-latus rectum in (\ref{conic}) must be modified as $p=\bar{a}(e^{2} -1) > 0$. Again $\cos \theta < - 1/e$.  The~orbit is a branch of a hyperbola with semi-major axis $\bar{a}$ and eccentricity $e$. The~center of attraction is at the interior focus of the branch, so that $r_{min}=\bar{a}(e - 1)$. In~much the same fashion that the eccentric anomaly is used in (\ref{ellipse}), we introduce a parameter $\psi$ related to the angle $\theta$ by $\tan(\theta/2)=[(e+1)/(e-1)]^{1/2} \tanh(\psi/2)$.  Here $\cosh \psi > 1/e$. Now the orbit equation in parametric representation is
\begin{equation}\label{hyper}
r = \bar{a}(e \cosh \psi - 1),
\end{equation}
which may further be written as
\begin{equation}\label{hyper2}
r = \bar{a}\left\{(e - 1) \cosh^{2}(\psi/2) +(e + 1)\sinh^{2}(\psi/2)\right\},
\end{equation}
whose minimum occurs when $\psi = 0$.
Correspondingly, $\rho=\sqrt{r/C_{2}\,}$ is expressed as
\begin{equation}\label{rhoh}
\rho = \left[\alpha^{2} \cosh^{2}(\psi/2) + \beta^{2}\sinh^{2}(\psi/2)\right]^{1/2},
\end{equation}
where
\begin{equation}\label{alphbet2}
\alpha = \sqrt{\bar{a}( e -1)/C_{2}\, }, \,\, \quad \, \, \beta = \sqrt{\bar{a}(e + 1)/C_{2}\, }.
\end{equation}

Hence $\rho_{min}=\sqrt{\bar{a}( e -1)/C_{2}\, }$.
Letting
\begin{equation}\label{uv2}
u = \alpha \cosh(\psi/2), \, \, \quad \, \, v=\beta \sinh(\psi/2),
\end{equation}
we obtain $\rho=\sqrt{u^{2} + v^{2}\, }$ and the equation for a hyperbola having two branches,
\begin{equation}\label{Hellip2}
\frac{u^{2}}{\alpha^{2}} - \frac{v^{2}}{\beta^{2}} = 1,
\end{equation}
which has the semi-major axis $\alpha = \sqrt{\bar{a}(e -1)/C_{2}\,}$ and the eccentricity $\epsilon=\sqrt{2e/(e-1)\,}$. Thus, the positive energy orbit in the attractive inverse potential, given by a branch of the hyperbola, maps to a positive energy orbit given by either branch of a hyperbola whose center coincides with the center of the repulsive Hooke~force.

(iii) \, For a repulsive potential with $\lambda_{a} > 0$ such as the repulsive Coulomb potential, the~orbit Equation~(\ref{conic}) describing a hyperbola holds true insofar as $E_{a} > 0$, i.e.,~$e > 1$. Since
$ p=-L_{a}^{2}/(m\lambda_{a}) < 0$ for $\lambda_{a} > 0$,  the~semi-lotus rectum must be replaced by
$\tilde{p}= -p$. At~the same time, the~angular variable has to be changed from $\theta$ to $\tilde{\theta}$ where $\cos \theta < -1/e$ and $\cos\tilde{\theta} > - 1/e$. The~conversion of the hyperbolic Equation~(\ref{conic}) for the attractive potential to the hyperbolic equation for the repulsive potential,
\begin{equation}\label{conic2}
\tilde{r} = \frac{\tilde{p}}{1 + e \cos \tilde{\theta}},
\end{equation}
is indeed the inversion process mentioned in {{}{Remark 8}
}. Since (\ref{conic}) and (\ref{conic2}) have the same form, we can follow the procedure given in (ii) to show that under $\tilde{r}=\sqrt{\rho/C_{2}\,}$ the positive energy orbit in the repulsive inverse potential, given by a branch of the hyperbola, maps to a negative energy orbit given by either branch of a hyperbola whose center coincides with the center of the repulsive Hooke~force.\\

\noindent {\bf Remark 16:}
In connection with Remark 14, we look at the self-dual pairs $(-1, -1)$ and
$(2, 2)$ which do not belong to {Class II}. Apparently the two pairs are closely related to each other via the Newton--Hooke pair $(-1, 2)$, so as to form a grand dual pair $((-1, -1), (2, 2))$. As~they are both on $\eta_{1}$, each of them is self-dual under scale changes and rotations. In~addition, $(-1, -1)$ is self-dual under the inversion. From~(iii) of {{}{Remark 15}
}, it is clear that due to the inversion the orbit equation takes the form (\ref{conic}). There the angular range for $\tilde{\theta}$ is $\theta_{e} < \tilde{\theta} < 2\pi - \theta_{e}$ where $\theta_{e}=\cos^{-1}(-1/e)$. Hence the resultant orbit has the center of orbit at the exterior focus. This means that a hyperbolic orbit in attraction with the center of force at the interior focus maps to the conjugate hyperbola in repulsion with the center of force at the exterior focus. In~contrast, any rotation maps a hyperbolic orbit under attraction (repulsion) into a hyperbolic orbit under attraction (repulsion). In~summary, the~inversion maps a hyperbolic orbit under attraction into a hyperbolic orbit under repulsion, whereas any rotation takes a hyperbolic orbit under attraction (repulsion) to a hyperbolic orbit under attraction (repulsion). According to Chandrasekhar's book~\cite{Chand}, what Newton established for $(-1, -1)$ and $(2, 2)$ are that the attractive inverse square force law is dual to the repulsive inverse square force law, and~that the repulsive linear force law is dual to itself. Thus, we are led to a view that Newton's $(-1, -1)$ is due to the inversion and their $(2, 2)$ is due to a rotation. Finally we wish to point out that by the mapping patterns (1) and (3) of $(-1, 2)$ a hyperbolic orbit of the attractive Newton system, whether attractive or repulsive, maps to a hyperbolic orbit of the repulsive Hooke system. In~other words, the~pair of forces $(attraction, repulsion)$ for $(-1, -1)$ goes to the pair of force $(repulsion, repulsion)$ for $(2, 2)$ with the help of $(-1, 2)$. This is compatible with the assertion that Newton's two self-dual pairs form the grand dual pair $((-1, -1), (2, 2))$ via $(-1, 2)$.

\subsection{Classical Energy~Formulas}\label{Section2.6}
We have used the energy-coupling exchange relations,
\begin{equation}\label{EF1}
\mathfrak{E}:  \, \, \, E_{b}=- \eta^{2} C^{a+2} \lambda_{a}, \, \quad \, \lambda_{b} = - \eta^{2} C^{2} E_{a},
\end{equation}
as essential parts of the power-duality operations. They demand primarily that the roles of energy and coupling be exchanged. Using these relations, we can also derive energy formulas which enable us to determine the energy value of one system from that of the other when two systems are power-dual to each~other.

In general $E_{a}$ depends on $\lambda_{a}$, $L_{a}$ and possibly other parameters. So let the energy function be $E_{a}=\mathcal{E}(\lambda_{a}, L_{a}, w_{a})$ where $w_{a}$ represents those additional parameters. Then we pull $\lambda_{a}$ out from the inside of $\mathcal{E}$ as
\begin{equation}\label{EF2}
\lambda_{a}=\mathcal{E}^{-1}(E_{a}, L_{a}, w_{a}).
\end{equation}

Now we insert this coupling parameter $\lambda_{a}$ into the first equation of (\ref{EF1}).  Substituting the second relation $E_{a}=-\lambda_{b} /(\eta^{2}C^{2})$ and the angular momentum transformation \mbox{$L_{a}=L_{b}/\eta$} to the right-hand side of (\ref{EF2}), we can convert the first relation of (\ref{EF1}) into an energy formula,
\begin{equation}\label{EF2a}
E_{b}(\lambda_{b}, L_{b}, w_{b})= - \eta^{2} C^{a+2} \mathcal{E}^{-1}(-\lambda_{b}/(\eta^{2} C^{2}), L_{b}/\eta, w_{a}(w_{b})).
\end{equation}

Thus, if~$E_{a}$ is known, then $E_{b}$ can be determined without solving the equations of motion for system $B$.  By~making an appropriate choice of $C$, the~value of $\lambda_{b}$ may be specified by the second relation of (\ref{EF1}).

Alternatively, let us combine the two relations in (\ref{EF1}) by eliminating the constant $C$ to get another energy formula,
\begin{equation}\label{EF3}
E_{b}=- \eta^{2} \lambda_{a} \left( -\frac{\lambda_{b}}{\eta^{2} E_{a}}\right)^{1/\eta}.
\end{equation}

This formula can be rearranged to the symmetric form,
\begin{equation}\label{FE4}
\left[4(a + 2)^{-2}|\lambda_{a}|^{-2/(a+2)}|E_{a}|\right]^{a} = \left[4(b+2)^{-2}|\lambda_{b}|^{-2/(b+2)}|E_{b}| \right]^{b}.
\end{equation}

Note that the signs of the energies and coupling constants are related via (\ref{EF1}). See also the four patterns discussed in {\it Statement 4} above.

When the parameters $w$ contained in $E_{a}$ are invariant, that is, $w_{a}=w_{b}$, under~the duality operations, the~last equation suggests that there is some positive function $\mathcal{F}(L, w)$, independent of $\lambda_{a}$ and $\lambda_{b}$, such that
\begin{equation}\label{EF5}
|E_{a}(\lambda_{a}, L_{a}, w)|=\frac{(a+2)^{2}}{4}|\lambda_{a}|^{2/(a+2)} \left\{\mathcal{F}\Bigl(\sqrt{2/(a+2)\,}L_{a}, w \Bigr)\right\}^{1/a},
\end{equation}
\begin{equation}\label{EF6}
|E_{b}(\lambda_{b}, L_{b}, w)|=\frac{(b+2)^2}{4}|\lambda_{b}|^{2/(b+2)} \left\{\mathcal{F}\Bigl(\sqrt{2/(b+2)\,}L_{b}, w \Bigr)\right\}^{1/b},
\end{equation}
where $L=\sqrt{(a+2)/2\,}L_{a}=\sqrt{(b+2)/2\,}L_{b}$. If~such a function is specified for $E_{a}$ by (\ref{EF5}), then $E_{b}$ can be determined by (\ref{EF6}) with the sign to be obtained via (\ref{EF1}). Notice that (\ref{EF6}) is useful as an energy formula to find $E_{b}$ only when $E_{a}$ has the form of (\ref{EF5}).\\

\noindent {\bf Remark 17:}
As an example, let us consider the Newton--Hooke dual pair for which $(a, b)=(-1, 2)$, $\eta=-b/a=2$ and $r=C\rho^{2}$.
Let system $A$ be consisting of a particle of mass $m$ moving around a large point mass $M \gg m$ under the influence of the gravitational force with $\lambda_{a}= - GmM < 0$.  Let system $B$ be an isotropic harmonic oscillator with $\lambda_{b}=\frac{1}{2}m\omega^{2} >0$.
Then, as~the exchange relations of (\ref{EF1}) demand,  $E_{a} < 0$ and $E_{b} > 0$. Hence the orbits of the two systems are bounded. This means that the Newton--Hooke duality occurs only when both systems are in bound~states.\\

Suppose the total energy of the particle is given in the form,
\begin{equation}\label{NH1}
E_{a} =  \frac{L_{a}^{2}}{2mr_{min}^{2}} + \frac{\lambda_{a}}{r_{min}} = \mathcal{E}(\lambda_{a}, L_{a}, r_{min}),
\end{equation}
where $r_{min}$ is the minimum value of the radial variable $r$ and $\lambda_{a}=-GmM$. Then we obtain the inverse function,
\begin{equation}\label{NH2}
\lambda_{a} = \mathcal{E}^{-1}\left( -\lambda_{b}/4C^{2},L_a,r_{min}\right)  = - \frac{L_{a}^{2}}{2mr_{min}} - \frac{\lambda_{b} r_{min}}{4C^{2}}.
\end{equation}

With this result, the~Formula (\ref{EF2}) immediately leads to the energy of the Hooke system in the form,
\begin{equation}\label{NH3}
E_{b} = \frac{L_{b}^{2}}{2m\rho_{min}^{2}} + \lambda_{b}\rho_{min}^{2}
\end{equation}
where $L_{b}=2L_{a}$ and $\rho_{min}=\sqrt{r_{min}/C\,}$.  Although~$\lambda_{b}$ may be interpreted as Hooke's constant, its detailed form $\frac{1}{2}m\omega^{2}$ cannot be determined by the energy formula. Noticing that $E_{a}$ is a constant, we let $\kappa =\sqrt{-2mE_{a}}$. If~we choose $C=m\omega /(2\kappa)$, then we have $\lambda_{b}=\frac{1}{2}m\omega^{2}$ from the second relation of (\ref{EF1}). With~the same choice of $C$, we have $m\omega \rho^{2}_{min}=2\kappa r_{min}$.

Suppose the energy of system $A$ is alternatively given in the form,
\begin{equation}\label{NH4}
E_{a}=- \frac{2 \pi^{2} m \lambda_{a}^{2}}{(J + 2\pi L_{a})^{2}},
\end{equation}
where $J$ is the radial action variable, $J=\oint \rmd r\, p_{r}$, or~more explicitly,
\begin{equation}\label{NH5}
J=2 \int_{_{r_{min}}}^{^{r_{max}}} \rmd r\, \sqrt{2m\left(E - \frac{\lambda_{a}}{r} - \frac{L^{2}}{2mr^{2}}\right)},
\end{equation}
which is a constant of motion. Let $E_{a}$ of (\ref{NH4}) be put into the form given via (\ref{EF5}) then we may identify
\begin{equation}\label{NH6}
\mathcal{F}\Bigl(\sqrt{2/(a+2)\,}L_{a}, J\Bigr)=  \left[J/(2\pi) + (\sqrt{2} (L_{a})/\sqrt{2}\right]^{2}/(2m).
\end{equation}

From this follows
\begin{equation}\label{NH7}
\left\{\mathcal{F}\Bigl(\sqrt{2/(b+2)\,}L_{b}, J\Bigr)\right\}^{1/2}= \left[J/(2\pi) + (L_{b})/2\right]/\sqrt{2m}
\end{equation}

Since the first relation of (\ref{EF1}) indicates that $E_{b}>0$ for $\lambda_{a} <0$, the~relation (\ref{EF6}) together with  $\lambda_{b}=\frac{1}{2}m\omega^{2}$ results in
\begin{equation}\label{NH8}
E_{b}=(\omega/2\pi) (2J + 2\pi L_{b}),
\end{equation}
which is an energy expression of the Hooke system obtainable from the Hamilton-\linebreak Jacobi~equation.

\subsection{Generalization to Multi-Term Power~Laws}\label{Section2.7}
In the following, on~a parallel with Johnson's treatment~\cite{John}, we examine how the duality can be realized with a sum of power potentials (i.e., a~multi-term potential) in the present~framework.

Let the potential $V_{a}$ be a sum of $N$ distinct power potentials as
\begin{equation}\label{sum}
V_{a}(r) = \sum_{i=1}^{N} \lambda_{a_{i}} r^{a_{i}}\,, \quad \, a_{i} > -2\,,\qquad (a_i\neq a_j \quad\mbox{for}\quad i\neq j)
\end{equation}
where $\lambda_{a_{i}}$ is the coupling constant of the $i$-th sub-potential in $V_{a}$.
Then $\mathfrak{R}$ and ${{\mathfrak{T}}}$ take the shifted potential in (\ref{gU2}) to
\begin{equation}\label{sumE}
gU_{a}(r)  =  \sum_{i=1}^{N} \lambda _{a_{i}} {C^{a_i+2}} \eta^{2} \rho^{2\eta -2 + a_{i}\eta} - {C^2}\eta^{2} \rho ^{2\eta - 2} E_{a}.
\end{equation}

Let us pick one of the terms in the sum in (\ref{sumE}), say, the~$i=k$ term, and~make its exponent zero by letting
\begin{equation}\label{eta}
\eta =\eta_{k} = \frac{2}{a_{k} + 2} \, ,\quad \, a_{k} > - 2\,,
\end{equation}
where $\eta$ is $k$-dependent. If~the exponent of the $i=k' $ term, instead of the $k \neq k'$ term, is made vanishing, then $\eta$ is to be given in terms of $a_{k'}$ where $a_{k'} \neq a_{k}$. Since $k=1, 2,\,\dots ,\,N$,  there are $N$ possible choices of $\eta$. Thus, it is appropriate to write $\eta$ in (\ref{eta}) with the subscript $k$ as $\eta_{k}$. Apparently, $\eta_{k}$ is a possible one of $\{\eta_{1}, \eta_{2},\,\ldots ,\,\eta_{N}\}$. Let the operations $\mathfrak{R}$ and ${{\mathfrak{T}}}$ for $\eta = \eta_{k}$ be denoted by $\mathfrak{R}_{k}$ and ${{\mathfrak{T}}_{k}}$, respectively.

For the remaining potential terms $(i\neq k)$ and the energy term in (\ref{sumE}), we rename the exponents of $\rho$ as
\begin{equation}\label{bkbi}
b_{k}=-\frac{2a_{k}}{a_{k} + 2}\,, \quad  b_{i} = \frac{2(a_{i} - a_{k})}{a_{k} + 2} \,, \quad  i \neq k \,,
\end{equation}
which can easily be inverted to express $a_{k}$ and $a_{i}$ in terms of $b_{k}$ and $b_{i}$ in the same form.
These relations are equivalent to the conditions on the exponents,
\begin{equation}\label{akbk}
\mathfrak{C}_{k}: \, \quad \, (a_{k} + 2)(b_{k} + 2) = 4, \, \quad \, (a_{i} - a_{k})(b_{i} - b_{k}) = a_{i} b_{i}.
\end{equation}

From (\ref{bkbi}) there also follows $b_{i} > -2$ for all $i$ if $a_{i} > -2$ for all $i$.
The first relation \mbox{of (\ref{akbk})} leads to alternative but equivalent expressions of $\eta$ in (\ref{eta}),
\begin{equation}\label{eta2}
\eta _{k}= - \frac{b_{k}}{a_{k}} = \frac{b_{k} + 2}{2}=\frac{2}{a_k +2}.
\end{equation}

To $\mathfrak{R}_{k}$ and ${{\mathfrak{T}}_{k}}$, we have to add two more operations,
\begin{equation}\label{LL}
\mathfrak{L}_{k}: \, \quad \, L_{b_{k}}=\eta_{k} L_{a_k},
\end{equation}
and
\begin{equation}\label{lamdaE}
\mathfrak{W}_{k}: \, \quad \, \lambda_{b_{k}} = - {C^2}\eta_{k}^{2} E_{a}, \, \quad \, E_{b_{k}} = -\eta_{k}^{2} {C^{a_k+2}}\lambda_{a_{k}}, \, \quad \mbox{and} \, \quad \,  \lambda_{b_{i}} = \eta_{k}^{2} {C^{a_i+2}}\lambda_{a_{i}}\,, \quad i \neq k.
\end{equation}

Then, we express the shifted potential of (\ref{sumE}) in the new notation as
\begin{equation}\label{sumE2}
gU_{a}(r) = V_{b_{k}}(\rho)  - E_{b_{k}} = U_{b_{k}}(\rho)
\end{equation}
where
\begin{equation}\label{sumE3}
V_{b_{k}}(\rho)  =\sum_{i=1}^{N} \lambda _{b_{i}}\rho^{b_{i}}.
\end{equation}

The set of operations ${\Delta}_{k} = \{\mathfrak{R}_{k}, {{\mathfrak{T}}_{k}}, \mathfrak{C}_{k}, \mathfrak{L}_{k}, \mathfrak{W}_{k} \}$ transforms the radial action of the $A$ system into
\begin{equation}\label{act5}
W_{\rho}(E_{b_{k}}) = \int_{\mathrm{I}_{\varphi}}\, \rmd\varphi \, {(\rmd s/\rmd\varphi)} \left\{\frac{m}{2}{(\rmd s/\rmd\varphi)}^{-2}\left(\frac{\rmd\rho}{\rmd\varphi}\right)^{2} - \frac{L_{b_{k}}^{2}}{2m \rho^{2}} - U_{b_{k}}(\rho) \right\}.
\end{equation}

Thus, we find the duality between the $A$-system and $B_{k}$-system with respect to ${\Delta}_{k}$.
Again, this duality is only one of the $N$ dualities;  there are $N$ pairs of dual systems,  $(a_k, b_{k})$ for $k=1, 2,\,...,\,N$.

\section{Power-Duality in the Semiclassical~Action}\label{Section3}
The power-duality argument made for the classical action in Section~\ref{Section2} can easily be carried over to the semiclassical action. In~semiclassical theory the power-duality is a relationship between two quantum systems which are not mutually interacting. In~studying such a relationship, there are two distinct approaches; one is to pay attention to a reciprocal relation between two systems, and~the other to pursue a deeper connection between the quantum states of two systems (see {{}{Remark 18}
}). Our power-duality argument is of the former approach, taking {reciprocity} as a heuristic guiding. Special care will have to be exercised though, when dealing with the quantum structure of each~system.

\subsection{Symmetry of the Semiclassical~Action}\label{Section3.1}
The action in semiclassical theory is of the form, $W = \int \rmd q \,p $, which is Hamilton's characteristic function and essentially the same as that in (\ref{Action2}). The~semiclassical action for the radial motion reads
\begin{equation}\label{G1}
W=\int \rmd r \sqrt{2m\left(E - V(r) - \hbar^{2}L^{2}/(2mr^{2})\right)}.
\end{equation}

Here the classical angular momentum $L$ is replaced by $\hbar L$. Customarily the semiclassical angular momentum (divided by $\hbar$) of (\ref{G1}) is given by the Langer-modified form,
\begin{equation}\label{G2}
L = \ell + (D-2)/2, \, \, \quad  \ell = 0, 1, 2, ...
\end{equation}
if it is defined in $D$ dimensions.
Let us write the semiclassical action for system $A$ as
\begin{equation}\label{G3}
W_{a}=\int \rmd r \sqrt{- 2m \left[\hbar^{2}L_{a}^{2}/(2mr^{2}) + U_{a}(r) \right]\, }
\end{equation}
where $U_{a}(r)=V_{a}(r) - E_{a}$. After~the change of variable $r=f(\rho)$, the~action (\ref{G3}) of system $A$ becomes
\begin{equation}\label{G4}
W_{a}=\int \rmd\rho \sqrt{-2m \left[\hbar^{2}L_{a}^{2}g/(2mf^{2}) + gU_{a}(f)\right]\,},
\end{equation}
where $f'=\rmd r/\rmd\rho$ and $g=f'^{2}$.  The~following substitutions
\begin{equation}\label{G5}
\mathfrak{R}: \, \, \, f(\rho)=C\rho^{\eta},
\end{equation}
\begin{equation}\label{G6}
\mathfrak{L}: \, \, \, L_{a}=L_{b}/\eta,
\end{equation}
\begin{equation}\label{G7}
gU_{a}=U_{b},
\end{equation}
lead the action (\ref{G4}) to
\begin{equation}\label{G8}
W_{b}=\int \rmd\rho \sqrt{-2m \left[\hbar^{2}L_{b}^{2}/(2m\rho^{2}) + U_{b}(\rho) \right]},
\end{equation}
which is taken as the action for system $B$. Here we have assumed $m_{a}=m_{b}=m$ (i.e., $\mu=1$).
We shall also assume that two mutually power-dual systems are {by definition} in the same dimensions (i.e., $D_{a}=D_{b}=D$).

Only when the potential of system $A$ is a power potential, $U_{b}(\rho)$ in (\ref{G8}) can be brought to the form $V_{b}(\rho) - E_{b}$. The~change of variable $\mathfrak{R}: \, r=C\rho^{\eta}$ with the choice $\mathfrak{C}_2:\, \eta=2/(a+2)$ gives $g(\rho)=\eta^{2}C^{2}\rho^{-a\eta}$. Hence, for~$V_{a}(f)=\lambda_{a}C^{a}\rho^{a\eta}$, we have $g(\rho)V_{a}(f)=\eta^{2}C^{a+2}\lambda_{a}$ and $gE_{a}=\eta^{2}C^{2}E_{a}\rho^{b}$ where $b=-a\eta=-2a/(a+2)$. After~performing the energy-coupling exchange,
\begin{equation}\label{G9}
 \mathfrak{E}: \, \, \, \lambda_{a}= - E_{b}/(\eta^{2}C^{a+2}), \, \quad \, E_{a} = - \lambda_{b}/(\eta^{2}C^{2}),
\end{equation}
we obtain
\begin{equation}\label{G10}
g(\lambda_{a}r^{a} - E_{a}) = \lambda_{b}\rho^{b} - E_{b}.
\end{equation}

In effect, under~the operation of $g$,  the~following transformations have taken place,
\begin{equation}\label{G11}
gV_{a}(r) \rightarrow -E_{b}, \, \, \quad \, gE_{a} \rightarrow - V_{b}(\rho),
\end{equation}
where $V_{a}=\lambda_{a}r^{a}$ and $V_{b}(\rho)=\lambda_{b}\rho^{b}$.

In this manner, transforming the action $W_{a}$ of (\ref{G3}) to $W_{b}$ of (\ref{G8}) by the duality operations, we have $W_{a}=W_{b}$, that is,
\begin{equation}\label{G12}
\int \rmd r \sqrt{2m(E_{a} - \lambda_{a}r^{a}) - \hbar^{2}L_{a}^{2}/r^{2}\,} = \int \rmd\rho \sqrt{2m(E_{b} - \lambda_{b}\rho^{b}) - \hbar^{2}L_{b}^{2}/\rho^{2}\,}.
\end{equation}

It is also apparent that $W_{a}=X(a, b)W_{b}$ with $\xi_{a}=r$ and $\xi_{b}=\rho$. Thus, we see that
the semiclassical action (\ref{G1}) is form-invariant under the set of duality operations, $\{\mathfrak{R}, \mathfrak{L}, \mathfrak{C}, \mathfrak{E}\}$.

Although we have presented in the above the power-duality features of the semiclassical action similar to those in the classical case, we have not taken account of the possibility that the angular momentum $L$ is a discretely quantized entity given in terms of the angular quantum number $\ell=0, 1, 2, ...$ by (\ref{G2}). It is natural to expect that the operation $\mathfrak{L}: \, L_{b}= \eta L_{a}$ of (\ref{G6}) implies the equality,
\begin{equation}\label{G13}
\ell_{b} + (D_{b} - 2)/2= \eta \ell_{a} + \eta(D_{a} -2)/2.
\end{equation}

In addition, if~we demand that $\ell_{a}=0$ corresponds to $\ell_{b}=0$, then (\ref{G13}) can be separated into two equalities,
\begin{equation}\label{G13a}
\ell_{b}=\eta \ell_{a}, \, \, \quad \, \, D_{b}= \eta(D_{a} - 2) + 2.
\end{equation}

Either (\ref{G13}) or (\ref{G13a}) suggests that the allowed values of $\ell_{b}$ differs from those of $\ell_{a}$ unless $\eta=1$.
This means that the condition $\ell=0, 1, 2, ...$ in $(\ref{G2})$ cannot be imposed on system $A$ and system $B$ at the same time. Although~the transformations in (\ref{G13}) and (\ref{G13a}) are invertible, they cannot preserve the Langer-form (\ref{G2}) of the angular momentum in the two systems. In~other words, they are not reciprocal relations between the two systems. Insofar as operation $\mathfrak{L}$ implies the equalities (\ref{G13}), the~semiclassical action with the Langer modification is not form-invariant under the set of operations
$\{\mathfrak{R}, \mathfrak{L}, \mathfrak{C}, \mathfrak{E}\}$. Then, we may have to draw a conclusion that the power-duality valid in the classical action breaks down in the semiclassical action due to the quantized angular momentum~term.

In the above we have observed that the power-duality is incompatible with the angular quantization.
By the same token, the~energy-coupling relations of $\mathfrak{E}$ in (\ref{G9}) may have to be examined.
In the semiclassical action, the~energy $E$ and the coupling $\lambda$ may be treated as parameters. However, the~implication of the exchange relations in (\ref{G9}) becomes ambiguous after quantization. It is not clear whether $E_{a}$ in (\ref{G9}) is one of the energy eigenvalues of system $A$ or it represents the energy spectrum of the system.
As an aid of clarification, we study one of the energy formulas resulting from combining the two relations in (\ref{G9}),
\begin{equation}\label{Gex1}
E_{b}=- \eta^{2} \lambda_{a} \left( -\frac{\lambda_{b}}{\eta^{2} E_{a}}\right)^{1/\eta},
\end{equation}
which has been given in Section~\ref{Section2} as a classical energy formula. To~see if it will work in quantum mechanics, let us employ, e.g.,~the Coulomb--Hooke duality, the~quantum counterpart of the Newton--Hooke duality, and~test (\ref{Gex1}).
We assume that $E_{a}$ and $E_{b}$ in (\ref{Gex1}) represent the spectra of system $A$ and system $B$, respectively. According to (\ref{Gex1}), the~energy spectrum $E_{b}$ of the hydrogen atom with the Coulomb coupling $\lambda_{b} = - e^{2}$ is expected to follow from the spectrum $E_{a}$ of the {three}-dimensional isotropic harmonic oscillator with frequency $\omega = \sqrt{2\lambda_{a}/m}$.  For~this pair of systems, $(a, b)=(2, -1)$ and $\eta=-b/a=1/2$. Given $E_{a}(n_{r}, \ell_{a})= \hbar \omega (2n_{r} + \ell_{a} + 3/2)$ with $n_{r}=0, 1, 2, ...$ and $\ell_{a}=0, 1, 2, ...$, the~\mbox{Formula (\ref{Gex1})} immediately yields $E_{b}= - (me^{4}/2\hbar^{2})(n_{r} + \ell_{a}/2 + 3/4)^{-2}$. Here $n=n_{r} + \ell_{a}/2 + 3/4= 3/4, 5/4, 7/4, ...$.  The~result is not the energy spectrum of the hydrogen atom that is commonly known. Evidently, a~naive application of the energy Formula (\ref{Gex1}) fails at the level of angular quantum numbers.
By contrast, if~we consider the states of a {four}-dimensional oscillator which possess $\ell_{a}=0, 2, 4, ...$, then $n=n_{r} + \ell_{b} + 1 = 1, 2, 3, ...$ via $\ell_{b}=\ell_{a}/2$, which matches the principal quantum number of the hydrogen atom.  In~other words, the~energy Formula (\ref{Gex1}) suggests that the spectrum of the hydrogen atom can be composed of ``half the states'' of the four dimensional isotropic harmonic oscillator ({{}{to} 
 be more precise, the~set $\{\ell_{a}=0, 2, 4, ..., \, D_{a}=4\}$ for the oscillator and the set $\{\ell_{b}=0, 1, 2, ..., \, D_{b}=3\}$ for the H-atom are in one-to-one correspondence}).  The relation between the oscillator in four dimensions and the hydrogen atom in three dimensions is not reciprocal in (\ref{Gex1}). The~alternative scheme is not the Coulomb--Hooke duality that we pursue (see {{}{Remark 18}
}). The~Coulomb--Hooke duality in quantum mechanics will be discussed again in Section~\ref{Section4.3}.

In an effort to make the power duality meaningful in semiclassical theory, we shall take a view that {the power duality is basically a classical notion}. Accordingly, for~the duality discussions, all physical objects such as $L$, $E$ and $\lambda$, should be treated as classical entities, i.e.,~continuous parameters. Then we consider quantization as a process separate from the duality operations. The~duality is a classical feature of the relation between two systems, whereas quantization is associated with the micro-structures of each system. None of duality operations can dictate how the quantum structure of each system should be. The~equality of (\ref{G9}) which is compatible with reciprocity must not imply the non-reciprocal equality of (\ref{G13}). It is necessary to dissociate duality operations from quantization. Technically, we deal only with those continuous parameters for the duality discussions, and~replace them as a post duality-argument activity by appropriately quantized counterparts when needed for characterizing each quantum system. From~this view, the~power duality of the semiclassical action has already been established at the equality (\ref{G12}) with follow-up substitutions $L_{a}=\ell_{a} + (D-2)/2, \, (\ell_{a}=0, 1, 2, ..)$ and $L_{b}=\ell_{b} + (D-2)/2, \, (\ell_{b}=0, 1, 2, ...)$. It is helpful to introduce the dot-equality $\doteq$ to signify the equality amended by substitutions of quantized entities. The~power-duality of the semiclassical action in the amended version may be exhibited by
\begin{eqnarray}\label{G12a}
&&\int \rmd r \sqrt{2m(E_{a} - \lambda_{a}r^{a}) - \hbar^{2}(\ell_{a} + (D-2)/2)^{2}/r^{2}\,} \nonumber \\
&&\qquad \doteq \int \rmd\rho \sqrt{2m(E_{b} - \lambda_{b}\rho^{b}) - \hbar^{2}(\ell_{b} + (D-2)/2)^{2}/\rho^{2}\,}.
\end{eqnarray}

\subsection{The Semiclassical Energy~Formulas}\label{Section3.2}
In the preceding section, we have adopted the Coulomb--Hooke duality to \mbox{test (\ref{Gex1})}, and~failed. However, it should be recognized that if the energy spectrum of the three dimensional radial oscillator is given in the form $E_{a}(n_{r}, L_{a})= \hbar \omega (2n_{r} + L_{a} + 1)$ without requiring $L_{a}=\ell_{a} + \,1/2$, then the energy formula (\ref{Gex1}) together with $L_{a}=2L_{b}$ yields $E_{b}(n_{r}, L_{b})= - (me^{4}/2\hbar^{2})(n_{r}+ L_{b} + \,1/2)^{-2}$ which reduces to the desired Coulomb spectrum $E_{b}(\nu, L_{b})= - (me^{4}/2\hbar^{2})(n_{r} + \ell_{b} + 1)^{-2}$ after ad hoc substitution of $L_{b}=\ell_{b} + \, 1/2$ with $\ell_{b} \in \mathbb{N}_{0}$.  So long as $L$, $E$ and $\lambda$ are treated as continuous parameters, the~energy formula (\ref{Gex1}) derived from the exchange relations (\ref{G9}) should work for semiclassical systems provided that those parameters are eventually replaced by their quantum~counterparts.

In semiclassical theory, the~bound state energy $E_{a}$ of system $A$ can be evaluated by carrying out the integration on the left-hand side of (\ref{G12}) between two turning points. Namely, we calculate for $E_{a}$ the integral
\begin{equation}\label{G14}
J_{a} = 2 \int_{r'}^{r''} \rmd r \sqrt{2m(E_{a} - \lambda_{a}r^{a}) - \hbar^{2}L_{a}^{2}/r^{2}\,},
\end{equation}
where $r'$ and $r''$ are the turning points of the orbit where the integrand vanishes. The~quantity $J_{a}$ is indeed an action variable defined for a periodic motion by $\oint \rmd q \, p $. It is a constant depending on $E_{a}$, $\lambda_{a}$, and~$L_{a}$.  By~letting it be a constant $N_{a}$ multiplied by $2\pi \hbar$,
\begin{equation}\label{G15}
J_{a}(E_{a}, \lambda_{a}, L_{a}) = 2\pi \hbar N_{a},
\end{equation}
and solving (\ref{G15}) for $E_{a}$, we obtain the classical bound state energy as a function of parameters $\lambda_{a}$, $L_{a}$ and $N_{a}$,
\begin{equation}\label{G16}
E_{a}= E_a(\lambda_{a}, L_{a}, N_{a}).
\end{equation}

Once the classical energy $E_{a}$ of system $A$ is given in terms of $\lambda_{a}$, $L_{a}$ and $N_{a}$, when system $A$ and system $B$ are power-dual to each other, we can determine the energy $E_{b}$ of system $B$, with~the help of the operations $\mathfrak{L}$ and $\mathfrak{E}$, as~a function of $\lambda_{b}$, $L_{b}$ and $N_{b}$. Since $W_{a}=W_{b}$ as shown in (\ref{G12}), it is obvious that $N_{a}=N_{b}$. As~the former equality is a consequence of the duality operations, so is the latter equality. Hence the equality $N_{a}=N_{b}$ is a consequence but not a part of duality operations. So, we let $N=N_{a}=N_{b}$. With~the energy function (\ref{G16}),  the~semiclassical energy formula stemming from (\ref{Gex1}) is
\begin{equation}\label{Gex2}
E_{b}(\lambda_{b}, L_{b}, N)=- \eta^{2} \lambda_{a} \left( -\frac{\lambda_{b}}{\eta^{2} E_{a}(\lambda_{a}, L_{b}/\eta, N)}\right)^{1/\eta},
\end{equation}
which can be rearranged as the classical case in the following form
\begin{equation}\label{Gex3}
|E_{a}(\lambda_{a}, L_{a}, N)|=\frac{1}{4}(a+2)^{2}|\lambda_{a}|^{2/(a+2)} \left\{\mathcal{F}\Bigl(\sqrt{2/(a+2)\,}L_{a}, N\Bigr)\right\}^{1/a}
\end{equation}
\begin{equation}\label{Gex4}
|E_{b}(\lambda_{b}, L_{b}, N)|=\frac{1}{4}(b+2)^{2} |\lambda_{b}|^{2/(b+2)} \left\{\mathcal{F}\Bigl(\sqrt{2/(b+2)\,}L_{b}, N\Bigr)\right\}^{1/b}
\end{equation}
where $\mathcal{F}(L, N)$ is a function common to both systems. The~signs for both energy relations are determined as in  the classical case via the signs of the coupling constants, i.e.,~${\rm sgn}\,E_{a} = - {\rm sgn}\,\lambda_{b}$ and ${\rm sgn}\,E_{b} = - {\rm sgn}\,\lambda_{a}$.

Alternatively, expressing an explicit form of the energy function (\ref{G16}) by $\mathcal{E}(\lambda_{a}, L_{a}, N)$~as
\begin{equation}\label{Gex5}
E_{a}= \mathcal{E}(\lambda_{a}, L_{a}, N_{a}),
\end{equation}
and inverting (\ref{Gex5}) to take $\lambda_{a}$ out, we have
\begin{equation}\label{G17}
\lambda_{a} = \mathcal{E}^{-1}(E_{a}, L_{a}, N).
\end{equation}

Then we use the angular momentum transformation $\mathfrak{L}$ of (\ref{G6}) and the energy-coupling exchange relations $\mathfrak{E}$ of (\ref{G9}) to write down the bound state energy formula for $E_{b}$ as
\begin{equation}\label{G18}
E_{b}(\lambda_{b}, L_{b}, N) = - \eta^{2}C^{a+2}\mathcal{E}^{-1}(-\lambda_{b}/(\eta^{2}C^{2}), L_{b}/\eta, N),
\end{equation}
which is essentially the same as the energy formula (\ref{Gex2}).

To convert the classical energy $E_{a}$ in (\ref{G16}) to a quantum spectrum, we replace the parameters $L_{a}$ and $N$ by their corresponding quantized entities. The~angular momentum is quantized in the Langer form $L_{a} \doteq \ell_{a} + (D-2)/2$.
The Wentzel--Kramers--Brillouin (WKB) quantization formula for the radial motion,
\begin{equation}\label{G19}
\oint \rmd r \,p_{r} = 2\pi \hbar \left(n_{r} + \frac{1}{2} \right),\,\quad \, n_{r}=0, 1, 2, ...
\end{equation}
asserts that
\begin{equation}\label{G20}
N \doteq n_{r} + 1/2 \, ,\, \quad \, n_{r}=0, 1, 2, ...
\end{equation}

Substitution of the Langer-modified angular momentum (\ref{G2}) and the WKB quantization (\ref{G20}) in the classical energy function of (\ref{G16}) yields the energy spectrum,
\begin{equation}\label{G21}
E_{a}(n_{r}, \ell_{a}) \doteq E_{a}(\lambda_{a}, \ell_{a} - 1 + D/2, n_{r} + 1/2),
\end{equation}
where $n_{r}=0, 1, 2, ...$ and $\ell_{a}=0, 1, 2, ...$  Similarly, after~substitution of the Langer form (\ref{G2}) to $L_{b}$ and the WKB quantization (\ref{G20}) to $N$, the~semiclassical energy formula (\ref{G18}) leads to the energy spectrum of system $B$,
\begin{equation}\label{G22}
E_{b}(n_{r}, \ell_{a}) \doteq - \eta^{2}C^{a+2}\mathcal{E}^{-1}\left(-\lambda_{b}/(\eta^{2}C^{2}), (\ell_{b} -1 + D/2)/\eta, n_{r} + 1/2 \right)
\end{equation}
where $n_{r}=0, 1, 2,\ldots$ and $\ell_{b}=0, 1, 2, \ldots$.

\subsection{A System with a Non-Integer Power Potential and Zero-Angular Momentum}\label{Section3.3}
As a simple but non-trivial example, we study a non-integer power potential system with $L^{2}=0$ (see {{}{Remark 22}
}).  Let system $A$ be the case.  Bound states of system $A$ occur only when (i) $\lambda_{a} < 0$, $a < 0$ with $E_a < 0$ or (ii) $\lambda_{a} >0$, $a >0$ with $E_a > 0$.  The~integral (\ref{G14}) with $L_{a}=0$, denoted $J_{a}(E_{a}, \lambda_{a}, 0)$, is reducible to a beta function under either condition (i) or (ii). Suppose system $A$ be under condition (i). Then it goes to a beta function as
\begin{equation}\label{Gx1}
J_{a}(E_{a}, \lambda_{a}, 0) = \displaystyle M(E_{a}, \lambda_{a}) \int_{0}^{1} \rmd z\, z^{-\frac{a+2}{2a}-1}(1-z)^{\frac{3}{2}-1}
 = M(E_{a}, \lambda_{a}) B\left(-\frac{a+2}{2a}, \frac{3}{2}\right)
\end{equation}
where we have let $z=(E_{a}/\lambda_{a})r^{-a}$ and $M(E_{a}, \lambda_{a})=
\sqrt{-2m\lambda_{a}/a^{2}}(E_{a}/\lambda_{a})^{(a+2)/2a}$. As~\linebreak \mbox{in (\ref{G15})}, we express the right-hand side of (\ref{Gx1}) by the parameter $N$ as
\begin{equation}\label{Gx2}
M(E_{a}, \lambda_{a}) B\left(-\frac{a+2}{2a}, \frac{3}{2}\right) = 2\pi \hbar N,
\end{equation}
which we solve for $E_{a}$ to find the energy function $E_{a}=\mathcal{E}(\lambda_{a},  0, N)$,
\begin{equation}\label{Gx3}
E_{a} =  - (- \lambda_{a})^{\frac{2}{a+2}} \left(\frac{\sqrt{2m}}{\hbar |a| \pi} B\left(- \frac{a+2}{2a}, \frac{3}{2}\right) \right)^{-\frac{2a}{a+2}} N^{\frac{2a}{a+2}}.
\end{equation}

Now the WKB condition (\ref{G20}) yields the energy spectrum of system $A$,
\begin{equation}\label{Gx4}
E_{a}(n) \doteq - (- \lambda_{a})^{\frac{2}{a+2}} \left(\frac{\sqrt{2m}}{\hbar |a| \pi} B\left(- \frac{a+2}{2a}, \frac{3}{2}\right) \right)^{-\frac{2a}{a+2}} \left(n + \frac{1}{2}\right)^{\frac{2a}{a+2}}, \, \, \, \,
\end{equation}
where $n=0, 1, 2,\ldots$.
The bound state energy spectrum of system $B$ can be independently calculated in a similar fashion, and~the WKB quantization (\ref{G20}), separately applied to system $B$, will lead to a spectrum similar to but different from the spectrum of system $A$ \mbox{in (\ref{Gx4})}.  Insofar as system $B$ is power-dual to system $A$, the~bound state energy spectrum of system $B$ can be obtained via the formula (\ref{G22}). Inverting the $\lambda_{a}$ dependent \mbox{function (\ref{Gx3})}, we obtain
\begin{equation}\label{Gx5}
\lambda_{a} = \mathcal{E}^{-1}(E_{a}, 0, N)
            = -(-E_{a})^{(a+2)/2}\left(\frac{\sqrt{2m}}{\hbar |a| \pi} B\left(- \frac{a+2}{2a}, \frac{3}{2}\right) \right)^{a} N^{-a}.
\end{equation}

Utilizing this inverted function and the WKB condition (\ref{G20}) in the energy \mbox{Formula (\ref{G22})}, we arrive at the energy spectrum of system $B$,
\begin{equation}\label{Gx6}
E_{b}(n) \doteq \lambda_{b}^{\frac{2}{b+2}} \left(\frac{\sqrt{2m}}{\hbar |b| \pi} B\left(\frac{1}{b}, \frac{3}{2}\right) \right)^{-\frac{2b}{b+2}} \left(n + \frac{1}{2}\right)^{\frac{2b}{b+2}}, \, \, \, n=0, 1, 2, ...
\end{equation}
which is independent of the arbitrary constant $C$ appearing in (\ref{G18}).  In~the above, we have also changed $a$ to $b$ by using the relations, $a=-2b/(b+2)$ and $\eta a = -b$. Apparently, the~spectrum (\ref{Gx6}) is very similar in form with the spectrum of system $A$ in (\ref{G18}) but is not identical. The~relations (\ref{G9}) suggest that $E_{b} >0$ for $\lambda_{a} < 0$ and  $\lambda_{b} > 0$ for $E_{a} < 0$. Hence system $B$ has bound states with $E_{b} > 0$ only when $b >0$. This means that system $B$ is under condition (ii) and that the energy spectrum (\ref{Gx6}) is for the case where $\lambda_{b} >0$, $b >0$ with $E_{b} >0$. In~particular, if~$V_{a}(r) = \lambda_{a}/\sqrt{r}$ with $\lambda_{a} < 0$, the~spectrum resulting from (\ref{Gx4}) is
\begin{equation}\label{Gx7}
E_{a=-1/2}(n) \doteq  \frac{\lambda_{a}}{2}\left(-\frac{m\lambda_{a}}{\hbar^{2}}\right)^{1/3}\left(n + \frac{1}{2}\right)^{-2/3}, \, \, n=0, 1, 2, \ldots\,.
\end{equation}

For the dual partner potential $V_{b}(\rho)=\lambda_{b}\rho^{2/3}$ with $\lambda_{b} > 0$, the~spectrum follows from (\ref{Gx6})~as
\begin{equation}\label{Gx8}
E_{b=2/3}(n) \doteq 2\lambda_{b}\left(\frac{8\hbar^{2}}{9m\lambda_b}\right)^{1/4} \left(n + \frac{1}{2}\right)^{1/2}, \, \, n=0, 1, 2, \ldots \,.
\end{equation}

\subsection{Duality in SUSY Semiclassical~Formulas}\label{Section3.4}
Let us begin this section with a brief comment on the semiclassical quantization in supersymmetric quantum mechanics (SUSYQM). In~SUSYQM,  there are semiclassical quantization formulas similar to WKB's.  A~unified form of them for a radial motion is
\begin{equation}\label{SS1}
\int_{r'}^{r''} \rmd r \sqrt{2m(E - \Phi^{2}(r))}  =\pi \hbar \left(\nu + \frac{1}{2} + \frac{\Delta}{2}\right), \, \quad \, \nu=0, 1, 2, \ldots \,,
\end{equation}
defined for the partner Hamiltonians $H_{\pm}$. In~(\ref{SS1}), $E$ is the eigenvalue of $H_{\pm}$, and~$\Phi(r)$ is the superpotential which is a solution of the Riccati equation in the form
\begin{equation}\label{SS2}
\Phi^{2}(r) \pm \frac{\hbar}{\sqrt{2m}}\frac{\rmd\Phi(r)}{\rmd r} - V(r) - \frac{\hbar^{2}(L^{2} - \frac{1}{4})}{2mr^{2}}=0
\end{equation}
where $V(r)$ is a potential function, $r'$ and $r''$ denote the turning points defined by $\Phi^2(r')=E=\Phi^2(r')$ with $r' \leq r''$, and~$L=\ell + (D-2)/2$ with $\ell =0, 1, 2, \ldots$. There, $\Delta$ is the Witten index whose values are $\Delta = \pm 1$ for good SUSY and $\Delta = 0$ for broken SUSY ({{}{SUSY} stands for supersymmetry. If~$H_{\pm}$ are the partner Hamiltonians, then spec$(H_{-})\setminus \{0\} =$ spec$(H_{+})$ for good SUSY, and~spec$(H{-})=$ spec$(H_{+})$ for broken SUSY}). The quantization condition for good SUSY was found by Comtet, Bandrauk, and~Campell~\cite{CBC}.  The~broken SUSY case and the general formulation of the form (\ref{SS2}) were derived by Eckhardt~\cite{Eck86} and independently by Inomata and Junker~\cite{IJ93}.  It is known that both the Comtet--Bandrauk--Campbell (CBC) formula for good SUSY and the Eckhardt--Inomata--Junker (EIJ) condition for broken SUSY yield the exact energy spectra for many shape-invariant potentials. For~detail, see reference~\cite{Junker2019}.

Now we wish to study the power-duality in SUSY semiclassical action on the left-hand side of (\ref{SS1}) only for the $H_{-}$ case.
Let us write the action of system $A$ as
\begin{equation}\label{SS3a}
W_{a}=\int_{r'}^{r''} \rmd r \sqrt{2m(E_{a} - \Phi_{a}^{2}(r))},
\end{equation}
where $E_{a}$ is the eigenvalue of $H_{-}$.
Suppose the superpotential in (\ref{SS3a}) has the form,
\begin{equation}\label{SS3}
\Phi_{a}(r) = \epsilon\sqrt{\lambda_{a}}r^{a/2} - \frac{\hbar}{\sqrt{2m}} \frac{\mu_{a}}{r},
\end{equation}
where $\epsilon = \pm 1$ and $a$ in the shoulder of $r$ is an arbitrary real number. The~potential term appearing in the SUSY semiclassical action (\ref{SS3a}) is the squared-superpotential rather than the usual potential $V(r)$. For~the superpotential (\ref{SS3}), it is
\begin{equation}\label{SS4}
\Phi_{a}^{2}(r) = \lambda_{a}r^{a} + \lambda_{a'}\, r^{a'} + \frac{\hbar^{2}\mu_{a}^{2}}{2mr^{2}},
\end{equation}
where
\begin{equation}\label{SS4a}
a'=(a-2)/2, \, \, \quad \, \lambda_{a'}=-\epsilon \hbar \mu_{a}\sqrt{2\lambda_{a}/m}.
\end{equation}

Then we have
\begin{equation}\label{SS5}
\Phi_{a}^{2}(r) - \frac{\hbar}{\sqrt{2m}}\frac{\rmd\Phi_{a}(r)}{\rmd r} = \lambda_{a}r^{a} + \left(1 + \frac{a}{4\mu_{a}}
\right)\lambda_{a'}\, r^{a'} + \frac{\hbar^{2}\mu_{a}(\mu_{a}-1)}{2mr^{2}}.
\end{equation}

Evidently, $\Phi_{a}(r)$ of  (\ref{SS3}) satisfies the Riccati Equation~(\ref{SS2}) with a two-term\linebreak power~potential,
\begin{equation}\label{SS6}
V_{a}(r) = \lambda_{a} r^{a} + \left(1 + a/(4\mu_{a}) \right)\lambda_{a'} r^{a'},
\end{equation}
provided that
\begin{equation}\label{SS7}
a'=(a-2)/2, \, \, \, \quad \, \mu_{a}=L_{a} + \, 1/2.
\end{equation}

In (\ref{SS6}), $a$ is arbitrary but $a'$ is dependent on $a$ as given by the first condition of (\ref{SS4a}). If~both $a$ and $a'$ are assumed to be independent and arbitrary, the~superpotential of the form (\ref{SS3}) cannot be a solution of the Riccati equation.
The quantity on the left-hand side of (\ref{SS5}) is a SUSY effective potential, denoted by $V_{a}^{(-)}(r)$, that belongs to the Hamiltonian $H_{-}$.  It is related to $V_{a}(r)$ of (\ref{SS6}) by
\begin{equation}\label{SS8}
V_{a}^{(-)}(r) =\Phi_{a}^{2}(r) - \frac{\hbar}{\sqrt{2m}}\frac{\rmd\Phi_{a}(r)}{\rmd r} = V_{a}(r) + \frac{\hbar^{2}(L^{2} - \frac{1}{4})}{2mr^{2}}.
\end{equation}

The superpotential (\ref{SS3}) works for the radial oscillator and the hydrogen atom in a unified manner as it contains the two as special cases:\\

\noindent (1) {{}{Radial} 
 harmonic oscillator} with $a=2$,  $a'=0$,\, $\lambda_{a}=\frac{1}{2}m\omega^{2}$, $\lambda_{a'}= - \hbar \omega \mu_{a}$, \mbox{$\mu_{a} =L_{a} + \frac{1}{2}$},  $\epsilon =1$ :
{\begin{eqnarray}\label{SS9}
  \Phi_{a}(r) &=& \sqrt{\frac{m}{2}} \omega r - \frac{\hbar}{\sqrt{2m}}\frac{\mu_{a}}{r}, \\
  V_{a}^{(-)}(r) &=& \frac{1}{2}\omega^{2} r^{2} + \frac{\hbar^{2}\mu_{a}(\mu_{a} -1)}{2mr^{2}} - \hbar \omega (\mu_{a} + \, 1/2),
\end{eqnarray}}
\begin{equation}\label{SS9a}
E_{a} - \Phi_{a}^{2} =(E_{a} + \hbar \omega \mu_{a}) - \frac{1}{2} m\omega^{2} r^{2} - \frac{\hbar^{2}\mu_{a}^{2}}{2mr^{2}} .
\end{equation}

The CBC quantization of (\ref{SS1}) with $\Delta = -1$ yields
{\begin{equation}\label{SS10}
E_{a} = 2\hbar \omega \nu, \, \quad \, \nu \in \mathbb{N}_{0},
\end{equation}}
which corresponds to the energy spectrum in quantum mechanics,
{\begin{equation}\label{SS10a}
E_{a}^{QM}(\nu, \ell) = E_{a} +  \hbar \omega \mu_{a} = \hbar \omega (2\nu + \ell + \,D/2 - \,1/2),
\end{equation}}
if $\mu_{a}=L_{a} + \,1/2 = \ell + D/2 -\,1/2 $ with $\ell \in \mathbb{N}_{0}$.\\

\noindent(2) {Hydrogen atom} with $a=0$,  $a'=-1$, $\epsilon =1$, $\lambda_{a}=me^{4}/(2\hbar^{2}\mu_{a}^{2})$, $\lambda_{a'}=-e^{2}$, $\mu_{a}=L_{a} + \frac{1}{2}$\,:
{\begin{eqnarray}\label{SS11}
  \Phi_{a}(r) &=& \frac{\sqrt{2m}}{2 \hbar \mu_{a}} e^{2} - \frac{\hbar}{\sqrt{2m}}\frac{\mu_{a}}{r}, \\
  V_{a}^{(-)}(r) &=& - \frac{e^{2}}{r}  + \frac{\hbar^{2}\mu_{a}(\mu_{a} -1)}{2mr^{2}} + \frac{me^{4}}{2\hbar^{2}\mu_{a}^{2}},
\end{eqnarray}}
\begin{equation}\label{SS11a}
E_{a} - \Phi_{a}^{2} =\left(E_{a} - \frac{m e^{4}}{2 \hbar^{2} \mu_{a}^{2}} \right)  + \frac{e^{2}}{r} - \frac{\hbar^{2} \mu_{a}^{2}}{2mr^{2}} .
\end{equation}

The CBC result is

\begin{equation}\label{SS11b}
  E_{a} = E_{a}^{\small{QM}}(\nu, \ell) + m e^{4}/(2 \hbar^{2} \mu_{a}^{2})
        = - \frac{m e^{4}}{2 \hbar^{2}(\nu + \ell + D/2 - \,1/2)^{2}} + \frac{m e^{4}}{2 \hbar^{2}(\ell + D/2 - \,1/2)^{2}},
\end{equation}
where $\nu, \ell \in \mathbb{N}_{0}$.

Next we change the radial variable $r$ by
\begin{equation}\label{SSB1}
\mathfrak{R}: \, \quad \, r=f(\rho) =C\rho^{\eta}, \, \quad \, \Leftrightarrow \, \quad \, \rho =f^{-1}(r)=C^{-1/\eta} r^{1/\eta}.
\end{equation}
and let the system described by the new variable be system $B$.
Upon application of (\ref{SSB1}), the~action $W_{a}$ of (\ref{SS3a}) transforms to
\begin{equation}\label{SSB2}
W_{b}=\int_{\rho'}^{\rho''}\, \rmd\rho \sqrt{2mf'^{2}(E_{a} - \Phi_{a}^{2})},
\end{equation}
where $f'=\rmd f(\rho)/\rmd\rho$ and
\begin{equation}\label{SSB3}
E_{a} - \Phi_{a}^{2}= E_{a} - \lambda_{a}r^{a} - \lambda_{a'}\, r^{a'} - \frac{\hbar^{2}\mu_{a}^{2}}{2mr^{2}}, \, \quad \, a'=(a-2)/2\,.
\end{equation}

Since $f'^{2}=\eta^{2}C^{2}\,\eta^{2\eta - 2}$,
\begin{equation}\label{SSB4}
f'^{2}(E_{a} - \Phi_{a}^{2})
 =\eta^{2}C^{2}E_{a}\rho^{2\eta - 2} - \eta^{2}C^{2+a} \lambda_{a} \rho^{a\eta + 2\eta -2} - \eta^{2}C^{2+a'}\lambda_{a'} \rho^{a'\eta + 2\eta - 2} - \frac{\hbar^{2} \eta^{2}\mu_{a}^{2}}{2m\rho^{2}}.
\end{equation}

If there is such a parameter $\eta$ that $f'^{2}(E_{a} - \Phi_{a}^{2})$ takes the form,
\begin{equation}\label{SSB5}
E_{b} - \Phi_{b}^{2} =E_{b} - \lambda_{b} \rho^{b} - \lambda_{b'} \rho^{b'} - \frac{\hbar^{2}\mu_{b}^{2}}{2m\rho^{2}},
\end{equation}
with
\begin{equation}\label{SSB5b}
b'=(b-2)/2,
\end{equation}
then the action is form-invariant under (\ref{SSB1}) and reciprocal, that is, $W_{a}=W_{b}$ and $W_{a}=\hat{X}(a, b) W_{b}$.
In the $\hat{X}(a,b)$-operation, we have temporarily let $r=\xi_{a}$ and $\rho=\xi_{b}$. We have also assumed that $\hat{X}(a, b)$ takes $b'=(b-2)/2$ to $a'=(a-2)/2$.  Furthermore, (\ref{SSB5}) together with (\ref{SSB5b}) implies that the new superpotential $\Phi_{b}(\rho)$ has the same form as that of $\Phi_{a}(r)$ in (\ref{SS3}), namely,
\begin{equation}\label{SSB5c}
\Phi_{b}(r) = \epsilon\sqrt{\lambda_{b}}r^{b/2} - \frac{\hbar}{\sqrt{2m}} \frac{\mu_{b}}{r}.
\end{equation}

If this were the case, we could establish the general power-duality of the \mbox{action (\ref{SS3a})} with the superpotential (\ref{SS3}). Unfortunately there is no way to transform system $A$ with an arbitrary power $a$ to system $B$ satisfying the conditions (\ref{SSB5}) and (\ref{SSB5b}).
Therefore, with~the superpotential (\ref{SS3}), we are unable to demonstrate in a general term the power-dual symmetry in SUSY semiclassical quantization. To~our knowledge, no qualified superpotential supporting the general power-duality in SUSY semiclassical action has ever been~reported.

Although we have to give up pursuing the general power-duality, we may find cases where duality occurs within the present scheme.  For~a dual symmetry, the~form-invariance of the superpotential $\Phi(r)$ is not an essential requirement, but~it is necessary that
$f'^{2}(E_{a} - \Phi_{a}^{2}(r))$ is reducible to the form $E_{b} - \Phi_{b}^{2}$ under the transformation $r=f(\rho)=C\rho^{\eta}$. There are two options for $\eta$ to reduce the left-hand side of (\ref{SSB4}) to the form of (\ref{SSB5}) under different conditions than (\ref{SSB5b}). Namely,
\begin{eqnarray*}
&(i)& \, \eta = 2/(a+2)=1/(a'+2), \, \quad \, a, a' \neq -2,\\
&(ii)& \, \eta  = 2/(a'+2) = 4/(a+2), \, \quad \, a, a' \neq -2.
\end{eqnarray*}

Let $\hat{D}(b, a)$ be such an operator that $\hat{D}(b, a)W_{a}=W_{b}$ under the change of \mbox{variable (\ref{SSB1})}. Since (\ref{SSB1}) with option (i) or (ii) is invertible, the~operator has an inverse. Hence $W_{a}=\hat{D}^{-1}(b, a) W_{b}$ in addition to $W_{a}=W_{b}$. Although~the strict reciprocity is broken, we can talk about the power-dual symmetry in this relaxed~sense.\\

Option (i):\,  Transformation $r=C\rho^{2/(a+2)}$ in (\ref{SSB1}) brings
\begin{equation}\label{SSC1}
E_{b} - \Phi_{b}^{2} =E_{b} - \lambda_{b} \rho^{b} - \lambda_{b'} \rho^{-1} - \frac{\hbar^{2}\mu_{b}^{2}}{2m\rho^{2}}.
\end{equation}
which contains a Coulomb-like potential in addition to a power potential for any value of $a$ other than $a=-2$ $(a'=-2)$. Option (i) must be associated with the substitutions,
\begin{equation}\label{SSC2}
E_{b}= - \eta^{2}C^{2+a} \lambda_{a}, \, \quad  \,  \lambda_{b}= - \eta^{2} C^{2} E_{a}, \, \quad \, \lambda_{b'} = \eta^{2} C^{2+a'} \lambda_{a'}, \, \, \quad \, \mu_{b}=\eta \mu_{a},
\end{equation}
and
\begin{equation}\label{SSC3}
\eta = \frac{2}{a+2}=\frac{1}{a'+2},\, \quad \,  b= - \frac{2a}{a+2}, \, \quad \, b'=-1.
\end{equation}

The second relation in (\ref{SSC2}) may be used to determine the constant $C$ of the\linebreak \mbox{transformation (\ref{SSB1})}. \\

Option (ii):\, Transformation $r=C\rho^{2/(a'+2)}$ yields
\begin{equation}\label{SSC4}
E_{b} - \Phi_{b}^{2} =E_{b} - \lambda_{b} \rho^{2} - \lambda_{b'} \rho^{b'} - \frac{\hbar^{2}\mu_{b}^{2}}{2m\rho^{2}},
\end{equation}
where  a Hooke potential appears in addition to a power potential for any $a \neq -2$.
Option~(ii) comes~with
\begin{equation}\label{SSC4a}
E_{b}= - \eta^{2}C^{2+a'} \lambda_{a'}, \, \quad  \,  \lambda_{b'}= - \eta^{2} C^{2} E_{a}, \, \quad \, \lambda_{b} = \eta^{2} C^{2+a} \lambda_{a}, \, \, \quad \, \mu_{b}=\eta \mu_{a},
\end{equation}
and
\begin{equation}\label{SSC5}
\eta = \frac{2}{a'+2}=\frac{4}{a+2}, \, \quad \, b=2, \, \quad \,  b'= - \frac{2a'}{a'+2} = -\frac{2(a-2)}{a+2}.
\end{equation}

Again, the~second relation of (\ref{SSC4a}) is able to fix the constant $C$. \\

\noindent {\bf Example 1.} The Coulomb--Hooke duality:

Option (i) is appropriate for the Hooke to Coulomb transition with $a=2$,  $a'=0$,  $b=-1$ and $b'=-1$.  By~$r= C\rho^{1/2}$,
\begin{equation}\label{SSE1a}
E_{a} - \Phi_{a}^{2} =\left(E_{a} + \hbar \omega \mu_{a}\right) - \frac{1}{2} m\omega^{2} r^{2} - \frac{\hbar^{2}\mu_{a}^{2}}{2mr^{2}} .
\end{equation}
transforms to
\begin{equation}\label{SSE1b}
E_{b} - \Phi_{b}^{2} =\left(E_{b} - \frac{m e^{4}}{2 \hbar^{2} \mu_{b}^{2}} \right)  + \frac{e^{2}}{\rho} - \frac{\hbar^{2} \mu_{b}^{2}}{2m\rho^{2}},
\end{equation}
where
\begin{equation}\label{SSE1c}
E_{b} - \frac{m e^{4}}{2 \hbar^{2} \mu_{b}^{2}}=- \frac{1}{8}m\omega^{2}C^{4}, \, \quad \, C^{2} = \frac{4e^{2}}{E_{a} + \hbar \omega \mu_{a}}, \, \quad \, \mu_{b}=\frac{1}{2}\mu_{a}.
\end{equation}

Combining the first and the second relation of (\ref{SSE1c}) gives
\begin{equation}\label{SSE1d}
E_{b}= - \frac{2m e^{4}}{ \hbar^{2} (E_{a}/\hbar \omega + \, \mu_{a})^{2}} + \frac{m e^{4}}{2 \hbar^{2} \mu_{b}^{2}}.
\end{equation}
which can be converted to the QM spectrum for the hydrogen atom
\begin{equation}\label{SSE1e}
E^{\small{QM}}_{b}(\nu, \ell) = E_{b} - \frac{m e^{4}}{2 \hbar^{2} \mu_{b}^{2}} = - \frac{m e^{4}}{2 \hbar^{2}(\nu + \ell + D/2 - \,1/2)^{2}},
\end{equation}
by substitution of $E_{a}=2\hbar \omega \nu $ and $\mu_{a}=2\mu_{b}= 2(\ell + D/2 - \,1/2)$.

Option (ii) is for the Coulomb to Hooke transition with $a=0$, $a'= -1$, $b=2$ and $b'=2$.  By~$\rho=C^{-1}r^{2}$, the~Equation~(\ref{SSE1b}) for the hydrogen atom transforms back to the Equation~(\ref{SSE1a}) for the radial oscillator. The~constant $C^{-1}$ appearing in the variable transformation is the inverse of $C$ obtainable from the second relation of (\ref{SSE1c}).
Obviously, for~the Coulomb--Hooke pair, option (ii) is the inverse of option (i). This confirms that the Coulomb--Hooke dual symmetry is valid in the SUSY semiclassical~action.\\

\noindent {\bf Example 2.} A~confinement problem: 

Option (i) and option (ii) may be used to study a confinement potential for which the superpotential (\ref{SS3}) is of the form,
\begin{equation}\label{SSE2a}
\Phi_{a}(r) = \epsilon\sqrt{\lambda_{a}}r^{1/2} - \frac{\hbar}{\sqrt{2m}} \frac{\mu_{a}}{r}, \, \quad \, (\epsilon =1, \, \, \lambda_{a} >0).
\end{equation}

Correspondingly, we have
\begin{equation}\label{SSE2b}
E_{a} - \Phi_{a}^{2}(r) =E_{a} - \lambda_{a} r  + \epsilon \hbar \mu_{a} \sqrt{\frac{2 \lambda_{a}}{m}} r^{-1/2} - \frac{\hbar^{2}\mu_{a}^{2}}{2mr^{2}}. \end{equation}

Option (i) with $a=1$ $(a'=-1/2)$ gives $\eta=2/3$.  By~$r=C\rho^{2/3}$, (\ref{SSE2a}) transforms to
\begin{equation}\label{SSE2c}
E_{b} - \Phi_{b}^{2}(\rho) =E_{b} - \lambda_{b} \rho^{-2/3}  -  \lambda_{b'} \rho^{-1} - \frac{\hbar^{2}\mu_{b}^{2}}{2mr^{2}} ,
\end{equation}
where
\begin{equation}\label{SSE2ca}
E_{b}= - \frac{4}{9}C^{3}\lambda_{a}, \, \,  \lambda_{b}= - \frac{4}{9}C^{2} E_{a}, \, \, \lambda_{b'} = - \epsilon \frac{4}{9}\hbar \mu_{b}C^{3/2}\sqrt{\frac{2\lambda_{a}}{m}}, \, \, \mu_{b}=\frac{2}{3} \mu_{a}.
\end{equation}

The result (\ref{SSE2c}) is not particularly interesting because it is not integrable. However, it is interesting that the limit $\lambda_{b} \rightarrow 0$ implies $E_{a} \rightarrow 0$. Hence the states in the vicinity of the zero-energy state of system $A$ may be approximated by a set of states of the hydrogen~atom.

Option (ii) with $a'=-1/2$ implies $\eta=4/3$.  The~transformation $r=C\rho^{4/3}$ reduces $E_{a} - \Phi_{a}^{2}(r)$ of (\ref{SSE2b})
to the form,
\begin{equation}\label{SSE2d}
E_{b} - \Phi_{b}^{2} =E_{b} - \lambda_{b} \rho^{2/3} - \lambda_{b'} \rho^{2} - \frac{\mu_{b}^{2} \hbar^{2}}{2m\rho^{2}},
\end{equation}
where
\begin{equation}\label{SSE2e}
E_{b}= \epsilon \frac{16}{9} \hbar \mu_{a} C^{3/2}\sqrt{\frac{2\lambda_{a}}{m}}, \, \, \lambda_{b}=- \frac{16}{9} C^{2}E_{a}, \, \,
\lambda_{b'}=  \frac{16}{9} C^{3} \lambda_{a}, \, \, \mu_{b}=\frac{4}{3} \mu_{a}.
\end{equation}

In the limit $\lambda_{b} \rightarrow 0$, system $B$ becomes a radial harmonic oscillator with the coupling constant, $\lambda_{b'} >0$.
Thus, the states of system $A$ in the vicinity of $E_{a}=0$ may be approximated by those of such a radial harmonic oscillator. The~confinement  problem will be revisited {}{in} Section~\ref{Section2.4}.\\

\noindent {\bf Remark 18:}
The duality relation between system $A$ and system $B$ is reciprocal in the sense that the two systems are bijectively mapped to each other. Hence, if~system $A$ is dual to system $B$ then system $B$ is dual to system $A$.  For~instance, the~Newton--Hooke duality in classical mechanics is reciprocal. The~Newton--Hooke duality is the Hooke--Newton duality. The~map from the Newton system to the Hooke system is bijective. By~contrast, it has been known~\cite{Louck, Bergmann, Kostelecky} that all the states of the hydrogen atom in three dimensions correspond to half the states of the isotopic harmonic oscillator in four dimensions. The~map from the three dimensional Coulomb system (of $\ell_{cou}=1, 2,3,\ldots$) to the four dimensional oscillator (of $\ell_{osc}=2, 4, 6, ...$) is injective. Hence all the states of the oscillator as a Hooke system (with $\ell_{osc}=0,1, 2, ...$) cannot be mapped back to the Coulomb system (with $\ell_{cou} = 0, 1, 2, ...$). The~relation between the Coulomb system and the Hooke system at the level of the quantum structures is not reciprocal~\cite{LouckShaffer1960,Bergmann}.\\

\noindent {\bf Remark 19:}
The Langer replacement, $\sqrt{\ell (\ell + D-2)} \rightarrow  \ell + (D- 2)/2$, is an ad hoc procedure introduced so as to be consistent with the quantum mechanical results~\cite{Langer37}.
In the literature~\cite{John}, it has been suggested to regard the angular momentum $L$ appearing in the Schr\"odinger equation as a continuous parameter since an arbitrary inverse square potential can be added to make the quantized angular momentum continuous. This reasoning, however, would make Langer's replacement~nonsensical.\\

\noindent {\bf Remark 20:}
Recall that $\eta=-b/a$ for a dual pair $(a, b)$ and that $\ell_{b}=\eta \ell_{a}$ and $D_{b}-2=\eta(D_{a}-2)$. Although~$\eta$ can be any positive real number,  in~the following, we list a few examples of relevant numbers and relations for integral values of $\eta$:
\begin{equation*}
\begin{array}{c|c|l|l|l}
\,\, \eta \,\,&(a, b)  & \ell_{a}=0, 1, 2, ... \qquad & \ell_{b}=0, 1, 2, ... \qquad &D_{a}= 2, 3, ... \nonumber \\
\hline
2 & (-1, 2) \, \, \, \, & \ell_{b}=0, 2, 4, ... \, \,  & \ell_{a}=0, 1/2, 1, ... \, \,  &D_{b}=2D_{a} -2 \nonumber \\
3 & (-4/3, 4) \, \, \, \, & \ell_{b}=0, 3, 6, ... \, \, & \ell_{a}=0, 1/3, 2/3, ... \, \,   &D_{b}=3D_{a} - 4  \nonumber \\
4 & (-3/2, 6) \, \, \, \,& \ell_{b}=0, 4, 8, ...\,   &\ell_{a} =0, 1/4, 1/2, ... \, \,  &D_{b}=4D_{a} - 6 \nonumber \\
\end{array}
\end{equation*}
For example, from~the line of $\eta=2$, we see that the states of the Coulomb system in $D_{a}=3$ correspond to half the states of the Hooke system in $D_{b}=4$. System $A$ and system $B$ cannot be reciprocal as long as the equality $\ell_{b}=\eta \ell_{a}$ is~assumed.\\

\noindent {\bf Remark 21:}
The time transformation ${{\mathfrak{T}}}$ has no role to play because the semiclassical action does not explicitly depend on time as a solution of the stationary Hamilton--Jacobi~equation.\\

\noindent {\bf Remark 22:}
The condition $L^{2}=0$ assumed for the example in (\ref{Gx4}), if~the Langer replacement (\ref{G2}) is employed, implies $\ell = 0$, which occurs only in two~dimensions.\\

\noindent {\bf Remark 23:}
The spectrum (\ref{Gx7}) for $a=-1/2$ is similar to the approximate result obtained from an exact solution of Schr\"odinger's equation in one dimension~\cite{Ishk16}.\\

\noindent {\bf Remark 24:}
The action on either side of (\ref{G12}) is not always integrable in closed form. Suppose the power $a$ of the potential $V_{a}$ be a non-zero integer. Then there are a few integrable examples. If~$a=2, -1$ or $-2$ then the action of system $A$ is reducible to an elementary function, and~if $a= 6, 4, 1, -3, -4$ or $-6$ then it can be expressed in terms of an elliptic function. Therefore, $(2, -1)$, $(-3, -6)$, $(-4, -4)$, $(1, -2/3)$, $(4, -1/3)$ and $(6, -3/2)$ are integrable dual pairs $(a,b)$ when $a$ is an integer other than $0$ and $-2$ though $b$ is not necessarily an integer. To~$a=-2$, there corresponds the self-dual pair $(-2, -2)$ with $\eta=1$.

\section{Power-Duality in Quantum~Mechanics}\label{Section4}
The main object to be studied for the power-duality in quantum mechanics is the energy eigenequation of the form $\hat{H}|\psi \rangle =E|\psi \rangle$ where $\hat{H}$ is the Hamiltonian operator for a system in a power-law potential.  Since one of the key operations in the power-duality transformation is the change of variable $r=C\rho^{\eta}$, we have to deal with the eigenequation in the radial coordinate representation, that is, the~radial Schr\"odinger equation.  In~the context of the duality argument, the~radial Schr\"odinger equation with power-law potentials have been exhaustively explored in the literature~\cite{QR,Gaze,John}. There is little room available to add something new.
The aim of this section is, however, to~present from the symmetry point of view the power-duality of the radial Schr\"odinger equation in parallel to the classical and semiclassical approaches. The~power-duality in the path integral formulation of quantum mechanics is important but is not included in the present~paper.

\subsection{The Action for the Radial Schr\"odinger~Equation}\label{Section4.1}
The stationary Schr\"odinger equation for a $D$ dimensional system in a central-force potential $V(r)$ can be separated in polar coordinates into a radial equation and an angular part. The~radial Schr\"odinger equation has the form,
\begin{equation}\label{H1a}
\left\{-\frac{\hbar^{2}}{2m}\left(\frac{\rmd^{2}}{\rmd r^{2}} + \frac{D-1}{r}\frac{\rmd}{\rmd r}\right) + \frac{\hbar^{2}\ell(\ell + D -2)}{2mr^{2}} + V(r) - E \right\}R_{\ell}(r) = 0.
\end{equation}

In the above equation, the~angular contribution appears in the third term, which stems from
$\hat{L}^{2} \mathcal{Y}_{\ell}^{m}(\mathbf{r}/r) = \ell(\ell + D - 2)\mathcal{Y}_{\ell}^{m}(\mathbf{r}/r)$ where $\hat{L}$ is the angular momentum operator and $\mathcal{Y}_{\ell}^{m}(\mathbf{r}/r)$ is the hyperspherical harmonics. Substituting $R_{\ell}(r)=r^{(1- D)/2}\psi_{\ell}(r)$ reduces it to a simplified differential equation on the positive half-line,
\begin{equation}\label{H1}
\left\{-\frac{\hbar^{2}}{2m}\frac{\rmd^{2}}{\rmd r^{2}} + \frac{\hbar^{2}(L^{2} - 1/4)}{2mr^{2}} + V(r) - E \right\} \psi_{\ell}(r) = 0,
\end{equation}
where
\begin{equation}\label{H2}
L = \ell + (D-2)/2, \, \, \quad \, \ell=0, 1, 2, ....
\end{equation}

For the sake of simplicity, we shall call Equation~(\ref{H1}) the {radial equation} and $\psi_\ell(r)$ the {wave function}.
The angular quantity $L$ in (\ref{H2}) is precisely the same as Langer's \mbox{choice (\ref{G2})} in the semiclassical action (see {{}{Remark 25}
}). Under~operation $\mathfrak{L}: L_{a}=L_{b}/\eta$, the~same problem that we have encountered in the semiclassical case should recur with the \mbox{equality (\ref{H2})}. Therefore, again, we adopt the view that the power-duality is basically a classical notion and follow the steps taken previously to circumvent the problem. Namely, for~the duality argument, we treat $L$ and $E$ in (\ref{H1}) as continuous parameters. Only after the duality is established, we replace the parameters by their quantized counterparts. We consider that operation $\mathfrak{L}$ applies only to the angular parameter and that $L_{a}=L_{b}/\eta$ does not imply $\ell_{a} + D/2 - 1= (\ell_{b} + D/2 -1)/\eta$. The~last equality breaks the reciprocity that $\ell_{a} \in \mathbb{N}_{0}$ and $\ell_{b} \in \mathbb{N}_{0}$.  The~relation (\ref{H2}) holds true for each quantum system as an internal structure being independent of duality~operations.

Suppose that system $A$ has a two-term power potential $V_{a}(r)=\lambda_{a}r^{a} + \lambda_{a'}r^{a'}$ where $a \neq a'$.  Defining the modified potential,
\begin{equation}\label{H3a}
U_{a}(r)=\lambda_{a}r^{a} + \lambda_{a'}r^{a'} - E_{a},
\end{equation}
we write the radial Equation~(\ref{H1}) as
\begin{equation}\label{H3}
\left\{ \frac{\rmd^{2}}{\rmd r^{2}} - \frac{L_{a}^{2} - 1/4}{r^{2}} - \frac{2m}{\hbar^{2}} U_{a}(r) \right\}\psi_{a}(r)=0.
\end{equation}

Since we ignore the relation (\ref{H2}) for a while, we have dropped the subscript $\ell$ of the state function $\psi_{a}(r)$.  The~radial Equation~(\ref{H3}) for system $A$ is derivable from the following action integral,
\begin{equation}\label{H4}
W_{a} = \int_{\sigma_{a}} \rmd r \, \mathcal{L}_{a}\left(\textstyle \frac{\rmd\psi_{a}^{\ast}(r)}{\rmd r},\frac{\rmd\psi_{a}(r)}{\rmd r}; \psi_{a}^{\ast}(r), \psi_{a}(r)\right),
\end{equation}
having a fixed range $\sigma_{a} \ni r$ and the Lagrangian of  the form,
\begin{eqnarray}\label{H5}
\mathcal{L}_{a} &= & \frac{\rmd\psi_{a}^{\ast}(r)}{\rmd r} \frac{\rmd\psi_{a}(r)}{\rmd r} + \left(\frac{L_{a}^{2}
- 1/4}{r^{2}} + \frac{2m}{\hbar^{2}}U_{a}(r)\right) \psi_{a}^{\ast}(r) \psi_{a}(r) \nonumber \\
~ && - \frac{1}{2}\frac{\rmd}{\rmd r}\left(\psi_{a}^{\ast}(r)\frac{\rmd\psi_{a}(r)}{\rmd r} + \psi_{a}(r)\frac{\rmd\psi_{a}^{\ast}(r)}{\rmd r}\right),
\end{eqnarray}
where $\psi_{a}^{\ast}(r)$ is the complex conjugate of $\psi_{a}(r)$.  Here we assume that the wave function $\psi_{a}(r)$ and its derivative are finite over the integration range $\sigma_{a}$.  The~last term of (\ref{H5}) is completely integrable, so that it contributes to the action as an unimportant additive constant. Use of the equality,
\begin{equation}\label{HA1}
\frac{\rmd\psi_{a}^{\ast}(r)}{\rmd r} \frac{\rmd\psi_{a}(r)}{\rmd r} = -  \psi_{a}^{\ast}(r) \frac{\rmd^{2}\psi_{a}(r)}{\rmd r^{2}}
+ \frac{\rmd}{\rmd r}\left(\psi_{a}^{\ast}(r)\frac{\rmd\psi_{a}(r)}{\rmd r}\right),
\end{equation}
enables us to put the Lagrangian (\ref{H5}) into an alternative form,
\begin{eqnarray}\label{HA2}
\mathcal{L}_{a}' &= & - \psi_{a}^{\ast}(r) \left\{\frac{\rmd^{2}\psi_{a}(r)}{\rmd r^{2}} - \left(\frac{L_{a}^{2} - 1/4}{r^{2}}
+ \frac{2m}{\hbar^{2}}U_{a}(r)\right) \psi_{a}(r) \right\} \nonumber \\
~ && + \frac{1}{2}\frac{\rmd}{\rmd r}\left(\psi_{a}^{\ast}(r)\frac{\rmd\psi_{a}(r)}{\rmd r} - \psi_{a}(r)\frac{\rmd\psi_{a}^{\ast}(r)}{\rmd r}\right).
\end{eqnarray}

The Euler--Lagrange equation, resulted from $\delta W/\delta \psi_{a}^{\ast}=0$,
\begin{equation}\label{H6}
\frac{\rmd}{\rmd r}\left\{\frac{\partial \mathcal{L}_{a}}{\partial \left(\frac{\rmd\psi_{a}^{\ast}}{\rmd r}\right)}\right\} - \frac{\partial \mathcal{L}_{a}}{\partial \psi_{a}^{\ast}} = 0,
\end{equation}
readily yields, with~either of $\mathcal{L}_{a}$ or $\mathcal{L}_{a}'$, the~radial Equation~(\ref{H3}). Since $\mathcal{L}_{a}$ is symmetric with respect to $\psi(r)$ and $\psi^{\ast}(r)$,  the~complex conjugate of (\ref{H3}) can be derived from it. However, $\mathcal{L}_{a}'$ is inappropriate for deriving the radial equation for $\psi_{a}^{\ast}(r)$.  For~now we put $\mathcal{L}_{a}'$ aside even though there is no need for complex conjugation of the radial equation. For~studying the power-duality in quantum mechanics, we focus our attention on the action $W_{a}$ of (\ref{H4}) with the Lagrangian (\ref{H5}) rather than the radial Equation~(\ref{H3}).

The symmetry operations that we consider for the power-duality in quantum mechanics are as follows
\begin{equation}\label{Sym1}
\mathfrak{R}: \, \,  r=f(\rho) = C\rho^{\eta} \, \, \quad (C > 0),
\end{equation}
\begin{equation}\label{Sym2}
\mathfrak{L}: \, \, L_{b}=\eta L_{a},
\end{equation}
\begin{equation}\label{Sym3}
\mathfrak{E}:  \, \, \, E_{b}=- \eta^{2} C^{a+2} \lambda_{a}, \, \quad \, \lambda_{b} = - \eta^{2} C^{2} E_{a},
\end{equation}
\begin{equation}\label{Sym4}
\mathfrak{C}: \, \, \eta=2/(a+2)=(b+2)/2, \, \quad \, (a \neq -2, \, \, b\neq -2),
\end{equation}
\begin{equation}\label{Sym5}
\mathfrak{B}: \, \, \lambda_{b'}=\lambda_{a'}\left(2/(a+2)\right)^{2} C^{a'+2}, \, \, \quad \, \, b'=2(a' - a)/(a +2),
\end{equation}
\begin{equation}\label{Sym6}
\mathfrak{F}: \, \,  \psi_{a}(r)=h(\rho)\psi_{b}(\rho).
\end{equation}

In (\ref{Sym6}), $h(\rho)$ is a continuous positive real function of $\rho$.

As $\rmd r$ goes to $\rmd\rho$, the~integration range of (\ref{H4}) changes from $\sigma_{a} \ni r$ to $\sigma_{b} \ni \rho$. \mbox{Under~(\ref{Sym1})}  and (\ref{Sym6}), the~first term of the Lagrangian (\ref{H5}) transforms as
\begin{equation}\label{H7b}
\frac{\rmd\psi_{a}^{\ast}(r)}{\rmd r} \frac{\rmd\psi_{a}(r)}{\rmd r} = \frac{h^{2}}{f'^{2}}\, \left\{\frac{\rmd\psi_{b}^{\ast}(\rho)}{\rmd\rho} \frac{\rmd\psi_{b}(\rho)}{\rmd\rho}
- \left[\frac{\rmd}{\rmd\rho}\left(\frac{h'}{h}\right) - \left(\frac{h'}{h}\right)^{2}\right]\psi_{b}^{\ast}\psi_{b} \right\}
 + \frac{h^{2}}{f'^{2}} \frac{\rmd}{\rmd\rho}\left(\frac{h'}{h}\psi_{b}^{\ast}\psi_{b}\right).
\end{equation}

By choice, we let $h^{2}(\rho)=f'(\rho)$. Then the second term on the right-hand side of (\ref{H7b}) reduces to the Schwarz derivative
\begin{equation}\label{Schwarz}
  \mathcal{S}[f]=\frac{f'''}{f'} - \frac{3}{2}\left(\frac{f''}{f'}\right)^{2}
\end{equation}
divided by $2f'$. The~third term of (\ref{H7b}) can be decomposed to two terms by using\linebreak the~relation,
\begin{equation}\label{H7c}
\frac{\rmd}{\rmd r}\left(\psi_{a}^{\ast}(r) \psi_{a}(r)\right)=\frac{h^{2}}{f'}\frac{\rmd}{\rmd\rho}\left(\psi_{b}^{\ast}(\rho) \psi_{b}(\rho)\right)
+ \frac{2hh'}{f'}\psi_{b}^{\ast}\psi_{b}.
\end{equation}

Therefore,
\begin{equation}\label{H7d}
\begin{array}{l}
\displaystyle
\frac{\rmd\psi_{a}^{\ast}(r)}{\rmd r} \frac{\rmd\psi_{a}(r)}{\rmd r}
- \frac{1}{2}\frac{\rmd}{\rmd r}\left[\frac{\rmd}{\rmd r} \left(\psi_{a}^{\ast}(r)\psi_{a}(r)\right)\right]
 = \\[4mm]
\displaystyle\qquad
\frac{1}{f'}\, \left\{\frac{\rmd\psi_{b}^{\ast}(\rho)}{\rmd\rho} \frac{\rmd\psi_{b}(\rho)}{\rmd\rho} -
 \frac{1}{2}\mathcal{S}[f] \psi_{b}^{\ast}\psi_{b} \right\} - \frac{1}{2f'}\frac{\rmd}{\rmd\rho}\left[\frac{\rmd}{\rmd\rho}\left(\psi_{b}^{\ast}(\rho) \psi_{b}(\rho)\right)\right].
\end{array}
\end{equation}

The angular term of the Lagrangian (\ref{H5}) transforms as
\begin{equation}\label{H7e}
\frac{L_{a}^{2} - 1/4}{r^{2}} \psi_{a}^{\ast}(r)\psi_{a}(r)=\frac{1}{f'}\frac{g(L_{a}^{2} - 1/4)}{f^{2}}\psi_{b}^{\ast}(\rho)\psi_{b}(\rho)
\end{equation}
where $g$ denotes $f'^{2}$ as in the classical and semiclassical cases.  The~energy-potential term \mbox{of (\ref{H5})} changes as
\begin{equation}\label{H7f}
\frac{2m}{\hbar^{2}} U_{a}(r) \psi_{a}^{\ast}(r)\psi_{a}(r) = \frac{2m}{\hbar^{2}} g U_{a}(f(\rho)) \psi_{b}^{\ast}(\rho)\psi_{b}(\rho).
\end{equation}

Moreover, we let $f(\rho)=C\rho^{\eta}$ as defined by (\ref{Sym1}). Then $\mathcal{S}[f]=-(\eta^{2}-1)/2$, $g=C^{2}\eta^{2}\rho^{2\eta - 2}$ and $g/f^{2}=C^{2}\eta^{2}\rho^{2}$. Hence, we have
\begin{equation}\label{Schwarz2}
g(L_{a}^{2} - 1/4)/f^{2} - (1/2)\mathcal{S}[f]= (\eta^{2} L_{a}^{2} - 1/4)/\rho^{2},
\end{equation}
which results in $(L_{b}^{2} - 1/4)/\rho^{2}$ under $\mathfrak{L}: \, L_{b}=\eta L_{a}$.  Changing the variable by (\ref{Sym1}) and making the energy-coupling exchange by (\ref{Sym3}) result in
\begin{equation}\label{H7g}
g(\rho) U_{a}(C\rho^{\eta})=- E_{b} \rho^{a\eta + 2\eta -2} + C^{a'+2} \lambda_{b'}\rho^{a'\eta + 2\eta - 2} + \lambda_{b}\rho^{2\eta -2},
\end{equation}
which is written as
\begin{equation}\label{H8}
U_{b}(\rho)=\lambda_{b}\rho^{b} + \lambda_{b'}\rho^{b'} - E_{b}
\end{equation}
with the help of (\ref{Sym4}) and (\ref{Sym5}). Namely, $U_{a}(r)$ goes to $U_{b}(\rho)$ by $U_{b}(\rho)=g(\rho)U_{a}(r)$.
Consequently, we obtain $W_{a}=W_{b}$ or, emphasizing the parameter dependence of the Lagrangian,
\begin{equation}\label{H9}
\int_{\sigma_{a}} \rmd r \, \mathcal{L}_{a}(\lambda_{a}, L_{a}, U_{a}) = \int_{\sigma_{b}} \rmd \rho \, \mathcal{L}_{b}(\lambda_{b}, L_{b}, U_{b}),
\end{equation}
where
\begin{equation}\label{H10}
\begin{array}{rl}
\mathcal{L}_{b} = & \displaystyle
\frac{\rmd\psi_{b}^{\ast}(\rho)}{\rmd\rho} \frac{\rmd\psi_{b}(\rho)}{\rmd\rho}
+ \left(\frac{L_{b}^{2} - 1/4}{\rho^{2}} + \frac{2m}{\hbar^{2}} U_{b}(\rho) \right)\psi_{b}^{\ast}(\rho) \psi_{b}(\rho)\\[4mm]
&\displaystyle
 - \frac{1}{2}\frac{\rmd}{\rmd\rho}\left(\psi_{b}^{\ast}(\rho)\frac{\rmd\psi_{b}(\rho)}{\rmd\rho} + \psi_{b}(\rho)\frac{\rmd\psi_{b}^{\ast}(\rho)}{\rmd\rho}\right).
\end{array}
\end{equation}

The last term of (\ref{H10}) is completely integrable and contributes to $W_{b}$ as an unimportant constant. We identify $\mathcal{L}_{b}$ of (\ref{H10}) with the Lagrangian of system $B$, use of which leads to the radial equation for system $B$,
\begin{equation}\label{H11}
\left\{ \frac{\rmd^{2}}{\rmd\rho^{2}} - \frac{L_{b}^{2} - 1/4}{\rho^{2}} - \frac{2m}{\hbar^{2}} U_{b}(\rho) \right\}\psi_{b}(\rho)=0.
\end{equation}

Apparently the form of the Lagrangian is preserved under the set of power-duality operations, $\{\mathfrak{R}, \mathfrak{L}, \mathfrak{C}, \mathfrak{E}, \mathfrak{B}, \mathfrak{F}\}$. Furthermore, with~the Lagrangians $\mathcal{L}_{a}$ of (\ref{H5}) and $\mathcal{L}_{b}$ \mbox{of (\ref{H10})}, the~equality (\ref{H9}) implies that the action $W$ of (\ref{H4}) is invariant under the same set of operations. By~(\ref{H9}) the complex conjugate of the radial Schr\"odinger Equation~(\ref{H1}) is as well assured to be~form-invariant.

To complete the procedure, as~we have done for the semiclassical case, we must replace in an ad hoc manner each of the angular momentum parameters by the quantized form $\ell + (D-2)/2$ with $\ell=0, 1, 2,\ldots$. Using the dot-equality introduced in Section~\ref{Section3.1}, we write the form-invariance of the action amended by the angular quantization with $\ell_{a}, \ell_{b} \in \mathbb{N}_{0}$,
\begin{equation}\label{H12}
\int_{\sigma_{a}} \rmd r \, \mathcal{L}_{a}(\lambda_{a}, \ell_{a} + (D-2)/2, U_{a}) \doteq \int_{\sigma_{b}} \rmd \rho \, \mathcal{L}_{b}(\lambda_{b},
\ell_{b} + (D-2)/2, U_{b}),
\end{equation}
which warrants that the radial Schr\"odinger Equation~(\ref{H1}) with the angular quantization (\ref{H2}) is form-invariant under the set of duality operations, $\{\mathfrak{R}, \mathfrak{L}, \mathfrak{C}, \mathfrak{E}, \mathfrak{B}, \mathfrak{F}\}$.  In~this modified sense we claim that two quantum systems with $V_{a}(r)=\lambda_{a}r^{a} + \lambda_{a'}r^{a'}$ and with $V_{b}(\rho)=\lambda_{b}\rho^{b}+ \lambda_{b'}\rho^{b'}$ are in power-duality provided that $(a+2)(b+2)=4$.

\subsection{Energy Formulas, Wave Functions and Green~Functions}\label{Section4.2}
In arriving at the invariance relation (\ref{H9}),  we have seen the equality $\rmd r\, \mathcal{L}_{a} = \rmd\rho\, \mathcal{L}_{b}$ under the duality operations. The~relation (\ref{H9}) is valid with the alternative Lagrangian $\mathcal{L}'$ of (\ref{HA2}), suggesting
$\rmd r \,\mathcal{L}_{a}' = \rmd\rho \,\mathcal{L}_{b}'$. The~last equality in turn leads to
\begin{equation}\label{QE1}
\left\{ \frac{\rmd^{2}}{\rmd r^{2}} - \frac{L_{a}^{2} - 1/4}{r^{2}} - \frac{2m}{\hbar^{2}} U_{a}(r) \right\}\psi_{a}(r)
= \frac{1}{h^{3}}\left\{ \frac{\rmd^{2}}{\rmd\rho^{2}} - \frac{L_{b}^{2} - 1/4}{\rho^{2}} - \frac{2m}{\hbar^{2}} U_{b}(\rho) \right\}\frac{1}{h}\psi_{a}(f(\rho))
\end{equation}
where $f'=h^{2}=C\eta\rho^{\eta-1}$.  Let $H_{a}(r)$ be the Hamiltonian for system $A$ in the $r$-representation, that is,
\begin{equation}\label{QE2}
H_{a}(r)= -\frac{\hbar^{2}}{2m}\frac{\rmd^{2}}{\rmd r^{2}} + \frac{\hbar^{2}(L_{a}^{2} - 1/4)}{2mr^{2}} + \lambda_{a}r^{a} + \lambda_{a'}r^{a'}.
\end{equation}

Similarly, we define $H_{b}(\rho)$ for system $B$.  By~using the exchange symbol $X(b, a)$, we have
$H_{b}(\xi_{b}) - E_{b}=X(b, a)\{H_{a}(\xi_{a}) - E_{a}\}$ where $\xi_{a}=r$ and $\xi_{b}=\rho$. Then the \mbox{equality (\ref{QE1})} may be put into the form,
\begin{equation}\label{QE3}
\{H_{a}(r) - E_{a}\}\psi_{a}(r)= \frac{1}{h^{3}}\{H_{b}(\rho) - E_{b}\}\psi_{b}(\rho),
\end{equation}
when $\psi_{a}(r)=h(\rho)\psi_{b}(\rho)$ with $r=f(\rho)$ and $f'=h^{2}$.
Evidently, the~radial Equation~(\ref{H3}), expressed as $\{H_{a}(r) - E_{a}\}\psi_{a}(r)=0$, implies $ \{H_{b}(\rho) - E_{b}\}\psi_{b}(\rho)=0$.

\subsubsection{Energy~Formulas}\label{Section4.2.1}
To find the energy spectrum of system $A$, we usually solve the radial Equation \mbox{of (\ref{H3})} by specifying boundary conditions on $\psi_{a}(r)$.  Suppose we found a solution $\psi_{a}(r; \nu)$ compatible with the given boundary conditions when the energy parameter took a specific value $E_{a}(\nu)$ characterized by a real number $\nu$. This solution may be seen as an\linebreak eigenfunction~satisfying
\begin{equation}\label{QE4}
H_{a}(r) \psi_{a}(r; \nu) = E_{a}(\nu) \psi_{a}(r; \nu).
\end{equation}

Since operation $\mathfrak{F}$ demands $\psi_{a}(r; \nu) = \psi_{a}(f(\rho); \nu) = h(\rho) \psi_{b}(\rho; \nu)$, the~Equation~(\ref{QE4}) should imply via the equality (\ref{QE3})
\begin{equation}\label{QE5}
H_{b}(\rho) \psi_{b}(\rho; \nu)\rangle  = E_{b}(\nu) \psi_{b}(\rho; \nu).
\end{equation}

This shows that the number $\nu$ is a dual invariant being common to $E_{a}(\nu)$ and $E_{b}(\nu)$.
As has been repeatedly mentioned earlier, the~duality operations cannot interfere the internal structure of each quantum system.
In general, there are a number of solutions for the given boundary conditions. Thus, $\nu$ may be representing a set of numbers. Then we understand that the value of $\nu$ is preserved by $\mathfrak{F}$.  For~a while, however, we treat $\nu$ as another parameter and express the energy $E_{a}$  as a function of $\lambda_{a}$, $L_{a}$ and $\nu$,
\begin{equation}\label{QE6}
E_{a}=E_{a}(\lambda_{a}, L_{a}, \nu).
\end{equation}

This corresponds to the energy function $E_{a}(\lambda_{a}, L_{a}, N)$ in the semiclassical case.  We convert this energy function to the energy spectrum of system $A$ by replacing the parameters $L_{a}$ and $\nu$ to their quantum counterparts.
If we restrict our interest to bound state solutions, the~parameter $\nu$ is to be replaced by a set of discrete numbers $\nu=0, 1, 2, \ldots$. Furthermore, putting the angular parameter $L_{a}$ into the Langer form (\ref{H2}), we obtain the discrete energy spectrum of system $A$,
\begin{equation}\label{QE7}
E_{a}(\ell_{a}, \nu)=E_{a}(\lambda_{a}, \ell_{a} + D/2 - 1, \nu),
\end{equation}
where $\ell_{a} \in \mathbb{N}_{0}$ and $\nu \in \mathbb{N}_{0}$.

Since the energy functions $E_{a}(\lambda_{a}, L_{a}, \nu)$ and $E_{b}(\lambda_{b}, L_{b}, \nu)$ are related by the classical energy formulas, (\ref{EF2}) and (\ref{EF5})--(\ref{EF6}), the~corresponding energy spectra $E_{a}(\ell_{a}, \nu)$ and $E_{b}(\ell_{b}, \nu)$ can be related by the same formulas provided the angular parameter and the quantum parameter are properly expressed in terms of quantum numbers.
Knowing the energy spectrum of the form $E_{a}(\ell_{a}, \nu)=\mathcal{E}(\lambda_{a}, L_{a}, \nu)$ for system $A$, we can determine the energy spectrum $E_{b}$ of system $B$ by
\begin{equation}\label{QE8}
E_{b}( \ell_{b}, \nu)= - \eta^{2} C^{a+2} \mathcal{E}^{-1}(-\lambda_{b}/(\eta^{2} C^{2}), L_{b}/\eta, \nu),
\end{equation}
where $L_{b}=\ell_{b} + D/2 -1$ with $\ell_{b} \in \mathbb{N}_{0}$.
For the bound state spectrum, $\nu =0, 1, 2, \ldots$.

If the energy spectrum of system $A$ is given in the form
\begin{equation}\label{QE9}
E_{a}(\ell_{a}, \nu) = \pm \frac{1}{4}(a+2)^{2} |\lambda_{a}|^{2/(a+2)}
\left[\mathcal{F}\Bigl(\sqrt{2/(a+2)\,}(\ell_{a} +D/2 -1), \nu\Bigr)\right]^{1/a}
\end{equation}
then the energy spectrum of system $B$ is given by
\begin{equation}\label{QE10}
E_{b}(\ell_{b}, \nu) = \pm \frac{1}{4}(b+2)^{2} |\lambda_{b}|^{2/(b+2)}
\left[\mathcal{F}\Bigl(\sqrt{2/(b+2)\,}(\ell_{b} + D/2 -1), \nu\Bigr)\right]^{1/b}.
\end{equation}

These relations are the same as the semiclassical relations (\ref{Gex2} and (\ref{Gex3}) where the signs are determined by the signs of the coupling constants, ${\rm sgn}\,E_{a} = - {\rm sgn}\,\lambda_{b}$ and ${\rm sgn}\,E_{b} = - {\rm sgn}\,\lambda_{a}$.

\subsubsection{Wave~Functions}
The wave function transforms as  $ \psi_{a}(r; L_{a}, \nu) = h(\rho) \psi_{b}(\rho; L_{b}, \nu)$. Therefore, if~an eigenfunction of system $A$ is given, then the corresponding eigenfunction of system $B$ can be determined by
\begin{equation}\label{QE11}
\psi_{b}(\rho; L_{b}, \nu)=\frac{1}{h(\rho)} \psi_{a}(C\rho^{\eta}; L_{b}/\eta, \nu),
\end{equation}
where $L_{b}=\ell_{b} + D/2 -1$ with $\ell_{b} \in \mathbb{N}_{0}$. Both $\psi_{a}(r)$ and $\psi_{b}(\rho)$ as eigenfunctions are supposed to be square-integrable, and~each of them must be normalizable to unity.  However, even if $\psi_{a}(r)$ is normalized to unity, it is unlikely that $\psi_{b}(\rho)$ constructed by (\ref{QE11}) is normalized to unity. This is because
\begin{equation}\label{QE12}
\int_{0}^{\infty} \rmd r\, |\psi_{a}(r)|^{2} =  \int_{0}^{\infty} \rmd\rho\, g(\rho) \,|\psi_{b}(\rho)|^{2} =1
\end{equation}
where $g(\rho)=[f'(\rho)]^{2}= [h(\rho)]^{4}=C^{2}\eta^{2} \rho^{2(\eta -1)} $. In~this regard, if~system $A$ and system $B$ are power-dual to each other, the~formula (\ref{QE11}) determines $\psi_{b}(\rho)$ of system $B$ out of $\psi_{a}(r)$ of system $A$ except for the~normalization.

\subsubsection{Green~Functions}\label{Section4.2.2}
The Green function $G(r, r'; z)=\langle r |\hat{G}(z)|r' \rangle$ is the $r$-representation of the resolvent $\hat{G}(z)= (z - \hat{H})^{-1}$ where $z \in \mathbb{C}\backslash {\rm spec}\, \hat{H} $ and $\hat{H}$ is the Hamiltonian operator of the system in question. Let $E(\nu)$ and $|\psi(\nu) \rangle $ be the eigenvalue of $\hat{H}$ and the corresponding eigenstate, respectively, so that $\hat{H} |\psi(\nu)\rangle = E(\nu) |\psi(\nu)\rangle $. For~simplicity, we consider the case where $\nu \in \mathbb{N}_{0}$. Assume the eigenstates are orthonormalized and form a complete set, that is,
\begin{equation}\label{Green1}
\langle \psi(\nu) |\psi(\nu')\rangle = \delta_{\nu,\nu'}\, , \, \, \quad \, \, \sum_{\nu\in \mathbb{N}_{0}}  |\psi(\nu) \rangle \langle \psi(\nu)| = 1.
\end{equation}

From the completeness condition in (\ref{Green1}), it is obvious that
\begin{equation}\label{Green2}
\hat{G}(z) = \sum_{\nu \in \mathbb{N}_{0}}  \frac{|\psi(\nu) \rangle \langle \psi(\nu)|}{z - E(\nu)}.
\end{equation}

Hence, the~Green function can be written as
\begin{equation}\label{Green3}
G(r, r'; z) = \sum_{\nu\in \mathbb{N}_{0}}  \frac{\psi^\ast(r'; \nu) \psi(r; \nu)}{z - E(\nu)}.
\end{equation}

Use of Cauchy's integral formula leads us to the expression,
\begin{equation}\label{Green4}
\psi^{\ast}(r, \nu) \psi(r' ; \nu) = \frac{1}{2\pi i} \oint_{C_{\nu}}\rmd z\, G(r, r'; z),
\end{equation}
where the closed contour $C_{\nu}$ counterclockwise encloses only the simple pole $z=E(\nu)$ for a fixed value of $\nu$. Note that we will deal only with radial, hence one-dimensional, problems where no degeneracies can occur.
Multiplying both sides of (\ref{Green4}) by two factors $v(r)$ and $v(r')$ yields
\begin{equation}\label{Green4a}
\tilde{\psi}^{\ast}(r, \nu) \tilde{\psi}(r'; \nu) = \frac{1}{2\pi i} \oint_{C_{\nu}}\rmd z\, \tilde{G}(r, r'; z),
\end{equation}
where $\tilde{\psi}(r; \nu) = v(r) \psi(r; \nu)$ and $\tilde{G}(r, r' ; z)=v(r) v(r')G(r, r' ; z)$.

For instance, the~Green function $\mathcal{G}(r, r'; E)$ for the radial Schr\"odinger Equation~(\ref{H1a}) is related to the Green function
$G(r, r'; E)$ for the simplified radial Equation~(\ref{H1}) by
\begin{equation}\label{Green4b}
\mathcal{G}(r, r'; E)= (r\,r')^{(1-D)/2} G(r, r'; E)
\end{equation}
as  the wave functions of (\ref{H1a}) and (\ref{H1}) are connected by $R_{\ell}(r)=r^{(1- D)/2}\psi_{\ell}(r)$.

Suppose the Green functions of system $A$ and system $B$ are given, respectively, by~\begin{equation}\label{Green5}
\psi_{a}^{\ast}(r; \nu) \psi_{a}(r'; \nu) = \frac{1}{2\pi i} \oint_{C_{\nu}} \rmd z\, G_{a}(r, r'; z) ,
\end{equation}
and
\begin{equation}\label{Green6}
\psi_{b}^{\ast}(\rho; \nu) \psi_{b}(\rho' ; \nu) = \frac{1}{2\pi i} \oint_{C_{\nu}} \rmd z\, G_{b}(\rho, \rho' ; z) .
\end{equation}

By comparing these two expressions, we see that
if $\psi_{a}(r; \nu) = h(\rho) \psi_{b}(\rho; \nu)$  then
\begin{equation}\label{Green7}
G_{a}(r, r'; E_{a}(\nu))= h(\rho) h(\rho')G_{b}(\rho, \rho' ; E_{b}(\nu)).
\end{equation}

The above result is obtained without considering the detail of the Hamiltonian.  In~the following, an~alternative account is provided for deriving the same result by using the Hamiltonian explicitly. Let $\hat{H}_{a}$ be the Hamiltonian operator of system $A$ such that
$\langle r |\hat{H}_{a} - E_{a}|r' \rangle = (H_{a}(r) - E_{a}) \langle r|r' \rangle $. Then it is obvious that
\begin{equation}\label{HG1}
\{H_{a}(r) - E_{a}\} G_{a}(r,  r';E_{a}) = - \delta( r - r').
\end{equation}

According to (\ref{QE1}),  Equation~(\ref{HG1}) implies
\begin{equation}\label{HG2}
\{H_{b}(\rho) - E_{b}\}\frac{1}{h} G_{a}( f(\rho),  f(\rho');E_{a}) = - h^{3}(\rho)\delta(f(\rho) - f(\rho')).
\end{equation}

From the relations,
\begin{equation}\label{HG2a}
\int \, \rmd r\, |r \rangle \langle r| = \int \, \rmd\rho \,f'(\rho) |f(\rho) \rangle \langle f(\rho')| = \int \, \rmd\rho \, |\rho \rangle \langle \rho'| =1,
\end{equation}
there follows $|\rho \rangle = h(\rho) |f(\rho) \rangle$. Hence we have, $\langle \rho|\rho' \rangle = h(\rho) h(\rho') \langle f(\rho)|f(\rho') \rangle$, that is, $\delta(r-r')=\delta(f(\rho)- f(\rho')) = [h(\rho) h(\rho')]^{-1}\delta(\rho - \rho')$.
Thus, we arrive at the radial equation satisfied by the Green function of system $B$,
\begin{equation}\label{HG3}
\{H_{b}(\rho) - E_{b}\}G_{b}(\rho, \rho';E_b)  = - \delta ( \rho - \rho'),
\end{equation}
if the Green function transforms as
\begin{equation}\label{HG4}
 G_{a}(r, r_{0}; E_a , L_{a}) = h(\rho)h(\rho') G_{b}(\rho, \rho' ; E_b , L_{b}).
\end{equation}

Substitution of $L_{a}=\ell_{a} + D/2 -1$ with $\ell_{a} \in \mathbb{N}_{0}$ and $L_{b}=\ell_{b} + D/2 -1$ with $\ell_{b} \in \mathbb{N}_{0}$ into (\ref{HG4}) results in
\begin{equation}\label{HG4b}
G_{a}(r, r' ; E_a , \ell_{a} + D/2 -1) \doteq h(\rho)h(\rho_{0}) G_{b}(\rho, \rho' ; E_b , \ell_{b} + D/2 -1),
\end{equation}
which is not an equality as $\ell_{a} \in \mathbb{N}_{0}$ and $\ell_{b} \in \mathbb{N}_{0}$ are assumed.
Insofar as system $B$ is power-dual to system $A$,
the Green function of system $B$ can be expressed in terms of the Green function of system $A$ as
\begin{equation}\label{HG5}
G_{b}(\rho, \rho'; E_b , \ell_{b} + D/2 -1,\lambda_b,\lambda_{b'})
= [(f'(\rho) f'(\rho')]^{-1/2} G_{a}\left(f(\rho), f(\rho' ); E_a ,(\ell_{b} + D/2 -1)/\eta , \lambda_a, \lambda_{a'} \right)
\end{equation}
where $f(\rho)=C\rho^\eta$ and the parameters $E_a$, $\lambda_a$ and $\lambda_{a'}$ are given via the relations (\ref{Sym3}) and (\ref{Sym5}) in terms of $E_b$, $\lambda_b$ and $\lambda_{b'}$. This relation is an equality even though (\ref{HG4b}) is a dot equality.  An~expression similar to but slightly different from (\ref{HG5}) has been obtained by Johnson~\cite{John} in much the same~way.

\subsection{The Coulomb--Hooke Dual~Pair}\label{Section4.3}
Again, we take up the Coulomb--Hooke dual pair to test the transformation properties shown in Section~\ref{Section3.1}.  Let system $A$ be the hydrogen atom with $\lambda_{a}=- e^{2} < 0$ and system $B$ a radial oscillator with $\lambda_{b}=\frac{1}{2}m \omega^{2} > 0$.  So $(a, b)=(-1, 2)$ and $\eta=-b/a=2$. Both systems are assumed to be in $D$ dimensional space. The~Coulomb system has the scattering states $(E_{a} > 0)$ as well as the bound states $(E_{a} < 0)$. However, the~exchange relations (\ref{Sym3}) prohibits the process $(E_{a} > 0, \lambda_{a} <0) \Rightarrow (E_{b} >0, \lambda_{b} > 0)$. The~Coulomb--Hooke duality occurs only when the Coulomb system is in bound~states.

\underline{{}{The}
 energy relations:} \,
Suppose we know that the energy spectrum of system $A$ has the form,
\begin{equation}\label{QCspec}
E_{a}(\lambda_{a}, L_{a}, \nu)= - \frac{me^{4}}{2\hbar^{2}(\nu + L_{a} +1/2)^{2}},
\end{equation}
where $\lambda_{a}=-e^{2}$, $\nu \in \mathbb{N}_{0}$ and $L_{a}=\ell_{b} + D/2 -1$ with $\ell_{a} \in \mathbb{N}_{0}$.  Then the formula (\ref{QE9}) leads~to
\begin{equation}\label{QCHe1}
\mathcal{F}\Bigl(\sqrt{2} L_{a}, \nu \Bigr)= \frac{\hbar^{2}}{2m}\left[\nu + (\sqrt{2} L_{a})/\sqrt{2} +1/2 \right]^{2}.
\end{equation}

Careful use of this result in the formula (\ref{QE10}) enables us to determine the energy spectrum of system $B$. Namely,
\begin{equation}\label{QCHe2}
E_{b}(\lambda_{b}, L_{b}, \nu)= 4\sqrt{\lambda_{b}} \sqrt{\frac{\hbar^{2}}{2m}}\left[\nu + (L_{b}/\sqrt{2})/\sqrt{2} + 1/2 \right]^{1/2}.
\end{equation}

Substituting $\lambda_{b}=m\omega^{2}/2$ and $L_{b}=\ell_{b} + D/2 - 1$  in (\ref{QCHe2}), we reach
the standard expression for the energy spectrum of the isotropic harmonic oscillator in $D$-dimensional space,
\begin{equation}\label{QOspec}
E_{b}(\ell_{b}, \nu)=\hbar \omega (2\nu + \ell_{b} + D/2) \, \quad \, (\ell_{b}, \, \nu \in \mathbb{N}_{0}).
\end{equation}
~

\underline{Wave functions:}\,
The radial Equation~(\ref{H1}) for the Coulomb potential $V(r)=-e^{2}/r$ can easily be converted to the Whittaker Equation \cite{WhiWat52}
\begin{equation}\label{QW1}
\left\{\frac{\rmd^{2}}{\rmd x^{2}} - \frac{L^{2} - 1/4}{x^{2}}  + \frac{k}{x} - \frac{1}{4} \right\} w(x) = 0,
\end{equation}
where $L=\ell + D/2 - 1\, \, (\ell \in \mathbb{N}_{0})$.
In the conversion, we have let $x=2\kappa r$, $k=me^{2}/(\hbar^{2}\kappa)=k_{a}$, $\hbar \kappa=\sqrt{-2mE}$, $L=L_{a}$ and $w(x)=\psi_{a}(x/(2\kappa))$. This set of replacements is indeed a duality map for the self-dual pair $(a, a)=(-1, -1)$.
The Whittaker functions, $M_{k, L}(x)$ and $W_{k, L}(x)$, are two linearly independent solutions of the Whittaker Equation~(\ref{QW1}).
For $|x|$ small, $M_{k, L}(x) \sim x^{L +\,1/2}$ and $W_{k, L}(x) \sim -\frac{\Gamma(2L)}{\Gamma(L-k+1/2)}x^{- L + \,1/2}$. If~$-\pi /2 < \arg x < 3\pi / 2$ and $|x|$ is large, then
\begin{equation}\label{Whitt1}
M_{k, L}(x) \sim \Gamma(2L +1) \left\{\frac{\rme^{i\pi(L - k + \frac{1}{2})} \rme^{-x/2} x^{k}}{\Gamma(L + k + \frac{1}{2})} + \frac{\rme^{x/2} x^{-k}}{\Gamma(L - k + \frac{1}{2})} \right\},
\end{equation}
and, if~$x \notin \mathbb{R}^{-}$ and $|x|$ is large,
\begin{equation}\label{Whitt2}
W_{k, L}(x) \sim \rme^{-x/2} x^{k} [ 1 + O(x^{-1})].
\end{equation}

The first solution $M_{k, L}(x)$ vanishes at $x=0$ as $L > -1/2$ but diverges as $|x| \rightarrow \infty$ unless $k -  L - \frac{1}{2} \in \mathbb{N}_{0}$, whereas the second solution $W_{k, L}(x)$ diverges at $x=0$ but converges to zero as $|x| \rightarrow \infty$.

The solution for the Coulomb problem is given in terms of the Whittaker function,
\begin{equation}\label{QCsol1}
\psi_{a}(r; L_{a}, \nu) = \mathcal{N}_{a}(L_{a})\, M_{\nu + L_{a} + \frac{1}{2},\, L_{a}}(2\kappa r),
\end{equation}
where $k_{a}$ is replaced by $\nu + L_{a} + \, 1/2$.  For~the bound state solution which vanishes at infinity,  we have to let $\nu = 0, 1, 2, \ldots$. In~this case,  $k_{a} = \nu + L_{a} + \, 1/2$ implies the discrete spectrum $E_{a}(\lambda_{a}, L_{a}, \nu)$ in (\ref{QCspec}).

Since the Whittaker function $M_{k, \mu}(z)$ is related to the Laguerre function $L_{\nu}^{2\mu}(z)$ as
\begin{equation}\label{Laguerre}
M_{\mu + \nu + \frac{1}{2}, \,\mu}(z) = \frac{\Gamma(2\mu + 1) \Gamma(\nu +1)}{\Gamma(2\mu + \nu + 1)} \, \rme^{-z/2} z^{\mu + \frac{1}{2}} \, L_{\nu}^{2\mu}(z),
\end{equation}
the eigenfunction may also be expressed in terms of the Laguerre function as
\begin{equation}\label{QCsol2}
\psi_{a}(r; L_{a}, \nu) = \mathcal{N}_{a}(L_{a})  \, \frac{\Gamma(2L_{a} + 1) \Gamma(\nu +1)}{\Gamma(\nu + 2L_{a} + 1)} \, \rme^{-\kappa r} (2\kappa r)^{L_{a} + \frac{1}{2}}\, L_{\nu}^{2L_{a}}(2\kappa r),
\end{equation}
which is normalized to unity with
\begin{equation}\label{QCsol3}
\mathcal{N}_{a}(L_{a})
= \frac{\hbar \kappa /\sqrt{me^{2}}}{\Gamma(2L_{a} + 1)}\sqrt{\frac{\Gamma(\nu + 2L_{a} + 1)}{\Gamma(\nu + 1)}}.
\end{equation}

The radial equation for the Hooke system with $V_{b}(\rho)=\frac{1}{2}m\omega^{2} \rho^{2}$, too, can be reduced to the Whittaker equation by letting
\begin{equation}\label{OW2}
y =(m\omega/\hbar) \rho^{2}, \, \quad \, L=L_{b}/2, \, \quad \, k=E_{b}/(2\hbar \omega)=k_{b}, \, \quad \,  w(y)=y^{1/4}\psi_{b}(y),
\end{equation}
which form a duality map for $(b, c)=(2, -1)$.
Here $L_{b}=\ell_{b} + D/2 - 1$ with $\ell_{b} \in \mathbb{N}_{0}$.
The bound state solution for the radial oscillator is given by
\begin{equation}\label{QOsol1}
\psi_{b}(\rho; L_{b}, \nu) = \mathcal{N}_{b}(L_{b})\, \frac{1}{\sqrt{\rho}}\, M_{\nu + \frac{1}{2}L_{b} + \frac{1}{2},\, \frac{1}{2}L_{b}}\left(\frac{m\omega}{\hbar} \rho^{2}\right).
\end{equation}

The choice $k_{b}= (2\nu + L_{b} + 1)/2$ with $\nu \in \mathbb{N}_{0}$ makes the solution (\ref{QOsol1}) the eigenfunction belonging to the energy $E_{b}(\nu, \ell_{b})$ in (\ref{QCHe2}). In~terms of the Laguerre function, it~reads
\begin{equation}\label{QOsol2}
\psi_{b}(\rho; L_{b}, \nu)=
\mathcal{N}_{b}(L_{b}) \,\frac{\Gamma(L_{b} + 1)\Gamma(\nu + 1)}{\Gamma(\nu + L_{b} +1)}
 \rme^{-(m\omega/2\hbar)\rho^{2}}\left(\frac{m\omega}{\hbar} \rho^{2}\right)^{(L_{b} + \frac{1}{2})/2}
L_{\nu}^{L_{b}} \left(\frac{m\omega}{\hbar} \rho^{2} \right),
\end{equation}
which is normalized to unity with
\begin{equation}\label{QOsol3}
\mathcal{N}_{b}(L_{b})= \frac{(4m\omega/\hbar)^{1/4}}{\Gamma(L_{b} + 1)}\, \sqrt{\frac{\Gamma(\nu + L_{b} +1)}{\Gamma(\nu + 1)}}.
\end{equation}

The process of going from (\ref{QCsol1}) to (\ref{QOsol1}) is rather straightforward.  First we notice that $\eta=-b/a=2$ for the Coulomb--Hooke pair $(a, b)=(-1, 2)$. Then we use the relation $\lambda_{b}=- \eta^{2} C^{2} E_{a}$, $\lambda_{b}= m\omega^{2}/2$ and $\hbar \kappa = \sqrt{-2mE_{a}}$ to get $C = m\omega/(2\hbar \kappa)$. Hence operation $\mathfrak{R}: \, r=C\rho^{\eta}$ with $\eta=2$ yields $2\kappa r = (m\omega/\hbar) \rho^{2}$. In~addition, we apply $\mathfrak{L}: \, L_{a}=L_{b}/2$.  Consequently, we have the right hand side of (\ref{QE11}) for $a=-1$, $\eta=2$ and $h(\rho)=\sqrt{m\omega/(\hbar \kappa)}\rho^{1/2}$ in the form,
\begin{equation}\label{QCOcomp1}
\sqrt{\hbar\kappa/m\omega}\, \frac{1}{\sqrt{\rho}}\, \psi_{a}((m\omega/2\hbar \kappa)\rho^{2};  L_{b}/2, \nu)
 = \tilde{\mathcal{N}}_{b}(L_{b})\, \frac{1}{\sqrt{\rho}}\, M_{\nu + \frac{1}{2}L_{b} + \frac{1}{2},\, \frac{1}{2}L_{b}}\left(\frac{m\omega}{\hbar} \rho^{2}\right),
\end{equation}
which coincides with the eigenfunction for the radial oscillator in (\ref{QOsol1}) except for the normalization factor. In~(\ref{QCOcomp1}),
\begin{equation}\label{QCOcomp2}
\tilde{\mathcal{N}}_{b}(L_{b}) = \sqrt{\hbar\kappa/m\omega} \mathcal{N}_{a}(L_{b}/2),
\end{equation}
{which differs from $\mathcal{N}_{b}(L_{b})$ of (\ref{QOsol3}) due to the difference of factors,\linebreak \mbox{$\sqrt{\hbar^{2}\kappa^{3}/(me^{2})} (m\omega/\hbar)^{-1/2} \neq \sqrt{2} (m\omega/\hbar)^{1/4}$}.
The wave function of the radial oscillator can be determined by the radial wave function of the hydrogen atom except for \mbox{its~normalization}.}

\underline{The Green functions}: \,
The Green function of interest, $G_a(r, r'; E, L)$, obeys the radial equation,
\begin{equation}\label{QCOgreen1}
\left\{ \frac{\rmd^{2}}{\rmd r^{2}} - \frac{L^{2} - 1/4}{r^{2}} - \frac{2m}{\hbar^{2}}V(r) + \frac{2m}{\hbar^{2}}E
\right\}G_{a}(r, r'; E, L)= -\frac{2m}{\hbar^{2}}\delta(r - r'),
\end{equation}
where $V_a(r)=\lambda_{a}r^{a} + \lambda_{a'}r^{a'}$.  The~boundary conditions we impose on it are
\begin{equation}\label{QCOgreen2}
\lim_{r \rightarrow 0}G(r, r'; E, L) = 0\qquad\mbox{and}\qquad
\lim_{r \rightarrow \infty}G(r, r'; E, L)  < \infty.
\end{equation}

Let $\psi^{(1)}(r)$ and $\psi^{(2)}(r)$ be two independent solutions of the radial Equation~(\ref{H1}). Let us
assume that $\psi^{(1)}(r)$ remains finite as $r \rightarrow \infty$ while the second solution obeys $\psi^{(2)}(0)=0$. With~these solutions, following the standard procedure~\cite{MoFesh53},  we can construct the Green function $G(r, r'; E, L)$ as
\begin{equation}\label{QCOgreen3}
G(r, r'; E, L) =\frac{2m}{\hbar^{2}\mathcal{W}[\psi^{(1)}, \psi^{(2)}]} \left\{\begin{array}{ll} \psi^{(1)}(r) \,\psi^{(2)}(r'), &\, \quad \, r > r'  \\
\psi^{(1)}(r')\, \psi^{(2)}(r), &\, \quad \, r' > r  \end{array} \right.
\end{equation}
where $\mathcal{W}[\cdot, \cdot]$ signifies the~Wronskian.

For the Coulomb problem with $V_{a}(r)= - e^{2} r^{-1}$, we let $\psi^{(1)}(r)=W_{k_{a}, \,L_{a}}(2\kappa r)$ and $\psi^{(2)}(r)=
M_{k_{a}, \,L_{a}}(2\kappa r)$. Then we calculate the Wronskian to get
\begin{equation}\label{QCwrons1}
(2\kappa)^{-1}\mathcal{W}[W_{k, L}(2\kappa r), M_{k, L}(2\kappa r)]= \mathcal{W}[W_{k, L}(x), M_{k, L}(x)] = -
\frac{\Gamma(2L +1)}{\Gamma(L - k +\frac{1}{2})},
\end{equation}
where we have use the property,
\begin{equation}\label{QCwrons2}
\mathcal{W}[W_{k, L}(x), \, M_{k, L}(x)]= (dy/dx)\mathcal{W}[W_{k, L}(y), \, M_{k, L}(y)].
\end{equation}

Substituting this result in the formula (\ref{QCOgreen3}), we obtain the radial Green function for the Coulomb problem,
\begin{equation}\label{QCgreen}
G_{a}(r, r'; E_{a},L_{a}) = -\frac{m}{\hbar^{2}\kappa} \frac{\Gamma(L_{a} - k_{a} + \frac{1}{2})}{\Gamma(2L_{a} +1)}
W_{k_{a}, \,L_{a}}(2\kappa r_{>})\,M_{k_{a}, \,L_{a}}(2\kappa r_{<}),
\end{equation}
where $r_> = \max\{r,r'\}$ and $r_< = \min\{r,r'\}$. We have also set $\kappa = \sqrt{-2mE_{a}}\,\hbar$ and $k_{a}=me^{2}/(\hbar \sqrt{-2mE_{a}})$, both of which are in general complex numbers. The~resultant Green function is a double-valued function of $E_{a}$. It contains the contribution from the continuous states (corresponding to the branch-cut along the positive real line on $E_{a}$) as well as the bound states (corresponding to the poles on the negative real axis).  The~poles of $G(r, r'; L_{a}, E_{a})$ on the $E_{a}$-plane occur when $L_{a} - k_{a} + \frac{1}{2}= - \nu$ with $\nu \in \mathbb{N}_{0}$, yielding the discrete energy spectrum (\ref{QCspec}).

Similarly, for~the radial oscillator with $V_{b}(\rho)= (m/2)\omega^{2} \rho^{2}$, we let\linebreak $\psi^{(1)}(\rho)=W_{k_{b}, L_{b}}((m\omega/\hbar)\rho^{2})$ and $\psi^{(2)}(\rho)=M_{k_{b}, L_{b}}((m\omega/\hbar)\rho^{2})$.  Use of the property,
\begin{equation}\label{QOwrons1}
\mathcal{W}[\chi(y) W_{k, L}(y), \, \chi(y) M_{k, L}(y)]=[\chi(y)]^{2}\mathcal{W}[W_{k, L}(y), \, M_{k, L}(y)],
\end{equation}
for a differentiable function $\chi(y)$, together with (\ref{QCwrons2}) and (\ref{QCwrons1}), enables us to evaluate the Wronskian and to get to the Green function for the radial oscillator,
\begin{equation}\label{QOgreen}
G_{b}(\rho, \rho'; L_{b}, E_{b}) = -\frac{1}{\hbar \omega \, \sqrt{\rho \rho'}} \frac{\Gamma(\frac{1}{2}L_{b} - k_{b} + \frac{1}{2})}{\Gamma(L_{b} +1)}\, W_{k_{b}, \frac{1}{2}L_{b}}\left(\frac{m\omega}{\hbar} \rho^{2}_{_{>}}\right)\,
                             M_{k_{b}, \frac{1}{2}L_{b}}\left(\frac{m\omega}{\hbar} \rho^{2}_{_{<}}\right),
 \end{equation}
 where $k_{b}=E_{b}/(2\hbar \omega)$. Since $G(\rho, \rho'; l_{b}, E_{b})$  is not a multi-valued function of $E_{b}$, it has no branch point on the $E_{b}$-plane and contains no contribution corresponding to a continuous spectrum, but~has poles at $k_{b}=\nu + \frac{1}{2}L_{b} + \frac{1}{2}$ with $\nu \in \mathbb{N}_{0}$ yielding the discrete energy spectrum (\ref{QOspec}).

Finally, we compare the Green function for the bound state of the Coulomb problem (\ref{QCgreen}) and the Green function for the radial oscillator (\ref{QOgreen}).  The~Gamma functions and the Whittaker functions in (\ref{QCgreen}) are brought to those in (\ref{QOgreen}) by transformations $r=C\rho^{2}$ with $C=m\omega/(2\hbar \kappa)$, $L_{a}=L_{b}/\eta$ with $\eta=2$, and~$k_{a} = k_{b}$. Although~the first two transformations are two of the dual operations,  the~last one must be verified. Since $k_{a}=me^{2}/(\hbar^{2} \kappa)= - m\lambda_{a}/(\hbar^{2}\kappa)$ and $\lambda_{a}=- E_{b}/(4C)$, it immediately follows that $k_{a}=E_{b}/(2\hbar \omega)=k_{b}$ provided $C=m\omega/(2\hbar \kappa)$. For~the bound state problem, $k_{a}=\nu + L_{a} + \frac{1}{2}$ and $k_{b} = \nu + \frac{1}{2}L_{b} + \frac{1}{2}$. Hence, it is apparent that $k_{a}=k_{b}$ when $L_{a}=L_{b}/2$. The~extra function in (\ref{HG4}) is now given by
$h(\rho) h(\rho') = \sqrt{m \omega/(\hbar \kappa)}\sqrt{\rho \rho'}$. Hence the prefactor $m/(\hbar^{2}\kappa)$ in (\ref{QCgreen}) divided by the extra function gives rise to the prefactor $(\hbar \omega \, \sqrt{\rho \rho'})^{-1}$ \mbox{in (\ref{QOgreen})}.  In~this fashion, $G_a(r, r'; L_{a}, E_{a})$ of (\ref{QCgreen}) is completely transformed into $G_b(\rho, \rho'; L_{b}, E_{b})$ by the duality procedures with $C=(m\omega/2\hbar \kappa)$. By~letting $L_{b}=\ell_{b} + D/2 -1$ with $\ell_{b} \in \mathbb{N}_{0}$, we can see that the formula (\ref{HG5}) works well for the Coulomb--Hooke~pair.

\subsection{A Confinement Potential as a Multi-Term Power-Law~Example}\label{Section4.4}
One of the motivations that urged the study of power-law potentials was the quark-antiquark confinement problem. See, for~instance, references~\cite{QR,Gaze,John}. Here we examine a two-term power potential as a model of the confinement~potential.

Let system $A$ consist of a particle of mass $m$ confined in a two-term power potential,
\begin{equation}\label{Conf1}
V_{a}(r) = \lambda_{a}r^{a} + \lambda_{a'}r^{a'},
\end{equation}
where $\lambda_{a} \neq 0$, $\lambda_{a'} \neq 0$, $a \neq a'$, $a \neq 0$, and~$a' \neq 0$.  Let system $B$ be power-dual to system $A$ and quantum-mechanically solvable. Then we expect that some quantum-mechanical information can be obtained concerning the confined system $A$ by analyzing the properties of system $B$.  As~we have seen earlier, when system $A$ and system $B$ are dual to each other, the~shifted potential of system $A$,
\begin{equation}\label{Conf2}
U_{a}(r)=\lambda_{a}r^{a} + \lambda_{a'}r^{a'} - E_{a},
\end{equation}
transforms to that of system $B$,
\begin{equation}\label{Conf3}
U_{b}(\rho) = \lambda_{b}\rho^{b} + \lambda_{b'}\rho^{b'} - E_{b},
\end{equation}
by
\begin{equation}\label{Conf4}
U_{b}(\rho)=g(\rho)U_{a}(f(\rho)).
\end{equation}

Here $r=f(\rho)=C\rho^{\eta}$, \, $g(\rho)=C^{2}\eta^{2}\rho^{2\eta - 2}$, \, $\eta=2/(a+2)=-b/a$, and~\begin{equation}\label{Conf5}
b'=2(a'-a)/(a+2) \, \, \quad \, \,    \lambda_{b'}=\lambda_{a'} \eta^{2}C^{a'+2}.
\end{equation}

Note also that the exchange relations,
\begin{equation}\label{Conf6}
E_{b}=- \eta^{2} C^{a+2} \lambda_{a}, \, \quad \, \lambda_{b} = - \eta^{2} C^{2} E_{a},
\end{equation}
play an essential role in verifying the equality (\ref{Conf4}).

First, we wish to tailor the potential of system $A$ to be a confinement potential. To~this end, we set the following~conditions.

(i) System $B$ behaves as a radial harmonic oscillator $\, (\lambda_{b}=0, \, \lambda_{b'} >0, \, b'=2)$

 (ii) System $A$ has a bound state with $E_{a} = 0$ and its potential is asymptotically linearly-increasing $\, (\lambda_{a'} >0, \, a'=1)$.

Since we are unable to solve analytically the Schr\"odinger equation for system $B$ \mbox{with (\ref{Conf3})} in general, we consider the limiting case for which
$\lambda_{b} \rightarrow 0$, that is, we employ for the potential of system $B$
\begin{equation}\label{Conf7}
\mathcal{U}_{b}(\rho) = \lim_{\lambda_{b} \rightarrow 0}U_{b}(\rho) = \lambda_{b'}\rho^{b'} - E_{b}.
\end{equation}

According to the second relation of (\ref{Conf6}), the~limit $\lambda_{b} \rightarrow 0$ implies $E_{a} \rightarrow 0$.  Hence we study only the zero-energy state of system $A$ by assuming that it exists and is characterized by an integral number $\nu_{0}$. We denote the zero-energy by $E_{a}(\nu_{0})$. There are only a few exactly soluble nontrivial examples with $\mathcal{U}_{b}$ of (\ref{Conf7}). Our choice is the one for the radial harmonic oscillator with $b'=2$ and $\lambda_{b'} >0$,
\begin{equation}\label{Conf8}
\mathcal{U}_{b}(\rho)=\lambda_{b'}\rho^{2} - E_{b} \, \, \, (\lambda_{b'} >0).
\end{equation}

Namely, we consider that system $B$ behaves as the radial harmonic oscillator with frequency $\Omega=\sqrt{2\lambda_{b'}/m}$ and angular momentum $L_{b}$.
Since $b'=2$ implies $2(a'-a)/(a+2)=2$ as obvious from (\ref{Conf5}), the~corresponding potential of system $A$ is
\begin{equation}\label{Conf9}
V_{a}(r) = \lambda_{a}r^{(a' -2)/2} + \lambda_{a'}r^{a'}.
\end{equation}

Next we assume that a possible confinement potential behaves asymptotically as a linearly increasing function.  Thus, letting $a'=1$ and $\lambda_{a'} >0$ in (\ref{Conf9}),  we have
\begin{equation}\label{Conf10}
V_{a}(r) = \lambda_{a}r^{-1/2} + \lambda_{a'}r, \, \, \quad \, (\lambda_{a} <0, \, \lambda_{a'} >0).
\end{equation}

\begin{figure}
\includegraphics{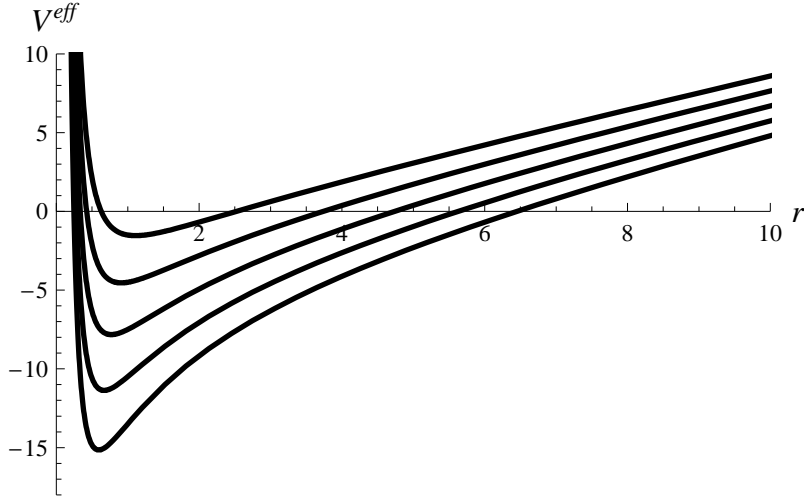}
 \caption{\label{Fig5}The effective potential (\ref{Conf11}) related to the eigenfunctions (\ref{Conff3}) for $\nu_0=0,1,2,3,4$ from top to bottom. The~parameters and units are set to $\ell_a=\lambda_{a'}=1$, $D=3$ and $2m=\hbar=1$, respectivly.}
\end{figure}

If $\lambda_{a} > 0$, then $V_{a}(r) > 0$ for all $r$, and~the assumed zero-energy state cannot exist.
For $\lambda_{a} < 0$, the~effective potential of system $A$,
\begin{equation}\label{Conf11}
V_{a}^{eff}(r) = \frac{(L_{a}^{2}-\frac{1}{4})\hbar^{2}}{2mr^{2}} - |\lambda_{a}|r^{-1/2} + |\lambda_{a'}|r,
\end{equation}
can accommodate the zero-energy state provided that $ \lambda_{a}$ and $\lambda_{a'}$ are so selected that $V_{a}^{eff}(r_{1}) < 0$ where $r_{1}$ is a positive root of $\rmd V_{a}^{eff}(r)/\rmd r=0$.  Here $L_{a}=L_{b}/\eta$ and $L_{a}=\ell_{a}+D/2 -1$ with $\ell_{a} \in \mathbb{N}_{0}$.
In this manner, we are able to obtain the confinement potential (\ref{Conf10}) which is asymptotically linearly increasing and may accommodate at least the assumed zero-energy state. Figure~\ref{Fig5} shows the effective potential (\ref{Conf11}) of system $A$ for $\ell_{a}=1$, $D=3$, $\lambda_{a'}=1$ and $\nu_0=0, 1, 2, 3, 4$ in units $2m=\hbar =1$.

Since $a'=1$, we have $a=(a'-2)/2=-1/2$, $\eta=2/(a+2)=4/3$ and $b=-a\eta =  2/3$. The~last information concerning $b$ is unimportant insofar as $\lambda_{b} \rightarrow 0$ is assumed. The~second relation of (\ref{Conf5}) demands that
\begin{equation}\label{ConfC}
C=\left(9\lambda_{b'}/16\lambda_{a'}\right)^{1/3}.
\end{equation}

Therefore, the~first relation of (\ref{Conf6}) yields
\begin{equation}\label{Conf12}
E_{b}=- \frac{4}{3} \lambda_{a}\,\sqrt{\frac{\lambda_{b'}}{\lambda_{a'}}}.
\end{equation}

On the other hand, since system $B$ behaves as a radial harmonic oscillator with frequency $\Omega=\sqrt{2\lambda_{b'}/m}$ and angular momentum $L_{b}$, its energy spectrum is given by
\begin{equation}\label{Conf13}
E_{b}(\nu_{0}, \ell_{b}) = \hbar \Omega \,(2 \nu_{0} + L_{b} + 1),
\end{equation}
where $\nu= \nu_{0}$ is fixed by $E_{a}(\nu_{0})$ and $L_{b}=\ell_{b} + D/2 -1$ with $\ell_{b}\in \mathbb{N}_{0}$. Letting $L_{b}=(4/3)L_{a}$ in (\ref{Conf13}) and interpreting that $E_{b}$ of (\ref{Conf12}) represents an allowed value in the spectrum (\ref{Conf13}),  we observe that the coupling constant $\lambda_{a}$ may take one of the values specified by the set of $(\nu_{0}, \ell_{a})$ via
\begin{equation}\label{Conf14}
\lambda_{a} = - \frac{3}{4} \sqrt{\frac{2\lambda_{a'}\hbar^2}{m}} (2\nu_{0} + (4/3)L_{a} +1),
\end{equation}
where $L_{a}=\ell_{a} + D/2 -1$ with $\ell_{a} \in \mathbb{N}_{0}$.

The energy eigenfunction of the radial oscillator has been given in (\ref{QOsol1}).
Replacing $(m\omega/\hbar)$ in the previous result by $\beta=m\Omega/\hbar=\sqrt{2m \lambda_{b'}\,}/\hbar$, we write down the eigenfunction of the present oscillator as
\begin{equation}\label{Conff1}
\phi_{b}(\rho; L_{b}, \nu_{0}) = \mathcal{N}_{b}(L_{b}, \nu_{0}, \beta)\, {\left( \beta\rho^2 \right)^{-1/4}}\, M_{\nu_{0} + \frac{1}{2}L_{b} + \frac{1}{2},\, \frac{1}{2}L_{b}}\left(\beta \rho^{2}\right),
\end{equation}
which is normalized to unity with
\begin{equation}\label{Conff2}
\mathcal{N}_{b}(L_{b}, \nu_{0}, \beta)= \frac{(4\beta)^{1/4}}{\Gamma(L_{b} + 1)}\, \sqrt{\frac{\Gamma(\nu_{0} + L_{b} +1)}{\Gamma(\nu_{0} + 1)}}.
\end{equation}

Moreover, utilizing the eigenfunction just obtained,
we construct the eigenfunction for the zero-energy state in the confinement potential (\ref{Conf10}) by following the simple prescription $\phi_{a}(r)=h(\rho)\phi_{b}(\rho)$. For~the pair $(a, b)=(-1/2, 2/3)$, the~two variables $r$ and $\rho$ are related by $r=C\rho^{4/3}$ with $C$ given in (\ref{ConfC}).  Since $\rho^{2}=C^{-3/2}r^{3/2}$ and  $C^{-3/2}=(4/3)\sqrt{\lambda_{a'}/\lambda_{b'}}$, we~let
\begin{equation}\label{Conff2a}
\alpha = \frac{4}{3}\beta \sqrt{\frac{\lambda_{a'}}{\lambda_{b'}}}=\frac{4}{3}\frac{\sqrt{2m\lambda_{a'}}}{\hbar}, \, \, \quad \, \beta=\frac{\sqrt{2m\lambda_{b'}}}{\hbar},
\end{equation}
and
\begin{equation}\label{Conff2b}
\beta \rho^{2}=\alpha r^{3/2}.
\end{equation}

Multiplying $\phi_{b}(\rho)$ of (\ref{Conff1}) by $h(\rho)=\sqrt{dr/d\rho}= \sqrt{4C/3}\rho^{1/6}$, and~substituting (\ref{Conff2a}) and $L_{b}=(4/3)L_{a}$ into $\phi_{b}(\rho)$, we arrive at the eigenfunction for the zero-energy state of system $A$,
\begin{equation}\label{Conff3}
\phi_{a}(\rho; L_{a}, \nu_{0}) = \mathcal{N}_{a}(L_{a}, \nu_{0}, \alpha)\, {\left( \alpha r^{3/2}\right)^{-1/6}}\, M_{\nu_{0} + \frac{2}{3}L_{a} + \frac{1}{2},\, \frac{2}{3}L_{a}}\left(\alpha r^{3/2}\right),
\end{equation}
where $L_{a}=\ell_{a} + D/2 -1$ with $\ell_{a} \in \mathbb{N}_{0}$. Here the factor $\mathcal{N}_{a}(L_{a}, \nu_{0}, \alpha)$ that normalizes $\phi_{a}(\rho)$ to unity cannot be determined by $\mathcal{N}_{b}((4/3)L_{a}, \nu_{0}, (3/4)\alpha \sqrt{\lambda_{b'}/\lambda_{a'}})$.
Corresponding to the value of $\lambda_{a}$ specified in (\ref{Conf14}) by the set $(\nu_{0}, \ell_{a})$, the~eigenfunction $\phi_{a}(\rho; \ell_{a}, \nu_{0})$ is characterized by the same set $(\nu_{0}, \ell_{a})$ of~numbers.

The Green function of system $A$ obeys the inhomogeneous radial equation,
\begin{equation}\label{ConfG1}
\left\{ \frac{\rmd^{2}}{\rmd r^{2}} - \frac{L_{a}^{2} - 1/4}{r^{2}} - \frac{2me^{2}}{\hbar^{2}}\left(\lambda_{a}r^{-1/2} + \lambda_{a'}r\right) +
\frac{2me^{2}}{\hbar^{2}} E_{a} \right\} G_{a}(r, r';E_{a}, L_{a})
 =  -\frac{2m}{\hbar^{2}}\delta(r - r').
\end{equation}

Since the Green function for the radial oscillator has been given in (\ref{QOgreen}), we can write down the Green function $G_{b}(\rho, \rho'; E_{b}(\nu_{0}))$ of system $B$ with $\lambda_{b}=0$ as
\begin{equation}\label{ConfG2}
G_{b}(\rho, \rho'; E_{b},L_{b}) =-\frac{m}{\hbar^{2}\beta} \,\frac{1}{\sqrt{\rho \rho'}} \frac{\Gamma(\frac{1}{2}L_{b} - k_{b} +
\frac{1}{2})}{\Gamma(L_{b} +1)} \, W_{k_{b}, \frac{1}{2}L_{b}}(\beta \rho_{_{>}}^{2})\,M_{k_{b}, \frac{1}{2}L_{b}}(\beta \rho^{2}_{_{<}}),
\end{equation}
where $k_{b}=E_{b}/(2\hbar \Omega)$.  The~pole of $G_{b}(\rho, \rho'; E_{b})$ that corresponds to $E_{b}(\nu_{0})$ occurs when $k_{b}(L_b,\nu_0)=\nu_{0} + \frac{1}{2}L_{b} + \frac{1}{2}$ where $\nu_{0}$ is a non-negative~integer.

The Green function $G_{a}(r, r'; E_{a},L_{a})$ of system $A$ at $E_a = 0$ can be found by substituting (\ref{Conff2b}) together with
\[
h(\rho)=\sqrt{4/3}C^{3/8} r^{1/8},\, \, \quad \frac{1}{2}L_{b}=\frac{2}{3}L_{a},\, \, \quad
\]
into $h(\rho)h(\rho')G_{b}(\rho, \rho'; E_b,L_{b})$. Namely,
\begin{equation}\label{ConfG3}
G_{a}(r, r'; E_a=0,L_{a})=\frac{4}{3} C^{3/4} (r r')^{1/8}\,G_{b}\left((r/C)^{3/4}, (r'/C)^{3/4}; E_b=\frac{16}{9}|\lambda_a| ,\frac{4}{3}L_{a}\right),
\end{equation}
where $C$ has been given in (\ref{ConfC}). Explicitly, we have
\begin{equation}\label{ConfG3a}
G_{a}(r, r'; E_{a},L_{a}) = -\frac{4m}{3\hbar^{2}\alpha} \,(rr')^{-1/4} \frac{\Gamma(\frac{2}{3}L_{a} - k_{a} +
\frac{1}{2})}{\Gamma(\frac{4}{3}L_{a} +1)}\, W_{k_{a}, \frac{2}{3}L_{a}}(\alpha r_{_{>}}^{3/2})\,M_{k_{a}, \frac{3}{2}L_{a}}(\alpha r^{3/2}_{_{<}}).
\end{equation}
where $\alpha$ and $\beta$ have been given by (\ref{Conff2a}).
The pole corresponding to $E_{a}(\nu_{0})=0$ occurs when $k_{a}=\nu_{0} + (2/3)L_{a} + \, 1/2$ and $L_{a}=\ell_{a} + D/2 -1$. We have to remember that the Green function (\ref{ConfG3}) is meaningful only in the vicinity of $E_{a}=0$.\\

\noindent{\bf Remark 25:}
The angular momentum $L$ in (\ref{H2}) is identical in form to that used in the semiclassical case (\ref{G2}). However, no Langer-like ad hoc treatment has been made in the Schr\"odinger equation. The~angular contribution $\ell(\ell + D-2)$ and an additional contribution $(D-1)(D-3)/4$ from the kinetic term due to the transformation of base function, $R_{\ell}(r)$ to $\psi_{\ell}(r)$, make up the term $L^{2} - 1/4$ in the effective centrifugal potential term of (\ref{H3}).\\

\noindent{\bf Remark 26:}
The~time transformation $\mathfrak{T}$ needed in classical mechanics takes no part in the power duality of the stationary Schr\"odinger equation. Instead, the~change of the base function plays an essential role. While  $\mathfrak{T}$ assumes $\rmd t=g(\rho)\rmd s$, the~state function changes as $\psi_{a}(r) = [g(\rho)]^{1/4}\psi_{b}(\rho)$. The~possible connection between the time transformation and the change of state function has been discussed in the context of path integration for the Green function in~\cite{Junk}.  So long as the stationary Schr\"odinger equation is concerned, there is no clue to draw any causal relation between $\mathfrak{T}$ and $\mathfrak{F}$.  However, one might expect that $\mathfrak{T}$ would play a role in the time-dependent Sch\"odinger equation. If~the energy-coupling exchange operation $\mathfrak{E}$ of (\ref{G9}) is formally modified as
\begin{equation}\label{HTE1}
\mathfrak{E}': gV_{a}(r) \rightarrow - i\hbar \frac{\partial}{\partial \bar{s}}, \, \, \quad \, g i\hbar \frac{\partial}{\partial \bar{t}} \rightarrow - V_{b}(\rho),
\end{equation}
then the time-dependent radial Schr\"odinger equation,
\begin{equation}\label{HTE2}
\left[-\frac{\hbar^{2}}{2m}\frac{\rmd^{2}}{\rmd r^{2}} + \frac{\hbar^{2}(L_{a}^{2} - 1/4)}{2mr^{2}} + V_{a}(r) \right] \psi_{a}(r) = i\hbar \frac{\partial \psi_{a}(r)}{\partial \bar{t}},
\end{equation}
transforms into
\begin{equation}\label{HTE2a}
\left[-\frac{\hbar^{2}}{2m}\frac{\rmd^{2}}{\rmd\rho^{2}} + \frac{\hbar^{2}(L_{b}^{2} - 1/4)}{2m\rho^{2}} + V_{b}(\rho) \right] \psi_{b}(\rho) = i\hbar \frac{\partial \psi_{b}(\rho)}{\partial \bar{s}},
\end{equation}
under the set of $\{\mathfrak{R}, \mathfrak{L}, \mathfrak{E}', \mathfrak{F}\}$. It is important that $\bar{t}$ and $\bar{s}$ are
not necessarily connected by $\mathfrak{T}$; they are basically independent time-like parameters. In~conclusion, the~time transformation $\mathfrak{T}$ has no role in the time-dependent Schr\"odinger~equation.\\

\noindent{\bf Remark 27:}
More on time transformations. Since we are dealing with the action integral (\ref{H4}) rather than the Schr\"odinger equation, it is easy to observe that the time transformation $\mathfrak{T}$ in the classical action in Section~\ref{Section2}  is closely related to the transformation $\mathfrak{F}$ of wave functions in  the quantum action (\ref{H4}). Recall that
$\mathfrak{T}: \, {(\rmd t/\rmd\varphi)} = g(\rho){(\rmd s/\rmd\varphi)}$ where $g=f'^{2}$ with $f=C\rho^{\eta}$, and~that
\begin{equation}\label{HTF1}
\rmd t\, U_{a} = \rmd s\, gU_{a} = \rmd s\, U_{b}.
\end{equation}

From (\ref{H5}) and (\ref{H9}), we have
\begin{equation}\label{HTF2}
\rmd r\, U_{a}\psi_{a}^{\ast}\psi_{a} = \rmd\rho \, f'h^{2} U_{a} \psi_{b}^{\ast}\psi_{b} = \rmd\rho \,U_{b} \psi_{b}^{\ast}\psi_{b},
\end{equation}
where $g=f'h^{2}=f'^{2}$. Comparing (\ref{HTF1}) and (\ref{HTF2}), we see that $\rmd t = g \rmd s$ in classical mechanics corresponds to $
\rmd r\, \psi_{a}^{\ast}\psi_{a} = g\, \rmd\rho\, \psi_{b}^{\ast}\psi_{b}$ in quantum mechanics. In~other words, $\rmd r\,\psi_{a}^{\ast}\psi_{a}$ has the same transformation behavior that $\rmd t$ does. In~this respect, we may say that the role of $\mathfrak{T}$ in classical mechanics is replaced by $\mathfrak{F}$ in quantum~mechanics.


\section{Summary and~Outlook}\label{Section5}
In the present paper we have revisited the Newton--Hooke power-law duality and its generalizations from the symmetry point of~view.

(1)  We have stipulated the power-dual symmetry in classical mechanics by form-invariance and reciprocity of the classical action in the form of Hamilton's characteristic function, and~clarified the roles of duality operations $\{\mathfrak{C}, \mathfrak{R}, \mathfrak{T}, \mathfrak{E}, \mathfrak{L}\}$. The~exchange operation $\mathfrak{E}$ has a double role; it may decide the constant $C$ appearing in the transformation $r=C\rho^{\eta}$, while it leads to an energy formula that relates the new energy to the old~energy.

(2)  We have shown that the semiclassical action is symmetric under the set of duality operations $\{\mathfrak{C}, \mathfrak{R}, \mathfrak{E}, \mathfrak{L}\}$ without $\mathfrak{T}$ insofar as angular momentum $L$ is treated as a continuous parameter, and~observed that the power-duality is essentially a classical notion and breaks down at the level of angular quantization. To~preserve the basic spirit of power-duality in the semiclassical action, we have proposed an ad hoc procedure in which angular momentum transforms as $L_{b} = \eta L_{a}$, as~the classical case, rather than $\ell_{b}=\eta \ell_{a}$; after that each of $L$ is quantized as $L=\ell + \,D/2 - 1$ with $\ell \in \mathbb{N}_{0}$.
As an example, we have solved by the WKB formula a simple problem for a linear motion in a fractional power~potential.

(3)  We have failed to verify the dual symmetry of the supersymmetric (SUSY) semiclassical action for an arbitrary power potential, but~have succeeded to reveal the Coulomb--Hooke duality in the SUSY~action.

(4)  To study the power-dual symmetry in quantum mechanics, we have chosen the action in which the variables are the wave function $\psi(r)$ and its complex conjugate $\psi^{\ast}(r)$ and from which the radial Schr\"odinger equation can be derived. The~potential appearing in the action is a two-term power potential. We have shown that the action is symmetric under the set of operations $\{\mathfrak{C}, \mathfrak{R}, \mathfrak{E}, \mathfrak{L}\}$ plus the transformation of wave function $\mathfrak{F}$ provided that angular momentum $L$ is a continuous parameter. Again the ad hoc procedure introduced for the semiclassical case must be used in quantum mechanics. Associated with $\mathfrak{F}$ is the transformation of Green functions from which we have derived a formula that relates the new Green function and the old one. We have studied the Coulomb--Hooke duality to verify the energy formula and the formula for the Green functions. We also discussed a confinement potential and the Coulomb--Hooke--Morse~triality.

There are more topics that we considered important but left out for the future work. They include the power-dual symmetry in the path integral formulation of quantum mechanics, the~Coulomb--Hooke duality in Dirac's equation, and~the confinement problem in Witten's framework of supersymetric quantum mechanics. Feynman's path integral is defined for the propagator (or the transition probability) with the classical action in the form of Hamilton's principal function, whereas the path integral pertinent to the duality discussion is based on the classical action in the form of Hamilton's characteristic function. Since the power-dual symmetry of the characteristic action has been shown, it seems obvious that the path integral remains form-invariant under the duality operations, but~the verification of it is tedious. As~is well-known, Dirac's equation is exactly solvable for the hydrogen atom. There are also solutions of Dirac's equation for the harmonic oscillator. However, the~Coulomb--Hooke duality of Dirac's equation has never been established. The~situation is similar to Witten's model of SUSYQM. Using the same superpotential as that used for the semiclassical case in Section~\ref{Section4}, we may be able to show the Coulomb--Hooke symmetry and handle the confinement problem in Witten's~framework.

\vspace{6pt}



\appendix
\section{The Coulomb--Hooke--Morse~Triality}\label{SectionAPP}
In this Appendix~\ref{SectionAPP}, we wish to present the Coulomb--Hooke--Morse triality that relates the Morse oscillator to the Coulomb--Hooke duality. Specifically, letting system $A$ be the hydrogen atom (for the Coulomb system), system $B$ be the radial harmonic oscillator (for the Hooke system) and system $C$ be the Morse oscillator, we deal with their triangular relation.
The Morse oscillator is a system obeying the one-dimensional Schr\"odinger Equation \cite{Morse},
\begin{equation}\label{A1}
- \frac{\hbar^{2}}{2m} \frac{\rmd^{2}\psi_{c}(\xi)}{\rmd^{2}\xi} + \left(V_{c}(\xi) - E_{c}\right) \psi_{c}(\xi) = 0, \, \quad \, \xi \in \mathbb{R},
\end{equation}
where
\begin{equation}\label{A2}
V_{c}(\xi) = D_1\,{\rm e}^{-2\alpha \xi} - 2D_2\,{\rm e}^{-\alpha \xi}, \, \qquad \alpha,\, D_1,\, D_2>0 ,
\end{equation}
which is the Morse potential in a slightly modified form.  The~potential (\ref{A2}), being not a power-law potential, is beyond the scope of the main text. It is yet interesting to observe how the Morse oscillator is related to the Coulomb--Hooke duality. It is straightforward, if~one follows the general transformation procedure~\cite{Junk} for the Schr\"odinger equation, to~transform (\ref{A1}) directly to the Schr\"odinger equation for each of the hydrogen atom and the radial harmonic oscillation. Here, to~focus our attention on their trial nature, we place the Whittaker function at the center of the triangular relation. In~fact,
the Schr\"odinger Equation~(\ref{A1}) is easily transformed to the Whittaker Equation~(\ref{QW1}) under the substitutions
\begin{eqnarray}\label{A3}
~&&x=\gamma \, \rme^{-\alpha \xi}, \, \quad \gamma =\frac{\sqrt{8mD_{1}}}{\hbar \alpha} \\
~&&L_{c} = \frac{\sqrt{-2m E_{c}}}{\hbar \alpha}, \, \quad k_{c}= \sqrt{\frac{2mD_{2}^{2}}{\hbar^{2} \alpha^{2} D_{1}}}, \\
~&&w(x)= x^{1/2}\psi_{c}(\xi).
\end{eqnarray}

Hence the bound state solution of (\ref{A1}) can be expressed in terms of the Whittaker function as
\begin{equation}\label{A4}
\psi_{c}(\xi)= \mathcal{N}_{c}\, \rme^{\alpha \xi/2} M_{k_{c}, L_{c}}\left(\gamma \, \rme^{-\alpha \xi}\right),
\end{equation}
subject to the condition
\begin{equation}\label{A4a}
k_{c}=\nu + L_{c} + \frac{1}{2} \, ,\, \quad \nu \in \mathbb{N}_{0}.
\end{equation}

The last condition yields the energy spectrum,{
\begin{equation}\label{A5}
E_{c}= - \frac{\hbar ^{2} \alpha ^{2}}{2m} \left\{ \sqrt{\frac{2mD_2^2}{\hbar ^{2}\alpha ^{2}D_1}} - \left(\nu + \frac{1}{2}\right)\right\}^{2}, \, \, \nu=0,1,2,\ldots <\sqrt{\frac{2mD_2^2}{\hbar ^{2}\alpha ^{2}D_1}}-\frac{1}{2}\,.
\end{equation}}

The Morse oscillator solution $\psi_{c}(\xi)$ in (\ref{A4}) may be compared with the Coulomb bound state solution $\psi_{a}(r)$ and the Hooke oscillator solution $\psi_{b}(\rho)$ given, respectively, by~\begin{equation}\label{A6}
\psi_{a}(r)= \mathcal{N}_{a}\, M_{k_{a}, L_{a}}\left(2\kappa r \right),
\end{equation}
with
\begin{equation}\label{A6a}
k_{a}=\nu + L_{a} + \frac{1}{2} \,\, \quad \nu \in \mathbb{N}_{0},
\end{equation}
and
\begin{equation}\label{A7}
\psi_{b}(\rho)= \mathcal{N}_{b}\, {\left(\frac{m\omega}{\hbar}\rho^2 \right)^{-1/4}} M_{k_{b}, \frac{1}{2}L_{b}}\left(\frac{m\omega}{\hbar}\rho^{2} \right),
\end{equation}
with
\begin{equation}\label{A7a}
k_{b}=\nu + \frac{1}{2}L_{b} + \frac{1}{2} \,\, \quad \nu \in \mathbb{N}_{0}.
\end{equation}

The bound state conditions (\ref{A6a}) and (\ref{A7a}) lead to the energy spectrum of the Coulomb system ($A$) and that of the Hooke system ($B$), respectively, when
\begin{equation}\label{A8}
k_{a}=me^{2}/(\hbar^{2}\kappa) \,, \quad \, \hbar \kappa=\sqrt{-2mE_{a}}\,, \quad \, L_{a}=\ell + \,1/2\,, \quad \ell \in \mathbb{N}_{0}\,,
\end{equation}
\begin{equation}\label{A9}
k_{b}=E_{b}/(\hbar \omega) \,, \quad \,  L_{b}=\ell + \,1/2\,, \quad \ell \in \mathbb{N}_{0}\,.
\end{equation}

The triality relations are schematically shown below,

\[
\begin{array}{ccccccc}&{\rm Morse}&~~~~~~&&~~~~~~~ &{\rm Morse}&\\
{}^{CA}\, \swarrow& &\nwarrow \, {}^{BC}&~~~~~&{}^{AC}\, \nearrow& &\searrow \, {}^{CB} \\
{\rm Coulomb}& \longrightarrow &{\rm Hooke} &~~~~~&{\rm Coulomb}& \longleftarrow &{\rm Hooke}\\
~&{}^{AB}&~ && ~&{}^{BA}&~
\end{array}
\]

\noindent and the dual transformations $AC$, $CB$ and $BA$ are given by
\[
\begin{array}{lllll}
AC: &2\kappa r = \gamma \rme^{-\alpha \xi}, &k_{a}=k_{c}, & L_{a}=L_{c}, & \psi_{a}(r) = \rme^{- \alpha \xi /2}\psi_{c}(\xi)  \\
CB: &\gamma \rme^{-\alpha \xi}=(m\omega/\hbar)\rho^{2}, & k_{c}=k_{b}, &L_{c} = (1/2)L_{b}, & \rme^{- \alpha \xi /2}\psi_{c}(\xi) = \rho^{-1/2}\psi_{b}(\rho) \\
BA: &(m\omega/\hbar)\rho^{2} =2 \kappa r, & k_{b} = k_{a},  &(1/2)L_{b} = L_{a}, &\rho^{-1/2}\psi_{b}(\rho)=\psi_{a}(r)
\end{array}
\]
which are all invertible.
Although none of the energy formulas discussed earlier for the power-duality works when the Morse (non-power-law) potential is involved,
transforming one of the bound state conditions to another suffices as each condition generates an energy spectrum.
Let $\chi(k_{s}, \eta_{s}L_{s})$ represent the condition $k_{s} - \eta_{s}L_{s}- \frac{1}{2}=\nu$ where $s=a, b, c$, and~$\eta_{a}=\eta_{c}=1$ and $\eta_{b}=1/2$. The~map $\, \chi(k_{s}, \eta_{s}L_{s}) \, \Rightarrow  \, \chi(k_{s'}, \eta_{s'}L_{s'}) \,$ induces $E_{s}\, \Rightarrow \, E_{s'}$.

\[
\begin{array}{ccccccc}
& ~~~\chi(k_{c}, L_{c})&~~~~&&~~~~~~~ &E_{c}&\\
~~{}^{CA}\,\swarrow  &&\nwarrow \, {}^{BC}&~~\Rightarrow ~~&{}^{CA}\, \swarrow& &\nwarrow \, {}^{BC} \\
\chi(k_{a}, L_{a}) & \longrightarrow &\chi(k_{b}, \frac{1}{2}L_{b})  &~~~~~&E_{a}& \longrightarrow &E_{b}\\
~&{}^{AB}&~ && ~&{}^{AB}&~
\end{array}
\]

Finally, it must be mentioned that this triangular relation has been discussed in the context of so-called shape invariant potentials in supersymmetric quantum mechanics~\cite{CooperKhareSukhatme}. It may also be worth pointing out that the three systems share the $SU(1,1)$ dynamical group~\cite{IKG1992,IJ1994}.

\end{document}